\documentclass[a4paper,11pt]{article}
\usepackage{jinstpub} 

\usepackage{lineno}
\usepackage[dvipsnames]{xcolor}
\usepackage{caption}

\usepackage{subcaption}
\usepackage{siunitx}
\usepackage{hyperref}
\usepackage{wrapfig}
\usepackage{amsmath}
\usepackage{nicefrac}

\usepackage{tikz}
\usepackage{tkz-euclide}

\title{The design and flight of the Payload for Ultrahigh Energy Observations (PUEO)}

\author[a,b]{Q.~Abarr}
\author[c]{S.~Agarwal}
\author[d,e,f]{J.~Alfaro}
\author[c]{S.~Ali}
\author[a,b]{P.~Allison}
\author[g]{J.~Alvarez-Mu\~niz}
\author[d,e,f]{T.~Anderson}
\author[h]{H.~Barnett}
\author[h]{A.~Basharina-Freshville}
\author[a,b]{J.~J.~Beatty}
\author[a,b]{L.~Beaufore}
\author[c]{D.~Z.~Besson}
\author[d,e,f]{M.~Betts}
\author[i]{R.~Bose}
\author[i]{D.~Braun}
\author[j]{B.~Chamanbahar}
\author[k]{P.~Chen}
\author[k]{Y.~Chen}
\author[l]{J.~M.~Clem}
\author[a,b]{T.~Coakley}
\author[a,b]{A.~Connolly}
\author[c]{K.~Couberly}
\author[m]{L.~Cremonesi}
\author[d,e]{A.~Cummings}
\author[a,b]{P.~Dasgupta}
\author[n,o]{C.~Deaconu}
\author[a,b]{J.~Flaherty}
\author[h]{C.~Godden}
\author[p]{P.~W.~Gorham}
\author[c]{C.~Hornhuber}
\author[i]{J.~Hoffman}
\author[a,b]{K.~Hughes}
\author[q]{A.~Hynous}
\author[p]{A.~Jung}
\author[o,r]{A.~M.~Kofman}
\author[d,e]{Y.~Ku}
\author[d,e]{D.~Kullgren}
\author[k]{C.-Y.~Kuo}
\author[a,b]{P.~Linton}
\author[i]{L.~Lisalda}
\author[s]{T.~C.~Liu}
\author[a,b]{W.~Luszczak}
\author[j,o,t]{S.~C.~Mackey}
\author[j,o,t]{Z.~Martin}
\author[j,o,t,1]{K.~McBride\note{Corresponding author.}}
\author[p]{C.~Miki}
\author[k]{J.~Nam}
\author[o]{D.~A.~Newmark}
\author[h]{R.~J.~Nichol}
\author[l]{A.~Novikov}
\author[c]{A.~Nozdrina}
\author[n,o,t]{E.~Oberla}
\author[h]{H.~Pumphrey}
\author[i]{D.~Radomski}
\author[i]{B.~F.~Rauch}
\author[j,o,t]{R.~Scrandis}
\author[l]{D.~Seckel}
\author[c]{M.~F.~H.~Seikh}
\author[k,u]{J.~Shiao}
\author[i]{G.~Simburger}
\author[f]{J.~Tutt}
\author[n,j,o,t]{A.~G.~Vieregg}
\author[k]{S.-H.~Wang}
\author[d,e,f]{D.~Washington}
\author[d,e,f]{C.~Welling}
\author[j,o,t]{P.~Windischhofer}
\author[d,e,f]{S.~A.~Wissel}
\author[h]{C.~Xie}
\author[a,b]{J.~Yao}
\author[c]{R.~Young}
\author[g]{E.~Zas}
\author[d,e]{A.~Zeolla}

\affiliation[a]{Department of Physics, Ohio State University, Columbus, OH 43210}
\affiliation[b]{Center for Cosmology and Particle Astrophysics, Ohio State University, Columbus, OH 43210}
\affiliation[c]{Department of Physics and Astronomy, University of Kansas, Lawrence, KS 66045}
\affiliation[d]{Pennsylvania State Physics Department, University Park, PA 16802}
\affiliation[e]{Institute for Gravitation and the Cosmos, University Park, PA 16802}
\affiliation[f]{Pennsylvania State Astronomy and Astrophysics Department, University Park, PA 16802}
\affiliation[g]{Instituto Galego de F\'isica de Altas Enerx\'ias (IGFAE), Universidade de Santiago de Compostela, 15782 Santiago de Compostela, Spain}
\affiliation[h]{Department of Physics and Astronomy, University College London, London, United Kingdom}
\affiliation[i]{Department of Physics, McDonnell Center for the Space Sciences, Washington University in St. Louis, MO 63130}
\affiliation[j]{Department of Physics, University of Chicago, Chicago, IL 60637}
\affiliation[k]{Department of Physics, Grad. Institute of Astrophysics, Leung Center for Cosmology and Particle Astrophysics, National Taiwan University, Taipei, Taiwan}
\affiliation[l]{Department of Physics, University of Delaware, Newark, DE 19716}
\affiliation[m]{Department of Physics, Imperial College London, London, United Kingdom}
\affiliation[n]{Department of Astronomy and Astrophysics, University of Chicago, Chicago, IL 60637}
\affiliation[o]{Kavli Institute for Cosmological Physics, University of Chicago, Chicago, IL 60637}
\affiliation[p]{Department of Physics and Astronomy, University of Hawaii, Manoa, HI 96822}
\affiliation[q]{NASA Wallops Flight Facility, Wallops Island, VA 23337}
\affiliation[r]{Fermi National Accelerator Laboratory, Batavia, IL 60510}
\affiliation[s]{Department of Applied Physics, National Pingtung University, Pingtung City, Taiwan}
\affiliation[t]{Enrico Fermi Institute, University of Chicago, Chicago, IL 60637}
\affiliation[u]{National Nano Device Laboratories, Hsinchu 300, Taiwan}

\collaboration{PUEO Collaboration}

\emailAdd{kmcbride@uchicago.edu}

\DeclareUnicodeCharacter{2009}{\,} 

\abstract{Sources of cosmic rays at ultrahigh energies (UHE, $\gtrsim$ 1 EeV) are anticipated to give rise to neutrinos at similar energies by production via propagation interactions either within the source region or on cosmological scales. Measurements of the neutrino flux, flavor, and energy provide significant insight into those underlying processes governing UHE cosmic ray acceleration and propagation. Detecting neutrinos at these energies requires immense detector volumes due to their low predicted flux. We present the balloon-borne mission PUEO (Payload for Ultrahigh Energy Observations) which is an antenna array designed to detect the Askaryan emission from UHE neutrino interactions in the Antarctic ice. This paper reports on the design considerations and implementations for constructing PUEO, from the design parameters to the as-built validations.}

\divide\pdfpxdimen by 300 
\begin{document}
\maketitle
\flushbottom
\section{Introduction}
The Payload for Ultrahigh Energy Observations (PUEO) is a NASA Astrophysics Pioneers long-duration balloon mission designed to study ultrahigh-energy (UHE) particles ($\gtrsim$ 1 EeV), UHE neutrinos in particular, from astrophysical and cosmogenic sources \cite{Abarr_2021}. PUEO completed its 23-day flight over the Antarctic continent in the 2025--2026 austral summer aboard a $34.43$-Heavy MCF (million cubic feet) high-altitude balloon. After briefly summarizing the case for radio searches for UHE neutrinos, this paper is organized as follows. The description of event backgrounds and neutrino signals is discussed in section \ref{sec:RFE}. The science instrument and gondola design including antenna validation and description of subsystems is presented in section \ref{sec:instrument}. Section \ref{sec:telem_hi} presents the telemetry pathways and section \ref{sec:preflight} describes the preflight validation tests, followed by the calibration systems for in-flight performance validation in section \ref{sec:calibration_hi}. Finally, section \ref{sec:flight_and_perform} presents the flight path and a first look at the quality of the flight data. 

\subsection{UHE Neutrinos}

UHE neutrinos serve as a unique probe of the highest energy phenomena in the Universe. As messengers, cosmic rays are susceptible to deflection in magnetic fields and to interactions with matter and radiation in the interstellar medium \cite{NuAnnualReview2000, GAISSER1995173}. Gamma rays are not susceptible to magnetic deflection, but at TeV energies and above, their horizon is limited by their $\lesssim$ Mpc attenuation lengths due to interactions with extragalactic background light and the cosmic microwave background \cite{2025dawnneutrinoastronomynew}.
It is expected that UHE neutrinos travel unimpeded by these effects and are capable of pointing back to their sources throughout the Universe. Instead, the difficulty in their use as messengers lies in their \textit{predicted} extremely low flux at ultrahigh energies, and their very low interaction cross section in a detector target, compared to hadrons or photons. This necessitates the use of large detector volumes on the order of hundreds of cubic kilometers water equivalent to achieve measurements of the UHE neutrino flux on reasonable timescales. 

IceCube has measured the neutrino spectrum up to PeV energies, but, due to the constraints of optical detection, its sensitivity is limited in the EeV energy range \cite{Askebjer:97, IceCubeLimit}. This range includes the expected energies of cosmogenic neutrinos -- neutrinos produced via interactions between UHE cosmic rays (UHECRs) and the cosmic microwave background \cite{GZK1, GZK2, CosmogenicNeutrinoEnergies}. The flux of astrophysical neutrinos, those produced directly from sources, is predicted to extend to these energies as well \cite{Kotera:2011cp}. 

\subsection{Radio Detection of UHE Neutrinos}

Optical neutrino detector experiments, such as IceCube and KM3Net, detect the Cherenkov light emitted by the particle showers resulting from neutrino interactions in their instrumented media. Rather than optical observations, PUEO detects the radio emission from these particle showers, which generated by the Askaryan effect \cite{askaryan1961excess}: neutrino interactions in the ice produce relativistic particle cascades that develop a negative charge excess, resulting in coherent broadband radio signatures emitted at and near the Cherenkov angle. The Askaryan effect has been observed in accelerator experiments using various forms of natural media, including ice, as the propagation medium \cite{beam1, beam2, beam3}. 

Unlike optical light, radio waves at frequencies below $\sim 1.2$~GHz have kilometer-scale attenuation lengths in suitably cold polar ice \cite{barwick2005south}, and this enables significantly larger detector volumes. Radio monitoring for Askaryan emission has been used in-ice by the RICE \cite{RICE}, ARA \cite{ARA}, ARIANNA \cite{ARIANNA}, and RNO-G \cite{RNOG} projects. Monitoring the Antarctic ice sheet for these radio signatures has also been achieved from a long-duration balloon platform by four flights of the ANITA project \cite{ANITA4}. 

Balloon-borne radio detectors trade limited flight times for the ability to observe large instantaneous volumes, $\mathcal{O}(1)$~million km$^{3}$, of ice from above. This large instantaneous volume necessitates low triggering thresholds, as signals can originate far from the instrument. PUEO and its predecessor ANITA were able to observe such large volumes by using large cylindrically-arranged broadband horn antennas, pointed in a manner that their fields-of-regard covered nearly the entire ice sheet below the payload from a suborbital vantage. The antennas had significant overlap in their views with many of their neighbors, which enabled interferometric imaging and geolocation of any impulse observed, to accuracies of fractions of a degree, far better than the 40$^\circ$ diameter beams of the individual antennas. 

\subsection{Key Technical Advances}

While the PUEO instrument inherits many of the proven system designs of the ANITA balloon program, it achieves significant improvement in sensitivity through some key technical advances. One such advance is the implementation of an interferometric phased-array trigger, which achieves lower triggering thresholds as required for optimal balloon-borne observations. This trigger is implemented on commercially available Xilinx Radio-Frequency System-on-Chip (RFSoC) boards, and uses highly optimized filtering and beamforming logic to achieve maximal sensitivity.

Another key design choice is restricting the Main Instrument (MI) antenna frequency range to 300-1200 MHz, which decreases the antenna aperture area by roughly a factor of two compared to ANITA. This allows PUEO to use twice the number of antennas and avoids continuous wave (CW) interference from anthropogenic sources, found by ANITA to limit the usefulness of the 200-300 MHz band \cite{A4Notches, ANITA4}. 

Another novel addition is the drop-down low frequency (LF) instrument dedicated to frequencies in the 50-500 MHz range. The LF instrument enhances PUEO's ability to measure cosmic-ray or tau lepton decay induced air-showers, both shown in figure~\ref{fig:radiosig}. With these and other innovations outlined in further detail in this paper, PUEO's design achieves higher sensitivities than previous missions within the constraints of the high-altitude balloon platform.

\section{Radio Frequency Events}
\label{sec:RFE}
\subsection{Classes of Signal}

The central scientific methodology of PUEO is the detection of the Askaryan emission from UHE neutrino interactions with matter \cite{askaryan1961excess}. Through either a charged or neutral current interaction, the neutrino collides with the ice and initiates a particle shower which develops an overall negative charge excess. This charged shower, taking the form of a compact bunch several cm in diameter, radiates coherently in radio wavelengths below $\sim 1$~GHz, visible to both PUEO instruments. Askaryan emission has been detected in the lab and in nature as a broadband and linearly polarized signal (along the Cherenkov cone)  \cite{beam1, beam2, beam3, ARA_CR, Auger_Askaryan}. The PUEO instruments are expected to primarily detect vertically-polarized Askaryan signals, as the top of the Cherenkov cone is more visible relative to the antennas floating above the ice.

\begin{figure}
\centering
\includegraphics[width=0.99\textwidth]{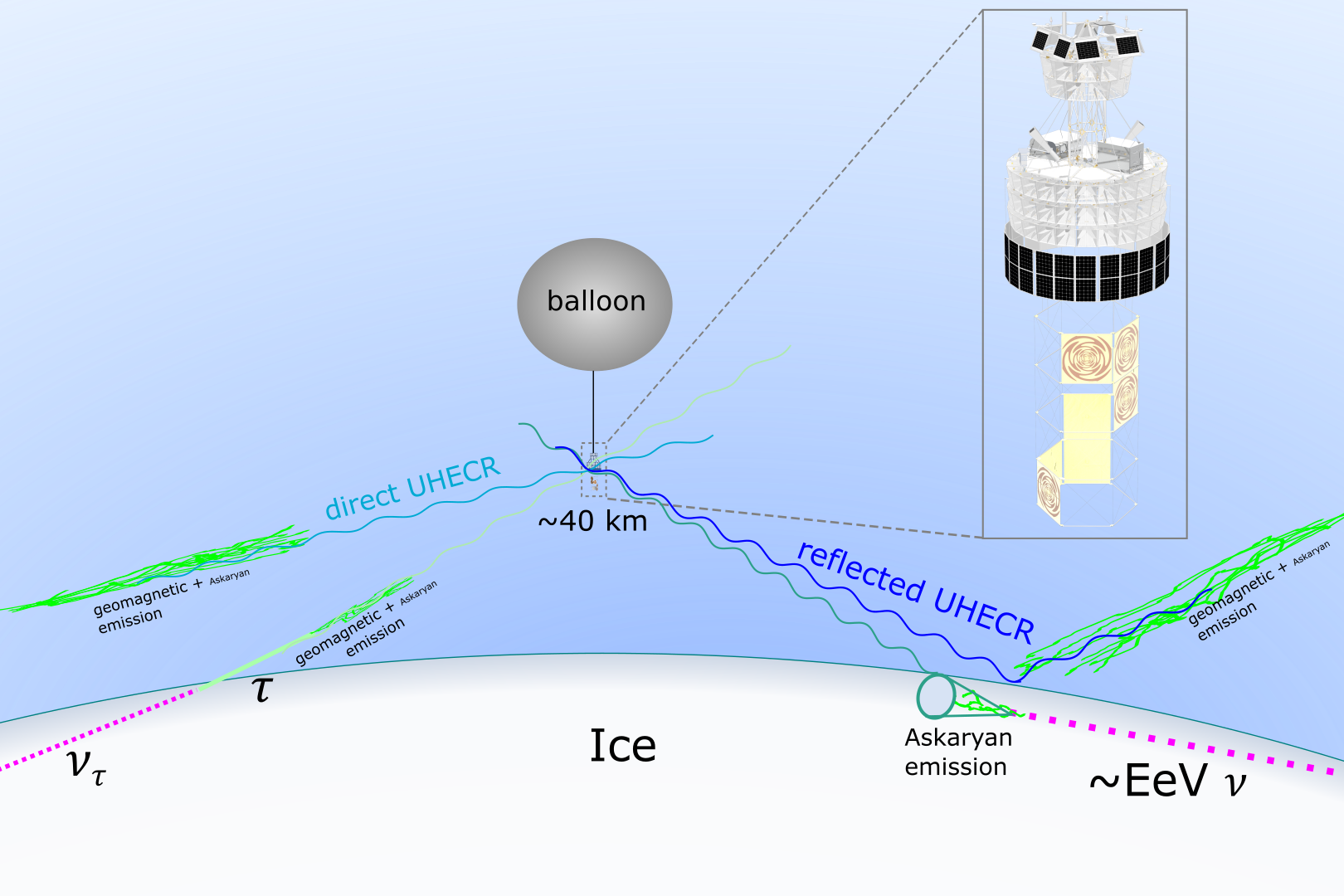}
\caption{Classes of particle-shower induced radio signatures observable by PUEO. UHECR showers can be observed through either the direct (left) or reflected (right) channels of geomagnetic and Askaryan emission in the air. Neutrinos can be detected via the channel of tau-decay on the left or in-ice Askaryan emission on the right. An inset diagram of the payload as flown is shown on the top right.}
\label{fig:radiosig}
\end{figure}

The PUEO instruments are also sensitive to extensive air showers (EAS) produced via two mechanisms. First, UHECRs interact in the upper atmosphere and generate geomagnetic radio emission \cite{Huege_2005}. These signals can be seen either directly from the showers with steeply inclined angles, or after a reflection of the radio signal off the surface of the Antarctic ice. The second mechanism for EAS production should be detectable from tau neutrino charged-current interactions, which would produce a tau lepton that decays in the atmosphere, and its EAS is again detectable via the geomagnetic radio emission \cite{EAS_Feng}. These are lower in frequency than the Askaryan channel but are expected to again be visible in both instruments. These different types of Askaryan and EAS signals are shown in figure~\ref{fig:radiosig}.

\subsection{Classes of Background Interference}

It is not feasible to continuously record radio signals from all of the PUEO antennas; PUEO must therefore use a {\it trigger}, based on active short-duration sensing of the ambient radio-frequency (RF) fields, to then enable recording of the signals for a period of time (a fraction of a microsecond) sufficient to capture the expected signal, when the trigger is activated. PUEO's trigger is sensitive to impulsive fluctuations in the ambient electric fields at levels close to the radio noise floor, which is set by thermal noise.
Thermal noise fluctuations within the ice itself are expected to make up the vast majority of all triggered events. The power within thermal noise events is generally spread evenly throughout the waveform, causing these events to have low signal-to-noise ratios (SNRs). These events do not tend to point back in any specific direction and are typically removed with a classifier like a linear discriminant.

RF triggers may also be caused by anthropogenic noise, but such noise can also produce a continuous level of contamination for all events. We group anthropogenic noise into three broad categories. The first is satellite noise, generally limited to specific frequencies below 400 MHz. Satellite signals are typically identified by their narrow communication bandwidths, and various filtering strategies are used to clean events without removing them from consideration as neutrino candidates \cite{ANITA3}. The second category is impulsive noise from large Antarctic bases, including the stations at the South Pole and McMurdo. These stations reduce the sensitivity of the RF trigger thresholds for the channels facing them, and the triggering system includes directional-masking techniques to allow the instrument to preserve its RF trigger on the quieter areas of ice. The third category is noise from smaller, isolated, and often transient anthropogenic sources, for example, aircraft in flight or vehicles on the ice surface. These sources create a smaller number of signals which must be removed through careful classification efforts.
In all cases, the ability to reconstruct anthropogenic events will determine which classification methods are most viable for each type of background.

\section{Instrument Design}
\label{sec:instrument}
\subsection{Design Considerations}
\label{sec:design}
The primary science goal is to detect and characterize the UHE neutrino spectra from both specific high-energy sources and from the expected diffuse cosmogenic flux. PUEO's design is optimized for exposure to this neutrino spectrum, and to measure the characteristic radio frequency signals from their interactions in the Antarctic ice with sufficient detail to reconstruct the properties of each originating neutrino.

\subsubsection{Objectives}

Increasing total exposure gives PUEO higher odds of observing neutrinos and, in the case of a non-detection, allows for more stringent limits to be set on the neutrino flux at ultrahigh energies. The two factors contributing to exposure are the detector acceptance (the combination of effective area and solid angle), and the experiment's livetime fraction, computed as the ratio of time when the instrument is alive and able to trigger at full efficiency, to the total flight time available. High acceptance requires an antenna array which views the ice sheet below out to the horizon. Acceptance can be increased by improving the instrument efficiency at lower SNR, which allows radio signals that are observed further off the Cherenkov cone to still be detected, thus increasing the acceptance solid angle. Higher efficiencies at lower SNRs also raise sensitivity at lower neutrino energies.

The anthropogenic noise discussed above negatively impacts acceptance by requiring either analysis stage geometric cuts in areas with high event density or by raising the SNR triggering threshold to protect the data pipeline from saturation; as such, resistance to anthropogenic noise is another critical design consideration. Similarly, designing an instrument with high pointing precision allows increased efficiency of geometric cuts of the noisy regions, and provides improved reconstruction of the original event in the case of a neutrino detection. 

In addition to the objectives above, PUEO's design is guided by consideration of the measurements needed to capture the properties of any detected neutrinos. RF polarization measurements are needed to reconstruct the class and direction of observed events. These reconstructions also require high-quality navigation information for the payload itself in order to distinguish between elevations of above- and below-horizon events and directions in azimuth. The measured frequency content is used to classify and reconstruct events and motivates instrument designs with greater bandwidths. 

\subsubsection{Additional Requirements}

A long duration and high-altitude balloon is an excellent platform for reaching these objectives, but it presents practical limitations on the instrument design. Notable requirements include keeping within the mass budget and physical envelope. The environmental conditions of operating in the upper atmosphere pose their own design requirements. For example, the payload must produce all the power required for operation from solar arrays during flight. The instruments must be able to maintain proper operating temperature while in direct sunlight and near-vacuum conditions, necessitating a design that radiates away its excess heat, operates down to cold temperatures, and toggles system power reliably as a contingency. Since the PUEO mission objective is to search for signals in the radio frequency regime, there are additional requirements for subsystems on the payload to maintain an RF-quiet environment, motivating the use of RF-tight Faraday enclosures for all electronics.

\subsection{Top-Level Design}
The requirements outlined in section \ref{sec:design} motivate the hardware-level design specifications highlighted below. Data obtained with the ANITA project indicated that a significant amount of anthropogenic noise dominates the lower-end of the Askaryan emission bandwidth, near $300$~MHz \cite{A4Notches}. ANITA and PUEO both utilized dual-polarized (VPol or vertical polarization and HPol or horizontal polarization) quad-ridged horn antennas due to their excellent impulse characteristic, relatively high gain, broadband response, and efficient packing characteristics in array form. These antennas are canted down by $10^{\circ}$ as the neutrino-induced radio signals should be coming from below the horizon. It was determined that excluding the lower-frequencies from the MI trigger would improve the sensitivity to simulated neutrino events in this noise environment. This is, in part, because raising the lower end of the frequency band reduces the required antenna face diameter to approximately $0.64$~m (rather than $0.96$~m on ANITA), allowing more antennas to fit within the array pattern allowed by the geometric constraints of the payload. 

PUEO uses two unique antenna arrays, the MI and the LF, where the LF antenna bandwidth overlaps with the MI but extends down to $50$~MHz to ensure the frequency content of the Askaryan events is still measured. The MI antennas are distributed at the maximum possible vertical separation, where the launch vehicle's hook-height constraint limits the maximum phase-center displacement vertically to be $\sim20$~ft between the top-most and bottom-most antennas. Farther spacing between antennas leads to higher precision reconstruction from timing delays. The LF instrument antennas require a greater size and so are stowed for launch. The payload's full hook-height is $\sim61$~ft when including the deployment of the LF. 

All electronics systems are housed inside RF-sealed (in most cases, using gasketing) metal Faraday enclosures that utilize shielded cables for input, output, and power. Most of these housings are installed inside of a main system enclosure called the Main Instrument Enclosure (MIE). The design philosophy aims to mitigate triggering from payload electronics noise, or Electromagnetic Interference (EMI, also called radio-frequency interference or RFI), that emits in the band of the MI antennas. To the extent possible, redundant interfaces, navigation systems, and storage disks are implemented to reduce the risk to achieving the science goals. 

\begin{figure}
    \centering
    \includegraphics[width=0.99\linewidth]{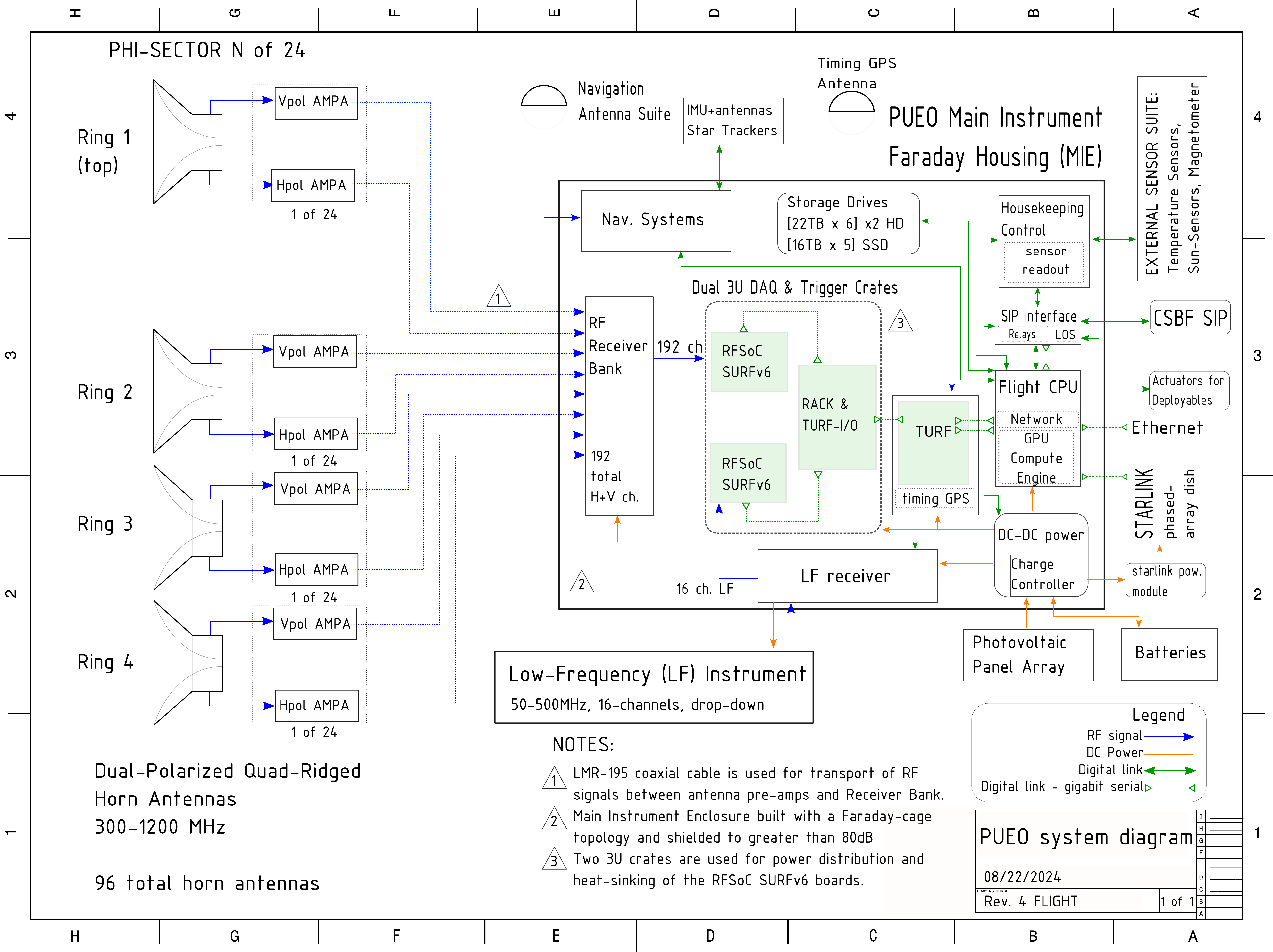}
    \caption{The system diagram of PUEO. RF signals are measured by the MI antennas on the left and by the LF antennas on the bottom. The diagram shows signal amplification and filtering before digitization and triggering. Most of the systems are inside the MIE, Faraday housing as shown in the center right. }
    \label{fig:systemdiagram}
\end{figure}

Figure~\ref{fig:systemdiagram} shows the system diagram, 
where the MIE houses the majority of the subsystems such as the MI and LF receivers, the Data Acquisition (DAQ) and flight computer, etc. The RF received by the MI and LF antenna arrays is amplified, filtered, and then digitized at the DAQ inside the MIE. The flight computer interfaces with the DAQ to command tunable parameters and record the data. This computer also records the navigation information and monitors the housekeeping of the entire payload, including power state of the subsystems, and writes to the storage disks. The telemetry options provide live commanding, such as firing the actuators, which enabled the successful deployment of the LF during the flight. In this scheme, the electronics noise is contained significantly in one Faraday cage, the MIE, and there is a clear division of the interfaces as external and internal, where external cables have strict requirements to respect the integrity of the Faraday cage.

\subsection{Gondola Structure}
\begin{figure}[htbp]
\centering
\includegraphics[scale = 0.15]{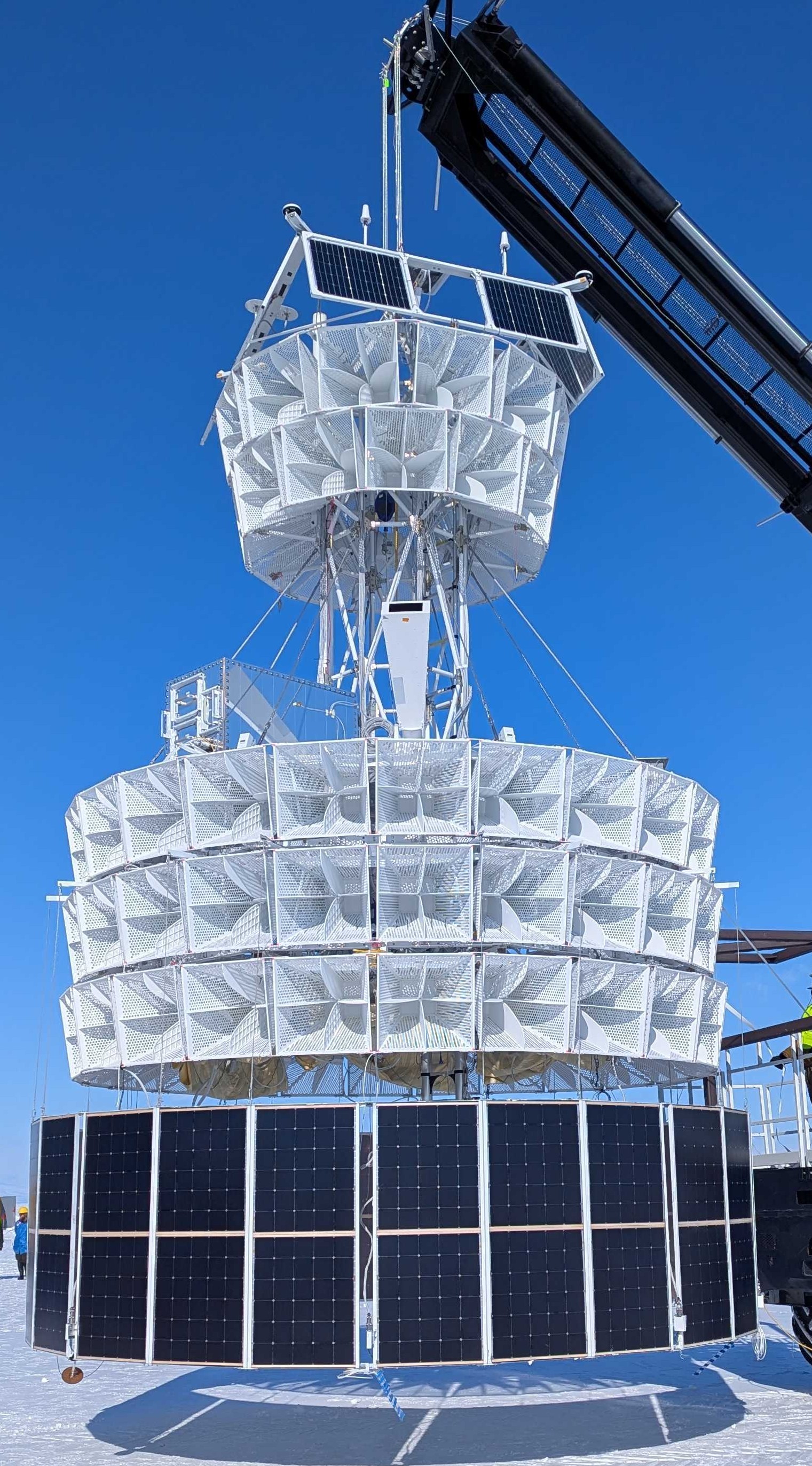}
\vskip 0cm
\caption{The payload on the launch vehicle with solar panels deployed. The MIE (visible on the center-left) and the NASA Support Instrumentation Package (SIP) enclosure (not visible) are on the deck level along with battery boxes and navigation systems. The payload is divided into 4 rings, with the top ring of antennas above the deck. Above the top ring of PUEO antennas is a suite of GPS and telemetry antennas. In this configuration, during climb out, the LF instrument is still stowed.}
\label{fig:payload}
\end{figure}

\begin{table}
    \centering
    \caption{PUEO Science Mass Table.}
    \begin{tabular}{| c | c |}
        \hline
        \textbf{Instrument Subsystem} & \textbf{Mass (lbs)}\\
        \hline
        \hline
        Gondola Structure & 1400\\
        \hline
        PUEO DAQ and SIP  & 1210\\
        \hline
        MI Antennas & 1056\\
        \hline
         External Power and Actuator System & 425\\
        \hline
        Cabling  & 300\\
        \hline
        Antenna-mounted front-end electronics & 224\\
        \hline
        External Navigation Suite & 125\\
        \hline
        LF Array & 115\\
        \hline
        \hline
        \textbf{Total} & \textbf{4855}\\
        \hline
    \end{tabular}
    \label{tab:mass_table}
\end{table}

The PUEO gondola design is based on the flight-proven ANITA gondola structure, updated to accommodate more instrumentation and deployable systems. The balloon used was an Aerostar 34.43-3H-29 34 MCF zero pressure balloon with a maximum capacity of 8000\,lbs, which drove total payload mass. PUEO comprises 96 quad ridge horn antennas and their front-end electronics, a photovoltaic (PV) array, an array of eight sinuous antennas (LF array), various navigation and communication systems, the PUEO data acquisition system, and the NASA Support Instrumentation Package (SIP). To maximize science on a mass and launch envelope constrained gondola, PUEO employs deployable PV arrays and an LF instrument separately actuated via Line of Sight (LOS) commanding. The design philosophy comprises three primary concepts: Structural integrity, staging and mobility, and recoverability. The gondola structure is designed to the NASA gondola engineering requirements directive 820-PG-8700.0.1 which ensures that the structure is able to endure multiple assembly and disassembly operations, truck, vessel, and flight transport, mobilization, launch, and termination. Hoist height limitations and high bay door openings at both the Columbia Scientific Balloon Facility (CSBF) in Palestine, TX and Payload Building 2 at the Long Duration Ballooning Facility (LDB) in Antarctica are additional factors driving the design. Furthermore, due to its size, the gondola is designed to be safe enough for personnel to climb and work on during assembly and field recovery.  Finally, the structure incorporates pins and a minimal variety of fastener types to allow for quick setup and teardown into subassemblies which are designed to fit into United States Antarctic Program (USAP) fixed-wing assets and be within typical allowable cabin load limits. A detailed breakdown of the payload weight distribution can be found in table \ref{tab:mass_table}.

\begin{figure}
    \centering
    \includegraphics[width=0.99\linewidth]{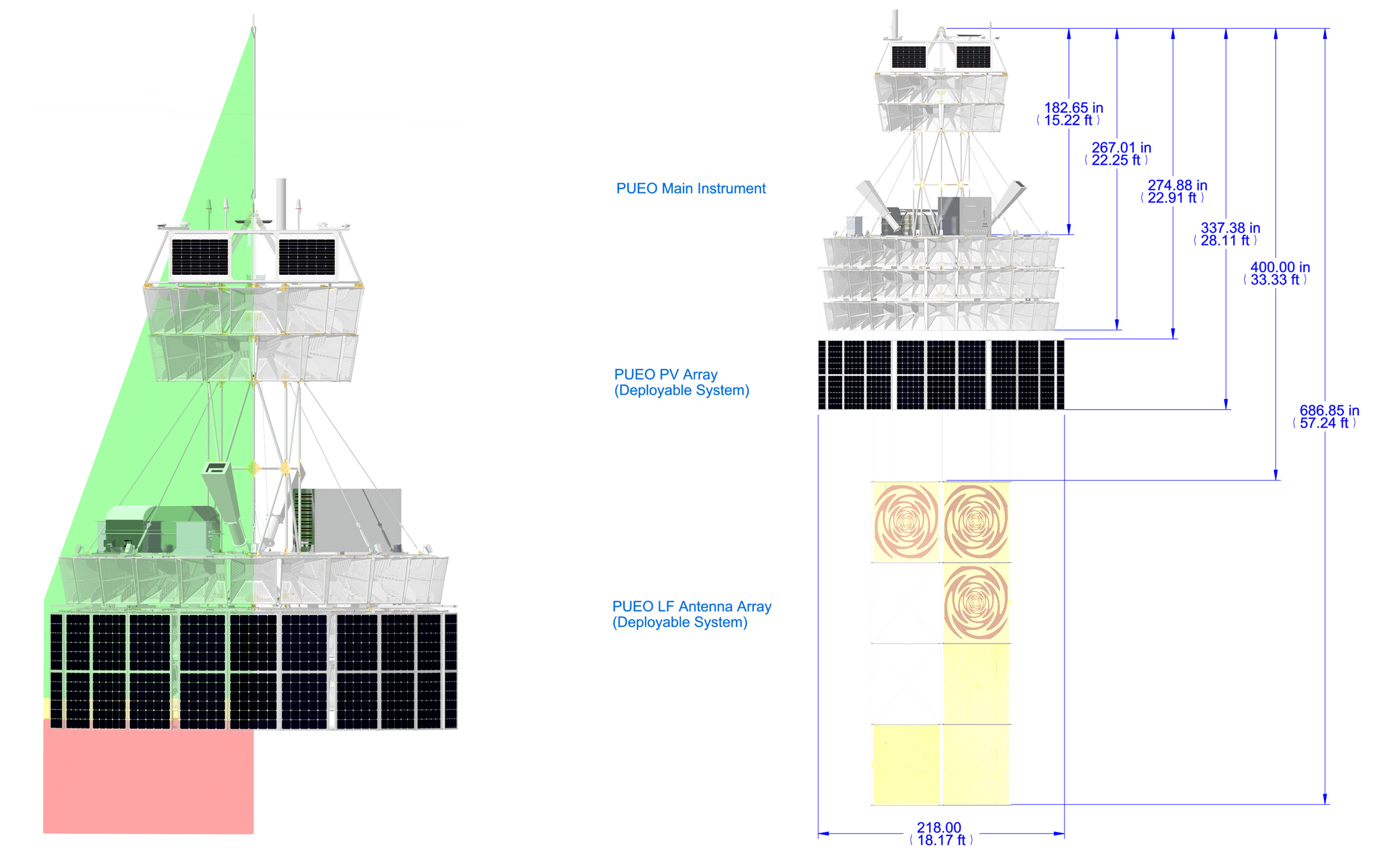}
    \caption{Left: PUEO in its launch configuration, with its PV array and  LF instrument stowed to fit within the launch vehicle's prescribed launch envelope. The green region indicates areas with safe clearance from the launch vehicle, the yellow and red areas denote areas with less than the required 6 and 5 feet of ground clearance, respectively. PUEO obtained waivers for the portions extending beyond the prescribed launch envelope. Right: PUEO in the fully deployed flight configuration.}
    \label{fig:gondola_dims}
\end{figure}

\begin{figure}[h!]
    
    \begin{subfigure}{0.5\textwidth}
        \centering\includegraphics[width=0.7\linewidth]{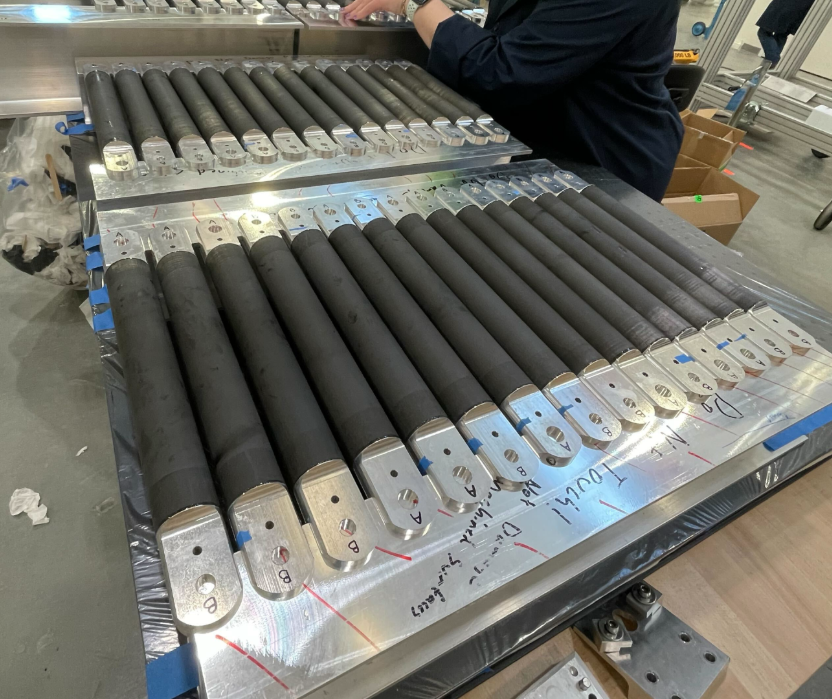} 
        \caption{Carbon fiber tubes curing}
        \label{fig:tube_asm2}
    \end{subfigure}
    \begin{subfigure}{0.5\textwidth}
        \centering\includegraphics[width=.85\linewidth]{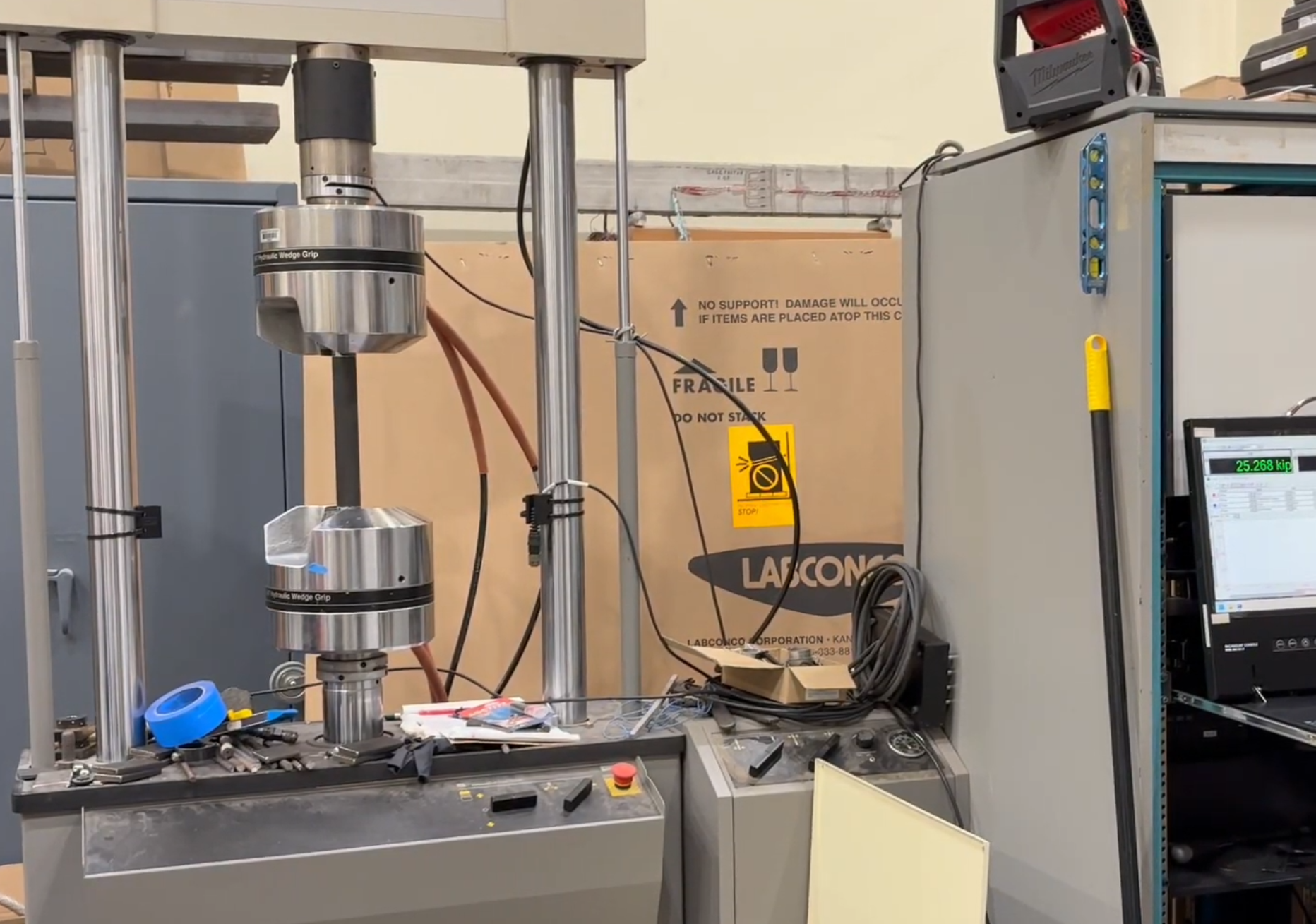}
        \caption{Tensile testing}
        \label{fig:kips}
    \end{subfigure}
    \caption{Carbon fiber structural tubes were bonded to their clevises then tested to ensure they met engineering requirements}
    \label{fig:tubes_tubes_tubes}
\end{figure}

The gondola primarily utilizes Rock West Composite 1.503” OD, .064” wall unidirectional carbon fiber tubes, Dragon Plate 0.875” OD, 0.0625” wall fiber roll-wrapped twill tubes, aluminum 7075-T6, aluminum 6061-T6, aluminum 6005-T5, and 6Al4V titanium. CNC-machined, threaded, and Iridited aluminum 7075-T6 clevises were bonded to the Rock West tubes using Loctite Hysol EA E-120HP epoxy, shown curing in figure~\ref{fig:tube_asm2}. Parallelism between opposing clevis ends was maintained by curing assembled tubes on parallel fly cut I-beams and second 0.501” clevis holes were machined last to ensure accurate hole-to-hole distances. Central column tube assembly plate hubs, plate clevises, and padeyes were machined out of aluminum 7075-T6 billet. Tubes are fastened to various other components utilizing 0.50" 6Al4V titanium pins for quick assembly and disassembly. To meet the structural integrity requirements, Finite Element Analysis (FEA) simulations were performed prior to the fabrication and tensile testing of sample components. These FEA simulations used the results of previous ANITA carbon fiber tube assembly tensile tests and mill testing certificates. The FEA simulations exceeded the requirements set in NASA directive 820-PG-8700.0.1, which prescribes the design load limits of structural flight hardware to be 8g vertical, 4g horizontal, and 4g at \ang{45}. Additionally, factors of safety (FoS) for structural metallic components must be greater than or equal to 1.4 and greater than or equal to 2 for bonded carbon fiber assemblies.

To ensure compliance, a representative subset of the epoxied carbon fiber support assemblies underwent a complete suite of ultimate tensile and proof testing to ensure the assemblies would meet the required specifications set by NASA Directive 820-PG-8700.0.1. Ultimate, or failure, tensile testing of the assemblies was conducted on an MTS 810 Material Test System outfitted with custom mating fixtures, shown in figure~\ref{fig:kips}. The nine samples were thermally cycled ten times from -60$^\circ$C to 60$^\circ$C. Three were cooled and tested at -60$^\circ$C, three were heated and tested at 60$^\circ$C, and three were tested at room temperature. Samples exceeded the minimum 12 kilopounds (kips), the strongest failing at 25 kips, for an FoS greater than 2. Proof testing was conducted on a larger subset of the epoxied carbon fiber support assemblies that underwent a single cycle in a thermal vacuum chamber that replicated flight-like conditions. Two samples of each of the 19 unique tube lengths were proof tested using an Instron machine outfitted with custom mating fixtures. All samples successfully exceeded the 6.6 kips proof stress requirement, equivalent to an FoS of 1.2. 
 
\subsubsection{Telemetry and Navigation Antennas}

Above the top ring of MI antennas (see figure \ref{fig:payload}), at the top of the payload are the GNSS (Global Navigation Satellite System) antennas and the majority of the telemetry antennas.
PUEO uses a total of eight GNSS antennas, detailed in table \ref{tab:gnss_antennas}, while CSBF uses four.

\begin{table}[h!]
\centering
  \caption{GNSS Antennas}
  \begin{tabular}{|c|c|c|}
    \hline
    PUEO subsystem & Antenna model & quantity \\
    \hline
    ABX-Two & Trimble LV59 & 3 \\
    \hline
    Boreas & Tallysman 33-VSP6337L & 2 \\
    \hline
    CPT7 & Novatel 42GNSSAV-XT-1 & 2 \\
    \hline
    TURF & Tallysman TW3885T & 1 \\
    \hline  
  \end{tabular}
  \label{tab:gnss_antennas}
\end{table}

CSBF-provided Iridium, omni TDRSS (Tracking and Data Relay Satellite System) and UHF (ultra high frequency) line-of-sight antennas are connected directly to the SIP. A Starlink Gen2 High-Performance dish and power-communication system, described in greater detail in section \ref{sec:starlink}, is also mounted near the top. The S-band line-of-sight transmit antenna is mounted near the bottom of the payload to a solar panel. 

\subsubsection{Photogrammetry}
Knowledge of the final payload geometry as built is required for accurate event reconstruction and background rejection during analysis. To get a first pass of the geometry necessary for coarse reconstruction, photogrammetry of the payload was performed prior to launch. The payload was suspended from a crane and rotated so that a photo could be taken with each column or phi-sector of antennas face-on. The camera was kept stationary throughout the process and allowed to cool to ambient temperature before taking photos to minimize any lens aberration effects. A total of 24 photos were used in the photogrammetry analysis, which were loaded into PhotoModeler\footnote{See \url{https://www.photomodeler.com}.} software. The corners of each antenna were manually identified using markings applied to the antennas during integration of the payload. The outer dimensions of the antennas provided a known scale for analysis. We achieved a root mean square point precision of 0.06 inches at a scale of 0.05 inches per pixel, as calculated from the average coordinate uncertainty across all images of all antennas. This exceeded our relative antenna position precision goals, set from the propagation of errors on the relative phases between neighboring antennas. The highest frequency sets the design goal, 1.2 GHz, such that approximately 0.2 inches is adequate to maintain system coherence. This resulted in four well-defined points per antenna, and surfaces of best-fit were used to determine the position and normal vector of each antenna's face. Overall, the antenna tilt direction is uniform and close to the design goal of \ang{10} with an average tilt angle of 9.89$\pm$\ang{0.04} below the horizontal axis, obtained from doing a least squares fit to each ring and averaging across the rings.

\subsection{Main Instrument Antennas and RF Front-End}
\subsubsection{Antennas}

\begin{figure}
    \centering
    \includegraphics[width=0.39\linewidth]{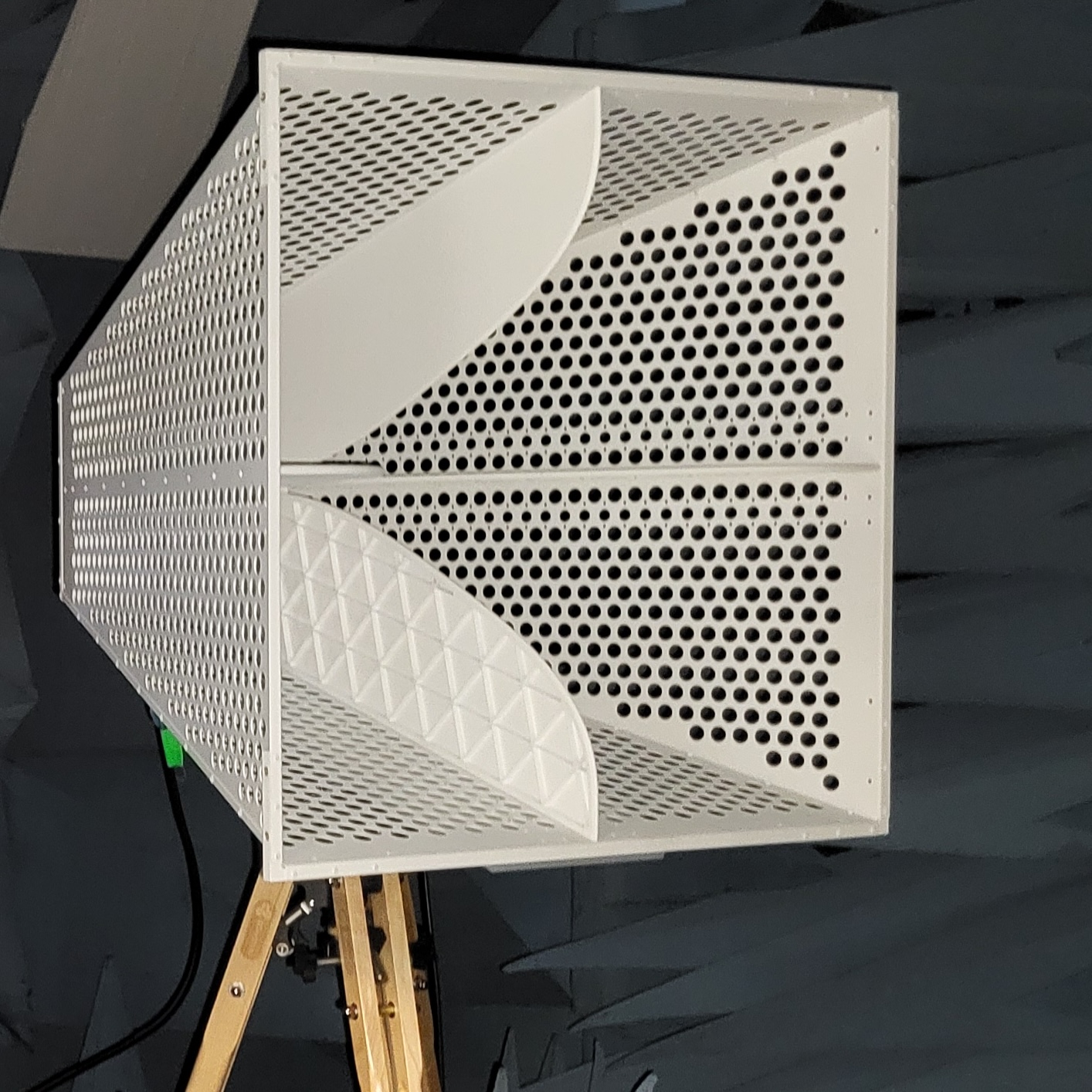}
    \includegraphics[width=0.59\linewidth]{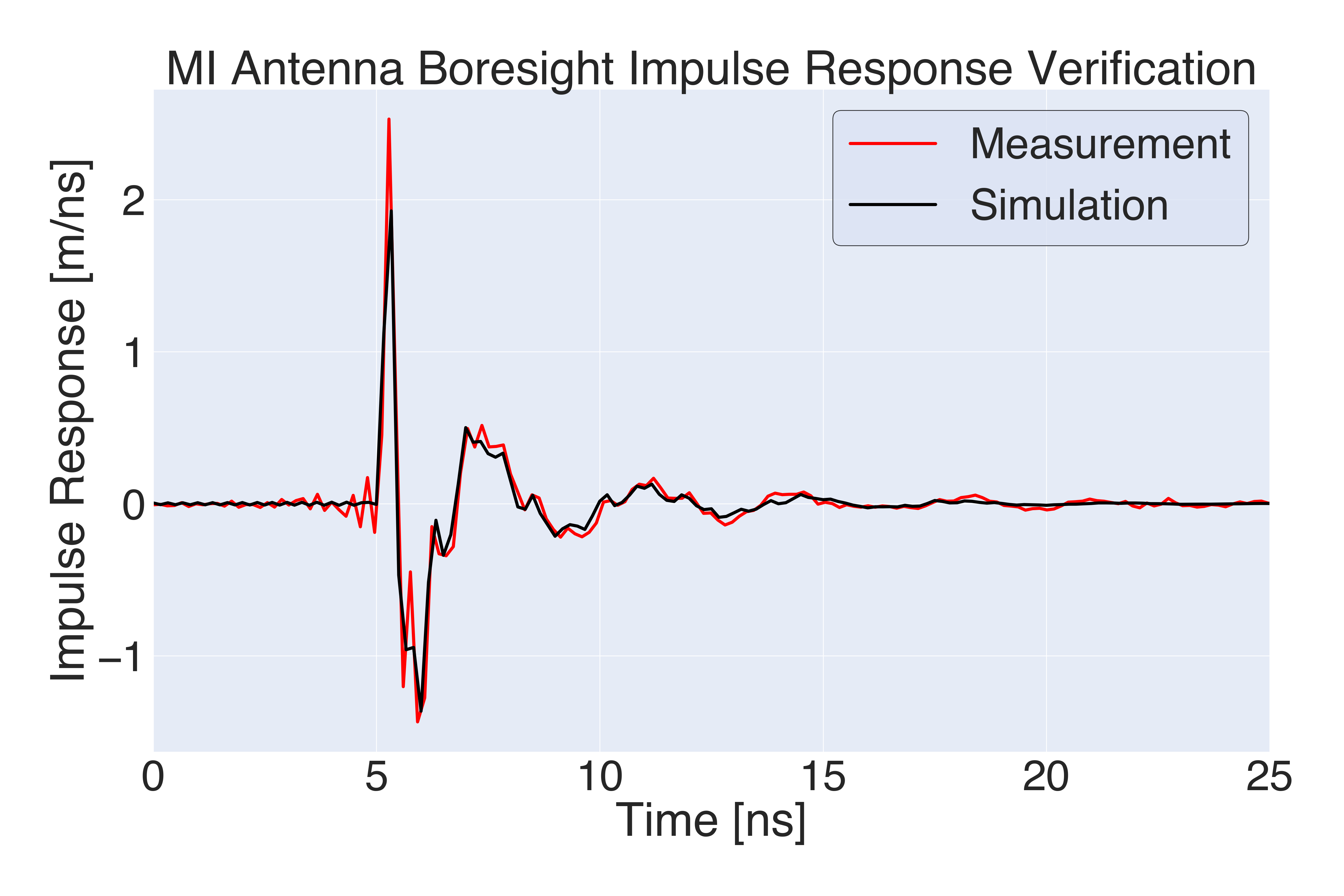}
    \caption{Left: A Toyon antenna being tested in an anechoic chamber. PUEO's antenna design inherits the quad-ridged horn design used on ANITA with light-weighting modifications and a length-wise reduction to 2/3 allowed by the higher low-frequency cutoff. Right: The antenna impulse response, as derived from processed time-domain pulse measurements compared to that from simulated incident E-fields in \texttt{XFdtd}. The non-causal features immediately prior to the pulse result from band-limited processing of the measurements. Note that the measurements are sampled at 20 GS/s and the simulated result is upsampled from 3 GS/s to match.}
    \label{fig:miant_impresp}
\end{figure}

\begin{figure}
    \centering
    \begin{subfigure}[b]{0.45\textwidth}
        \centering
        \begin{tikzpicture}[>=stealth,scale=1,line cap=round, bullet/.style={circle,inner sep=1.5pt,fill},font=\sffamily]

\path[use as bounding box] (-2.5,-3) rectangle (5.5,5);

\def\x{3}
\def\y{4}
\def\z{0} 


\def\B{3}

\def\S{0.5}

\def\L{4}

\pgfmathsetmacro{\rr}{sqrt(\x*\x+\y*\y+\z*\z)}
\pgfmathsetmacro{\rx}{\y/\rr}
\pgfmathsetmacro{\ry}{\z/\rr}
\pgfmathsetmacro{\rz}{\x/\rr}

\pgfmathsetmacro{\ax}{0}
\pgfmathsetmacro{\ay}{0}
\pgfmathsetmacro{\az}{1}

\pgfmathsetmacro{\ux}{\ay*\rz-\az*\ry}
\pgfmathsetmacro{\uy}{\az*\rx-\ax*\rz}
\pgfmathsetmacro{\uz}{\ax*\ry-\ay*\rx}

\pgfmathsetmacro{\unorm}{sqrt(\ux*\ux+\uy*\uy+\uz*\uz)}

\pgfmathsetmacro{\ux}{\ux/\unorm}
\pgfmathsetmacro{\uy}{\uy/\unorm}
\pgfmathsetmacro{\uz}{\uz/\unorm}

\pgfmathsetmacro{\vx}{\ry*\uz-\rz*\uy}
\pgfmathsetmacro{\vy}{\rz*\ux-\rx*\uz}
\pgfmathsetmacro{\vz}{\rx*\uy-\ry*\ux}


\coordinate (B1) at
({0.5*\B*(\ux+\vx)},
{0.5*\B*(\uy+\vy)},
{0.5*\B*(\uz+\vz)});

\coordinate (B2) at
({0.5*\B*(\ux-\vx)},
{0.5*\B*(\uy-\vy)},
{0.5*\B*(\uz-\vz)});

\coordinate (B3) at
({-0.5*\B*(\ux+\vx)},
{-0.5*\B*(\uy+\vy)},
{-0.5*\B*(\uz+\vz)});

\coordinate (B4) at
({-0.5*\B*(\ux-\vx)},
{-0.5*\B*(\uy-\vy)},
{-0.5*\B*(\uz-\vz)});


\coordinate (T1) at
({-\L*\rx + 0.5*\S*(\ux+\vx)},
{-\L*\ry + 0.5*\S*(\uy+\vy)},
{-\L*\rz + 0.5*\S*(\uz+\vz)});

\coordinate (T2) at
({-\L*\rx + 0.5*\S*(\ux-\vx)},
{-\L*\ry + 0.5*\S*(\uy-\vy)},
{-\L*\rz + 0.5*\S*(\uz-\vz)});

\coordinate (T3) at
({-\L*\rx - 0.5*\S*(\ux+\vx)},
{-\L*\ry - 0.5*\S*(\uy+\vy)},
{-\L*\rz - 0.5*\S*(\uz+\vz)});

\coordinate (T4) at
({-\L*\rx - 0.5*\S*(\ux-\vx)},
{-\L*\ry - 0.5*\S*(\uy-\vy)},
{-\L*\rz - 0.5*\S*(\uz-\vz)}); 

\coordinate (TC) at
({-\L*\rx},
{-\L*\ry},
{-\L*\rz});


\draw (B1)--(T1);
\draw (B2)--(T2);
\draw (B3)--(T3);
\draw (B4)--(T4);   

\path (B1)--(B2) coordinate[midway] (MB12); 
\path (B2)--(B3) coordinate[midway] (MB23); 
\path (B3)--(B4) coordinate[midway] (MB34); 
\path (B4)--(B1) coordinate[midway] (MB41); 

\path (T1)--(T2) coordinate[midway] (MT12); 
\path (T2)--(T3) coordinate[midway] (MT23); 
\path (T3)--(T4) coordinate[midway] (MT34); 
\path (T4)--(T1) coordinate[midway] (MT41);    

\path (B1)--(B2) coordinate[midway] (MB12); 
\path (B2)--(B3) coordinate[midway] (MB23); 
\path (B3)--(B4) coordinate[midway] (MB34); 
\path (B4)--(B1) coordinate[midway] (MB41); 
\path (T1)--(T2) coordinate[midway] (MT12); 
\path (T2)--(T3) coordinate[midway] (MT23); 
\path (T3)--(T4) coordinate[midway] (MT34); 
\path (T4)--(T1) coordinate[midway] (MT41); 
\coordinate (C1) at ($(MT12)!0.6!(0,0,0)$); 
\coordinate (C2) at ($(MT23)!0.6!(0,0,0)$); 
\coordinate (C3) at ($(MT34)!0.6!(0,0,0)$); 
\coordinate (C4) at ($(MT41)!0.6!(0,0,0)$); 


\draw (B1)--(T1);
\draw (B2)--(T2);
\draw (B3)--(T3);
\draw (B4)--(T4);   

\filldraw[ fill=white, draw=black, opacity=0.9,line width=0.01mm] ($(B2)$) -- ($(T2)$) -- ($(T3)$) -- ($(B3)$) -- cycle; 
\filldraw[fill=red!50,draw=black,line width=0.01mm] (MB23) -- (MT23) -- (C2) -- cycle;

\filldraw[fill=blue!25,draw=black,line width=0.01mm] (MB12) -- (MT12) -- (C1) -- cycle;
\filldraw[ fill=white, draw=black,opacity=0.9,line width=0.01mm] ($(B1)$) -- ($(T1)$) -- ($(T2)$) -- ($(B2)$) -- cycle;  

\filldraw[fill=white, draw=black, opacity=0.9,line width=0.01mm] ($(B3)$) -- ($(T3)$) -- ($(T4)$) -- ($(B4)$) -- cycle; 
\filldraw[fill=blue!25,draw=black,line width=0.01mm] (MB34) -- (MT34) -- (C3) -- cycle;

\draw[thin, dashed, black] (0,0,0)  -- (TC);
\filldraw[fill=red!50,draw=black,line width=0.01mm] (MB41) -- (MT41) -- (C4) -- cycle;  
\filldraw[ fill=white, draw=black, opacity=0.9,line width=0.01mm] ($(B4)$) -- ($(T4)$) -- ($(T1)$) -- ($(B1)$) -- cycle;      

\draw[draw=black, thin]
(B1)--(B2)--(B3)--(B4)--cycle;    

\draw[draw=black, thin]
(T4)--(T1)--(T2);  

\draw[draw=black,thin] (T4) -- (B4);
\draw[draw=black,thin] (T1) -- (B1);
\draw[draw=black,thin] (T2) -- (B2);


\def\vecmag{sqrt(\x*\x+\y*\y+\z*\z)}
\def\rtheta{acos(\z/\vecmag)}
\def\rphi{atan(\y/\x)}

\draw[thick, ->, green!50!black!100!] (0,0,0)  -- (\y,\z,\x)node[below]{$\hat{b}$};

\draw[thick, ->, blue] (0,0,0) -- ({-\vecmag/2*cos(\rphi)}, 0, {0+\vecmag/2*sin(\rphi)})node[below]{$\hat{n}_E$};

\draw[thick, ->, red] (0,0,0) -- ({0+\vecmag/2*cos(\rtheta)*sin(\rphi)},{0-\vecmag/2*sin(\rtheta)},{0+\vecmag/2*cos(\rtheta)*cos(\rphi)})node[right]{$\hat{n}_H$};    

\def\WaveAmp{1.25} 
\def\WaveLen{3} 
\def\WaveOffset{0} 
\def\WaveNum{1.5} 
\def\polVecOrigin{0.25} 

\def\polVecOrigin{0.25}
\draw[thin, ->, black] 
({((\WaveOffset+\polVecOrigin*\WaveLen)*\rx)},
{((\WaveOffset+\polVecOrigin*\WaveLen)*\ry)},
{((\WaveOffset+\polVecOrigin*\WaveLen)*\rz}) 
-- 
({((\WaveOffset+\polVecOrigin*\WaveLen)*\rx +\WaveAmp*sin(deg(2*pi*\polVecOrigin*\WaveLen/\WaveLen))*(-cos(\rtheta)*sin(\rphi)))},
{((\WaveOffset+\polVecOrigin*\WaveLen)*\ry +\WaveAmp*sin(deg(2*pi*\polVecOrigin*\WaveLen/\WaveLen))*(sin(\rtheta)))},
{((\WaveOffset+\polVecOrigin*\WaveLen)*\rz +\WaveAmp*sin(deg(2*pi*\polVecOrigin*\WaveLen/\WaveLen))*(-cos(\rtheta)*cos(\rphi)))}
)node[above]{$\hat{p}$};

\foreach \vecOrigin in {1,2,3,5,6,7,9,10,11} 
{
\def\thisVecOrigin{\vecOrigin/8}
\draw[thin, ->, black] 
({((\WaveOffset+\thisVecOrigin*\WaveLen)*\rx)},
{((\WaveOffset+\thisVecOrigin*\WaveLen)*\ry)},
{((\WaveOffset+\thisVecOrigin*\WaveLen)*\rz}) 
-- 
({((\WaveOffset+\thisVecOrigin*\WaveLen)*\rx +\WaveAmp*sin(deg(2*pi*\thisVecOrigin*\WaveLen/\WaveLen))*(-cos(\rtheta)*sin(\rphi)))},
{((\WaveOffset+\thisVecOrigin*\WaveLen)*\ry +\WaveAmp*sin(deg(2*pi*\thisVecOrigin*\WaveLen/\WaveLen))*(sin(\rtheta)))},
{((\WaveOffset+\thisVecOrigin*\WaveLen)*\rz +\WaveAmp*sin(deg(2*pi*\thisVecOrigin*\WaveLen/\WaveLen))*(-cos(\rtheta)*cos(\rphi)))}
);

}

\draw[ black, thick, samples=200, smooth, variable=\t, domain=0:\WaveNum*\WaveLen] 
plot({(\WaveOffset+\t)*\rx +\WaveAmp*sin(deg(2*pi*\t/\WaveLen))*(-cos(\rtheta)*sin(\rphi))},
{(\WaveOffset+\t)*\ry +\WaveAmp*sin(deg(2*pi*\t/\WaveLen))*(sin(\rtheta))},
{(\WaveOffset+\t)*\rz +\WaveAmp*sin(deg(2*pi*\t/\WaveLen))*(-cos(\rtheta)*cos(\rphi))});

\end{tikzpicture}
        \caption{Vpol}
    \end{subfigure}
    \begin{subfigure}[b]{0.45\textwidth}
        \centering
        \begin{tikzpicture}[>=stealth,scale=1,line cap=round, bullet/.style={circle,inner sep=1.5pt,fill},font=\sffamily]

\path[use as bounding box] (-3.5,-3) rectangle (5.5,5);

\def\x{3}
\def\y{4}
\def\z{0} 


\def\B{3}

\def\S{0.5}

\def\L{4}

\pgfmathsetmacro{\rr}{sqrt(\x*\x+\y*\y+\z*\z)}
\pgfmathsetmacro{\rx}{\y/\rr}
\pgfmathsetmacro{\ry}{\z/\rr}
\pgfmathsetmacro{\rz}{\x/\rr}

\pgfmathsetmacro{\ax}{0}
\pgfmathsetmacro{\ay}{0}
\pgfmathsetmacro{\az}{1}

\pgfmathsetmacro{\ux}{\ay*\rz-\az*\ry}
\pgfmathsetmacro{\uy}{\az*\rx-\ax*\rz}
\pgfmathsetmacro{\uz}{\ax*\ry-\ay*\rx}

\pgfmathsetmacro{\unorm}{sqrt(\ux*\ux+\uy*\uy+\uz*\uz)}

\pgfmathsetmacro{\ux}{\ux/\unorm}
\pgfmathsetmacro{\uy}{\uy/\unorm}
\pgfmathsetmacro{\uz}{\uz/\unorm}

\pgfmathsetmacro{\vx}{\ry*\uz-\rz*\uy}
\pgfmathsetmacro{\vy}{\rz*\ux-\rx*\uz}
\pgfmathsetmacro{\vz}{\rx*\uy-\ry*\ux}


\coordinate (B1) at
({0.5*\B*(\ux+\vx)},
{0.5*\B*(\uy+\vy)},
{0.5*\B*(\uz+\vz)});

\coordinate (B2) at
({0.5*\B*(\ux-\vx)},
{0.5*\B*(\uy-\vy)},
{0.5*\B*(\uz-\vz)});

\coordinate (B3) at
({-0.5*\B*(\ux+\vx)},
{-0.5*\B*(\uy+\vy)},
{-0.5*\B*(\uz+\vz)});

\coordinate (B4) at
({-0.5*\B*(\ux-\vx)},
{-0.5*\B*(\uy-\vy)},
{-0.5*\B*(\uz-\vz)});


\coordinate (T1) at
({-\L*\rx + 0.5*\S*(\ux+\vx)},
{-\L*\ry + 0.5*\S*(\uy+\vy)},
{-\L*\rz + 0.5*\S*(\uz+\vz)});

\coordinate (T2) at
({-\L*\rx + 0.5*\S*(\ux-\vx)},
{-\L*\ry + 0.5*\S*(\uy-\vy)},
{-\L*\rz + 0.5*\S*(\uz-\vz)});

\coordinate (T3) at
({-\L*\rx - 0.5*\S*(\ux+\vx)},
{-\L*\ry - 0.5*\S*(\uy+\vy)},
{-\L*\rz - 0.5*\S*(\uz+\vz)});

\coordinate (T4) at
({-\L*\rx - 0.5*\S*(\ux-\vx)},
{-\L*\ry - 0.5*\S*(\uy-\vy)},
{-\L*\rz - 0.5*\S*(\uz-\vz)}); 

\coordinate (TC) at
({-\L*\rx},
{-\L*\ry},
{-\L*\rz});


\draw (B1)--(T1);
\draw (B2)--(T2);
\draw (B3)--(T3);
\draw (B4)--(T4);   

\path (B1)--(B2) coordinate[midway] (MB12); 
\path (B2)--(B3) coordinate[midway] (MB23); 
\path (B3)--(B4) coordinate[midway] (MB34); 
\path (B4)--(B1) coordinate[midway] (MB41); 

\path (T1)--(T2) coordinate[midway] (MT12); 
\path (T2)--(T3) coordinate[midway] (MT23); 
\path (T3)--(T4) coordinate[midway] (MT34); 
\path (T4)--(T1) coordinate[midway] (MT41);    

\path (B1)--(B2) coordinate[midway] (MB12); 
\path (B2)--(B3) coordinate[midway] (MB23); 
\path (B3)--(B4) coordinate[midway] (MB34); 
\path (B4)--(B1) coordinate[midway] (MB41); 
\path (T1)--(T2) coordinate[midway] (MT12); 
\path (T2)--(T3) coordinate[midway] (MT23); 
\path (T3)--(T4) coordinate[midway] (MT34); 
\path (T4)--(T1) coordinate[midway] (MT41); 
\coordinate (C1) at ($(MT12)!0.6!(0,0,0)$); 
\coordinate (C2) at ($(MT23)!0.6!(0,0,0)$); 
\coordinate (C3) at ($(MT34)!0.6!(0,0,0)$); 
\coordinate (C4) at ($(MT41)!0.6!(0,0,0)$);  

\filldraw[fill=white, draw=black, opacity=0.9,line width=0.01mm] ($(B2)$) -- ($(T2)$) -- ($(T3)$) -- ($(B3)$) -- cycle; 
\filldraw[fill=blue!25,draw=black,line width=0.01mm] (MB23) -- (MT23) -- (C2) -- cycle;

\filldraw[fill=red!50,draw=black,line width=0.01mm] (MB12) -- (MT12) -- (C1) -- cycle;
\filldraw[fill=white, draw=black,opacity=0.9,line width=0.01mm] ($(B1)$) -- ($(T1)$) -- ($(T2)$) -- ($(B2)$) -- cycle;  

\filldraw[fill=white, draw=black, opacity=0.9,line width=0.01mm] ($(B3)$) -- ($(T3)$) -- ($(T4)$) -- ($(B4)$) -- cycle; 
\filldraw[fill=red!50,draw=black,line width=0.01mm] (MB34) -- (MT34) -- (C3) -- cycle;

\draw[thin, dashed, black] (0,0,0)  -- (TC);
\filldraw[fill=blue!25,draw=black,line width=0.01mm] (MB41) -- (MT41) -- (C4) -- cycle; 
\filldraw[fill=white, draw=black, opacity=0.9,line width=0.01mm] ($(B4)$) -- ($(T4)$) -- ($(T1)$) -- ($(B1)$) -- cycle;      

\draw[draw=black, thin]
(B1)--(B2)--(B3)--(B4)--cycle;     

\draw[draw=black, thin]
(T4)--(T1)--(T2);  

\draw[draw=black,thin] (T4) -- (B4);
\draw[draw=black,thin] (T1) -- (B1);
\draw[draw=black,thin] (T2) -- (B2);


\def\vecmag{sqrt(\x*\x+\y*\y+\z*\z)}
\def\rtheta{acos(\z/\vecmag)}
\def\rphi{atan(\y/\x)}

\draw[thick, ->, green!50!black!100!] (0,0,0)  -- (\y,\z,\x)node[below]{$\hat{b}$};

\draw[thick, ->, blue] (0,0,0) -- ({0-\vecmag/2*cos(\rtheta)*sin(\rphi)},{0+\vecmag/2*sin(\rtheta)},{0-\vecmag/2*cos(\rtheta)*cos(\rphi)})node[right]{$\hat{n}_E$};


\draw[thick, ->, red] (0,0,0) -- ({-\vecmag/2*cos(\rphi)}, 0, {0+\vecmag/2*sin(\rphi)})node[below]{$\hat{n}_H$};

\def\WaveAmp{1.25} 
\def\WaveLen{3} 
\def\WaveOffset{0} 
\def\WaveNum{1.5} 
\def\polVecOrigin{0.25} 

\def\polVecOrigin{0.25}
\draw[thin, ->, black] 
({((\WaveOffset+\polVecOrigin*\WaveLen)*\rx)},
{((\WaveOffset+\polVecOrigin*\WaveLen)*\ry)},
{((\WaveOffset+\polVecOrigin*\WaveLen)*\rz}) 
-- 
({((\WaveOffset+\polVecOrigin*\WaveLen)*\rx +\WaveAmp*sin(deg(2*pi*\polVecOrigin)) *(cos(\rphi))},
{((\WaveOffset+\polVecOrigin*\WaveLen)*\ry},
{((\WaveOffset+\polVecOrigin*\WaveLen)*\rz -\WaveAmp*sin(deg(2*pi*\polVecOrigin)) *(sin(\rphi))}
)node[above right]{$\hat{p}$};

\foreach \vecOrigin in {1,2,3,5,6,7,9,10,11} 
{
\def\thisVecOrigin{\vecOrigin/8}
\draw[thin, ->, black] 
({((\WaveOffset+\thisVecOrigin*\WaveLen)*\rx)},
{((\WaveOffset+\thisVecOrigin*\WaveLen)*\ry)},
{((\WaveOffset+\thisVecOrigin*\WaveLen)*\rz}) 
-- 
({((\WaveOffset+\thisVecOrigin*\WaveLen)*\rx +\WaveAmp*sin(deg(2*pi*\thisVecOrigin)) *(cos(\rphi))},
{((\WaveOffset+\thisVecOrigin*\WaveLen)*\ry},
{((\WaveOffset+\thisVecOrigin*\WaveLen)*\rz -\WaveAmp*sin(deg(2*pi*\thisVecOrigin)) *(sin(\rphi))}
);

}

\draw[ black, thick, samples=200, smooth, variable=\t, domain=0:\WaveNum*\WaveLen] 
plot({(\WaveOffset+\t)*\rx +\WaveAmp*sin(deg(2*pi*\t/\WaveLen)) *(cos(\rphi))},
{(\WaveOffset+\t)*\ry},
{(\WaveOffset+\t)*\rz -\WaveAmp*sin(deg(2*pi*\t/\WaveLen)) *(sin(\rphi))});

\end{tikzpicture}
        \caption{Hpol}
    \end{subfigure}
    \caption{Illustration of the E- and H-planes for the orthogonal (a) Vpol and (b) Hpol polarization states measured by a PUEO antenna. The antenna boresight vector $\hat b$ and polarization vector $\hat p$ define the E-plane normal $\hat n_E = \hat b \times \hat p$,  and the H-plane normal $\hat n_H = \hat b \times (\hat b \times \hat p)$. The E-plane and H-plane are shaded blue and red, respectively. Representative linearly polarized plane waves are shown propagating along the boresight direction. In general, a measured waveform may be expressed as a superposition of the Vpol and Hpol polarization states.}
    \label{fig:EHPlanes}
\end{figure}
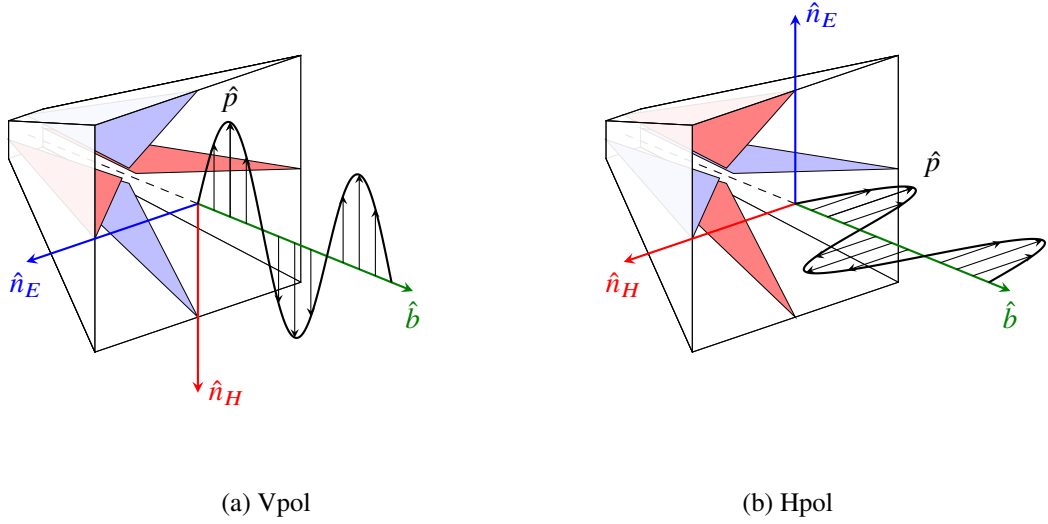

\begin{figure}[tbh]
    \centering
    \includegraphics[width=0.49\linewidth]{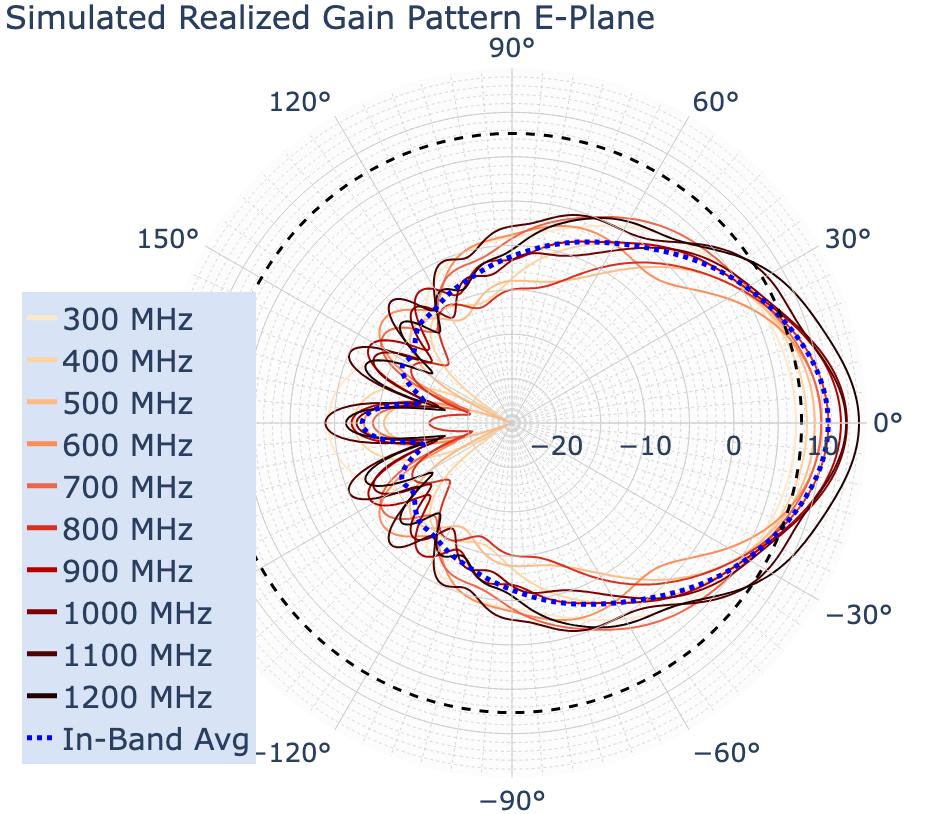}
    \includegraphics[width=0.49\linewidth]{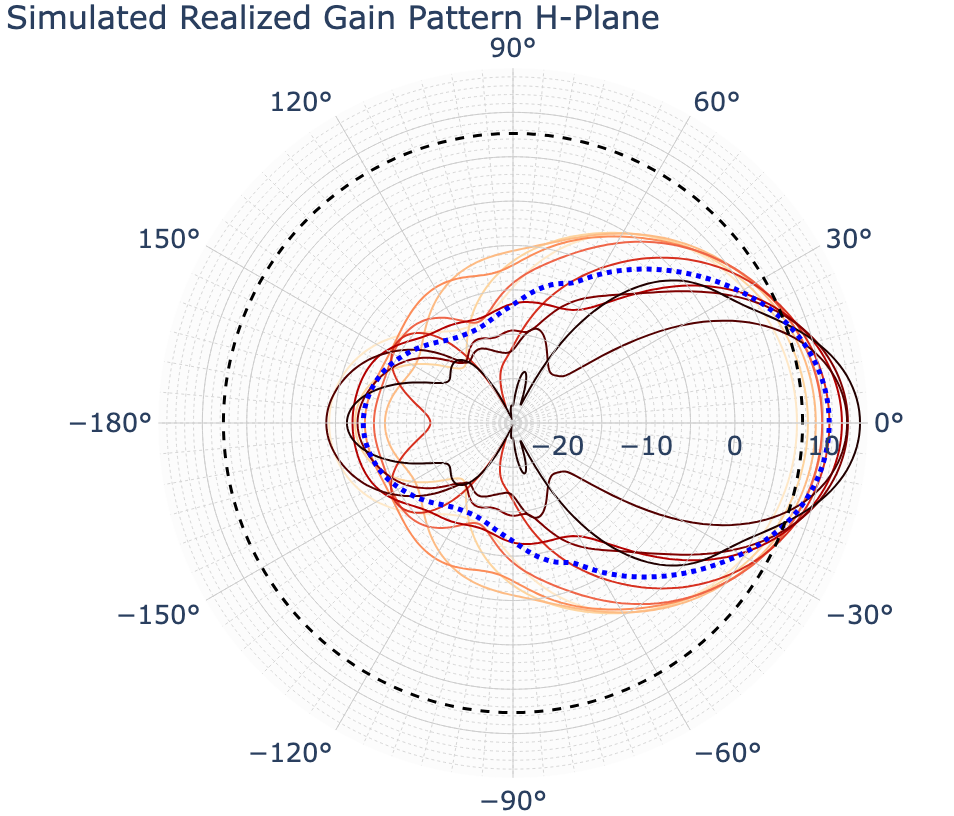}
    \caption{The E-plane (left) and H-plane (right) slices of the gain pattern for the Toyon antenna. The blue dashed line gives the average in-band gain, and the black dashed line indicates 3 dB down (from maximum gain, or boresight), which gives the full-width half-maximum (FWHM) of the gain response. The intersections of the two lines define the beam width: $40^{\circ}$ total, or $20^{\circ}$ in any direction from boresight. Note that the antenna has identical E-plane and H-plane patterns for both the HPol and VPol ports. Asymmetrical features appear in the arrayed configuration case, shown in figure \ref{fig:miant_arraypatt}, in which the HPol and VPol ports are no longer identical.
    \label{fig:miant_beampattern}}
\end{figure}

\begin{figure}[tbh]
    \centering
    \includegraphics[width=0.42\linewidth]{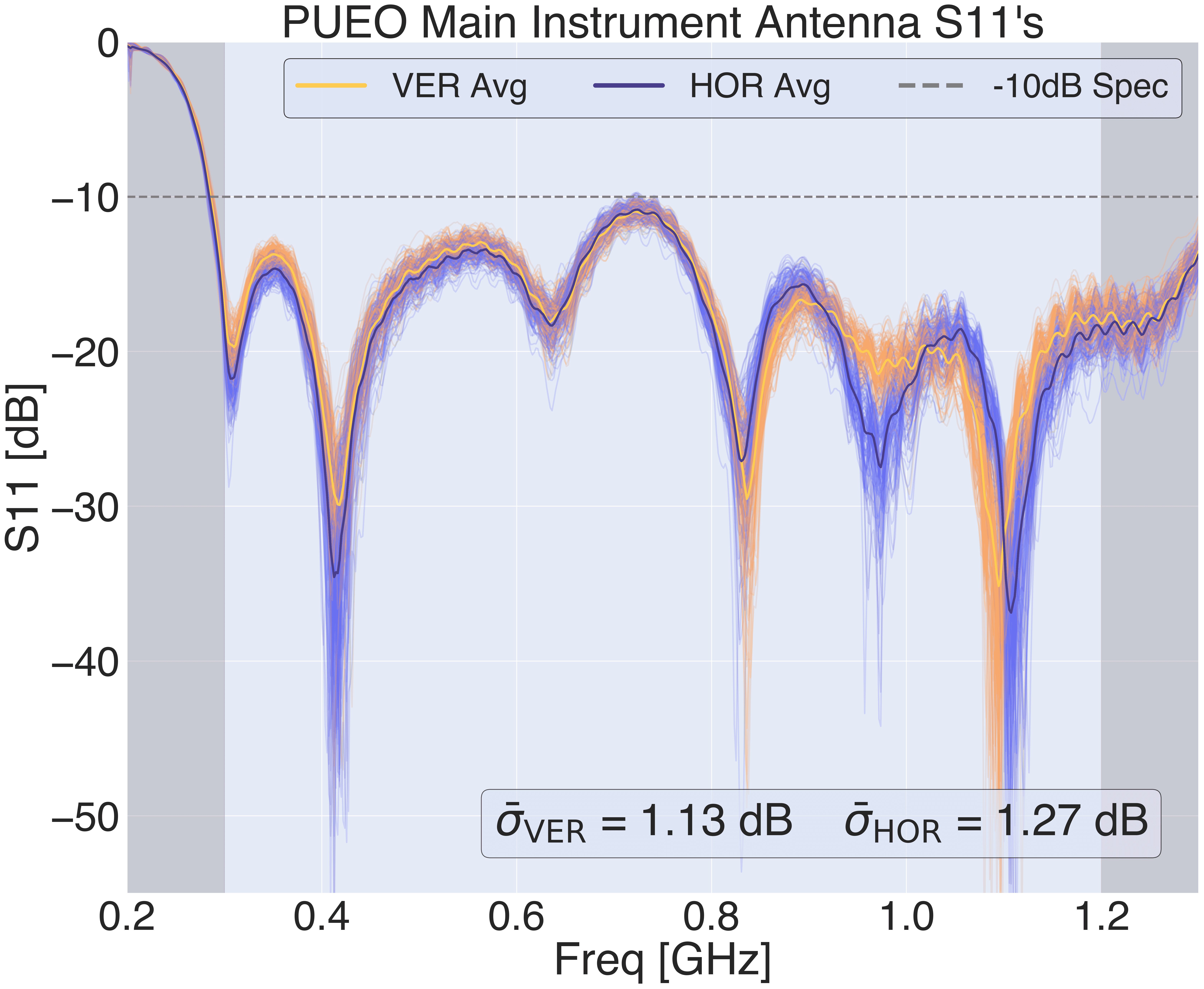}
    \includegraphics[width=0.57\linewidth]{S21_0deg_smooth_v3.jpg}
    \caption{Measured reflection, or S11 (left), and boresight transmission, or S21 (right), coefficients for all production antennas. The average is represented by solid lines. The specifications are indicated by gray dashed or dotted lines. The S11's are below the -10 dB specification, and the S21's are above the requisite gain specifications. The in-band standard deviations of all measurements are also given, and show agreement on the order of $\sim$1 dB or less in both S11 and S21.}
    \label{fig:miant_sparm}
\end{figure}

\begin{figure}
    \centering
    \includegraphics[width=0.99\linewidth]{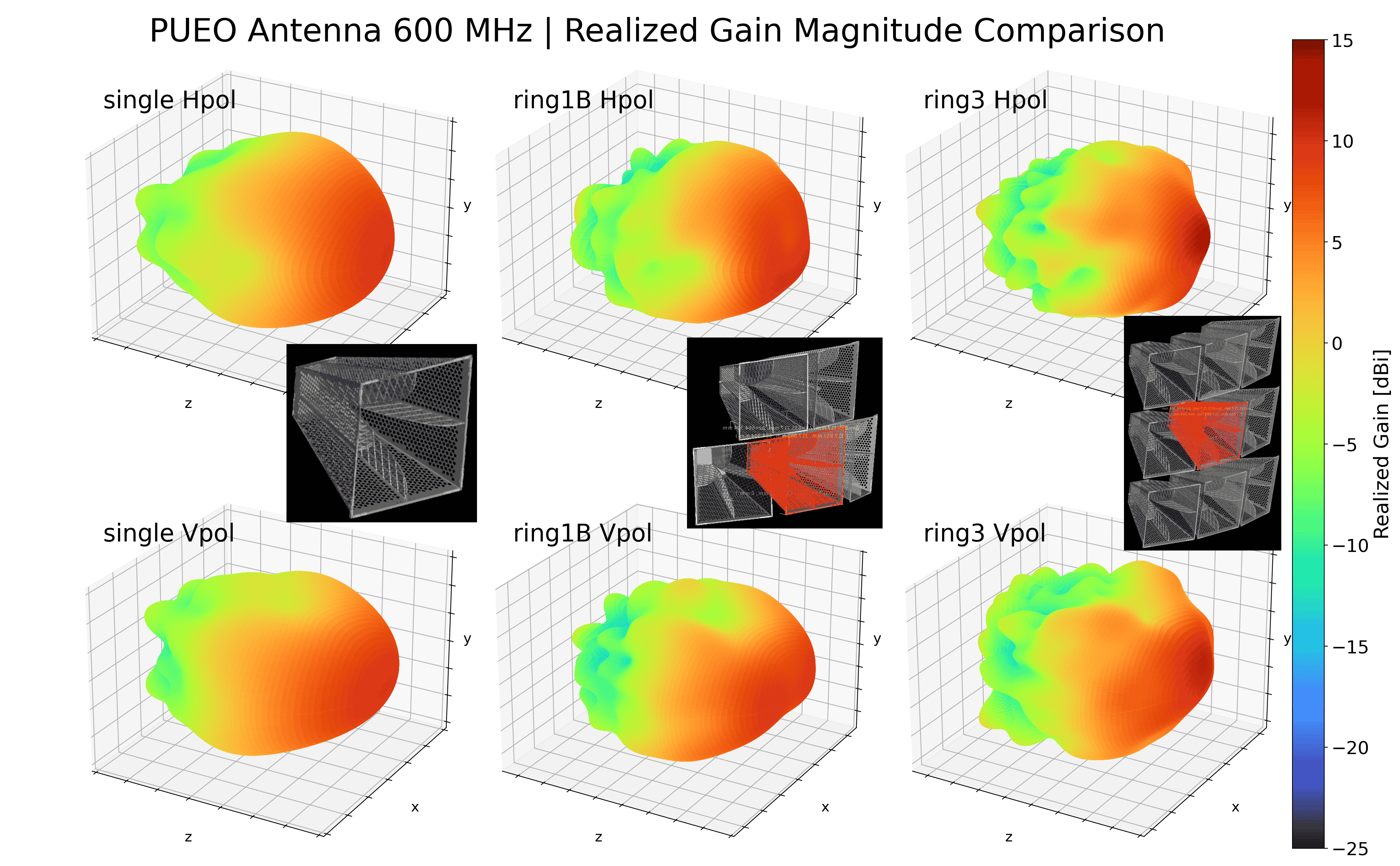}
    \caption{The realized gain 3D patterns comparing a single antenna (left column) to arrayed configurations for a bottom top-ring antenna, or ring1B (center column), and a center bottom-ring antenna, or ring3 (right column) at 600 MHz. The physical configurations of each column are shown in the center row, where the red-highlighted antenna (in the arrayed cases) indicates the antenna whose response is being simulated. Not shown are the results for the uppermost top-ring (ring1T), the uppermost bottom-ring (ring2), and the lowest bottom-ring (ring4). Note that the HPol and VPol are symmetrical by a 90$^{\circ}$ rotation around boresight. This symmetry breaks in the arrayed cases. The arrayed cases also showcase a sharper gain decrease around 90$^{\circ}$ in any direction when the adjacent antennas are in view, but the average gain within the 40$^{\circ}$ beam width is not significantly affected.}
    \label{fig:miant_arraypatt}
\end{figure}

PUEO consists of 96 custom horn antennas, one of which is shown on the left in figure \ref{fig:miant_impresp}. These horn antennas, developed by the Toyon Research Company, are specialized to the 300 - 1200 MHz band, which allows for a reduction to 2/3 of the ANITA antennas' length. The lowest sensitive frequency was 200 MHz for the ANITA missions, corresponding to $1.5$ m wavelength, whereas PUEO's is 300 MHz, corresponding to $1$ m wavelength. This allows twice the number of antennas to fit within the same payload envelope, which improves the RF beamforming SNR by a factor of $\sqrt{2}$.

The top-level design goal of measuring the impulsive radio emission from neutrino-induced showers in ice motivates the antenna specifications. Foremost, the antenna performance should be no worse than the previous antenna design used for ANITA. Among these specifications are electrical characterizations in the 300 - 1200 MHz band. The return loss specified is -10 dB or better. The realized gain at boresight is greater than 6.5 dBi from 300 - 500 MHz, 7 dBi from 500 - 700 MHz, and 9 dBi from 700 - 1200 MHz, with variations less than 0.5 dBi. This boresight gain specification is imperative to high pointing resolution in beamforming; PUEO uses highly directional, high-gain antennas encircling the payload instead of antennas with more isotropic beam patterns. The port isolation between polarizations is better than -30 dB. The antenna output ports are matched to $50$~$\Omega$, using N-type connectors. The antenna mass and size are as low as possible without detriment to its electrical characteristics. The antennas weigh 4.9 kg, and have a 0.636 x 0.636 m$^2$ face with a 0.661 m depth. The light weight is achieved by using aluminum material, fin-like ridges, and coarse-mesh side-panels. The design remains rigid and allows the antenna to be joined with adjacent antennas at face corner brackets, and mounted onto the gondola using custom brackets that align with central holes along the side-panels.

The electrical specifications are verified against lab measurements and simulations in \texttt{XFdtd}, a 3D EM Simulation Software by Remcom \cite{RemcomXFdtd}. The impulse response of the antenna describes its electrical response to an incident electric field or voltage excitation. Expressed in the frequency domain, the relationship is
\begin{equation}
    \tilde{V}(f) = \sqrt{\frac{Z_c}{Z_0}}~ \tilde{\mathbf{H}}(f)\cdot\tilde{\mathbf{E}}^{\text{inc}}(f)~, \label{eqn:miant_VHE}
\end{equation}
where $Z_0 = 120\pi~\Omega$ is the characteristic impedance of free space, and $Z_c = 50~\Omega$ is the impedance of the feed.

Here, we use the definition of the normalized impulse response, as suggested by \cite{farr1998timedomain}, to eliminate dependencies on impedances and transmission coefficients that are not constant with time or frequency. For impulsive events, as expected for the neutrino signals, it is ideal if the impulse response is close to a Dirac delta function in order to avoid signal distortions. Indeed, the impulse response for the Toyon antenna is shown on the right in figure \ref{fig:miant_impresp}, and the integrated power reaches 90\% within 2 ns. 

The simulated response is calculated using a plane wave excitation and equation \ref{eqn:miant_VHE}. In order to measure the response, a well characterized antenna transmits a high voltage ns pulse to the Toyon antenna, whose voltage response is measured using an oscilloscope. A vector network analyzer (VNA) is used to measure the frequency response, or transfer function, of the intermediate electrical components (i.e. cables, splitters, and attenuators) and thus, the source pulse and received voltage can be isolated and related by the antenna impulse responses. Additionally, a Wiener filter is applied to account for out-of-band noise measurements where signal divisions will result in 0 divided by 0 issues \cite{miocinovic2006timedomain, wiener1949extrapolation}. This result is described by the equation,
\begin{equation}
    \tilde{H}_\text{Rx}(f) = \Phi(f)\left(\frac{c~r(f)}{i~f}\right) \left(\frac{\tilde{V}_\text{received}(f)}{\tilde{V}_\text{pulse}(f)}\right) \left(\frac{1}{\tilde{H}_\text{Tx}(f)}\right) \label{eqn:miant_measresp}
\end{equation}
where $\tilde{H}_\text{Rx}(f)$ and $\tilde{H}_\text{Tx}(f)$ are the impulse response of the receiver (Rx) or transmitter (Tx), $\Phi(f)$ is the appropriate Wiener filter as described in \cite{Press2007NumericalRecipes3}, $r(f)$ is the distance between phase centers per frequency (approximated), and $\tilde{V}_\text{received}(f)$ and $\tilde{V}_\text{pulse}(f)$ are the measured received signal and transmitted pulse signal which have had intermediate component responses removed. The conversion to the time domain impulse response from the frequency domain follows the steps outlined in \cite{Scheers2000TEMHorns}, which gives $h_N(t) = 4~f_{\text{max}}~\text{Re}[\text{iDFT}(\tilde H(f))]$.

It is important to note that the impulse response is a complex-valued matrix for multi-polarization antennas, and can describe the response at all angles surrounding the antenna. In cases of polarized incident E-fields that are strongly aligned with an antenna polarization plane and incident near boresight, the impulse response can be approximated as a scalar value. This is what was measured and simulated in the verification of the antenna impulse response. However, when far from the antenna's 3 dB beam width, this approximation breaks down, and it should be noted that equation \ref{eqn:miant_VHE} becomes
\begin{equation}
    \begin{bmatrix}
        V^\text{H} \\ V^\text{V}
    \end{bmatrix}
    = \sqrt{\frac{Z_c}{Z_0}}~
    \begin{bmatrix}
        \tilde{\mathbf{H}}^\text{H} \\ \tilde{\mathbf{H}}^\text{V}
    \end{bmatrix}
    \cdot~\tilde{\mathbf{E}}^{\text{inc}} \label{eqn:miant_matrixh}~.
\end{equation}

The response at all angles can be simulated in \texttt{XFdtd} by using the reciprocity theorem, whereby a feed voltage excitation will radiate an E-field to the far-zone, which gives
\begin{equation}
        \tilde{\mathbf{H}}(f; \theta,\phi) = -i\frac{R~c}{f}\sqrt{\frac{Z_c}{Z_0}}\frac{\tilde{\mathbf{E}}_{\text{rad}}(f; \theta,\phi)}{\tilde V_s} \label{eqn:miant_xfH}
\end{equation}
where $R = 1$~m for the near-to-far-field transformation in \texttt{XFdtd}.

Another characteristic quantity of the antenna is its beam pattern, or gain pattern, which describes its gain response at all angles. This can be simulated as a direct result in \texttt{XFdtd} using the radiated E-field as described above. It can also be obtained from equation \ref{eqn:miant_xfH} by
\begin{equation}
    G(f; \theta,\phi) = \frac{4\pi f^2}{c^2}|\tilde{\mathbf{H}}(f; \theta,\phi)|^2~. \label{eqn:miant_gain}
\end{equation}
The beam pattern can be summarized from two principal planes. The E-plane is defined by the boresight and polarization vectors, $\mathbf{\hat{n}}_\text{E} = \mathbf{\hat{b}}\times\mathbf{\hat{p}}$ , and the H-plane is defined by the boresight vector and the perpendicular-polarization vector, $\mathbf{\hat{n}}_\text{H} = \mathbf{\hat{b}}\times(\mathbf{\hat{b}}\times\mathbf{\hat{p}})$. Figure \ref{fig:EHPlanes} illustrates these planes, while figure \ref{fig:miant_beampattern} shows the beam patterns. A 3 dB line below the average gain at boresight returns the full-width at half-max of the average beam pattern. This defines a beam width of 40$^{\circ}$, in both E and H planes. The beam patterns were verified using VNA measurements at Penn State's Radar and Communications Laboratory Anechoic Chamber in Warminster, PA.

After verifying the response for the Toyon antenna, all production antennas were verified by their reflection and transmission coefficients, corresponding to the S11 and S21 parameters on a VNA. The S11 is measured with the antenna pointing at an open sky. The S21 measurement is made by transmission between two Toyon antennas, calibrated using one of the antennas measured in detail at the anechoic chamber. The results of all antenna measurements are in figure \ref{fig:miant_sparm}.

Additional studies of the Toyon in \texttt{XFdtd} simulations include phase center estimation and effects of arraying. The phase center was approximated by fitting circles within the beam width to the wavefront emitted by the antenna on its E-plane at far distances. This resulted in an average phase center of $0.30\pm0.08$ meters from the front face.

The effects of arraying are relevant due to the structure of antenna mounting on the gondola. The response measurements described for a single antenna were repeated in a 3x3 array for the lower ring antennas and a 2+3 array for the top ring antennas, which allow the response to include effects of adjacent antennas. Figure \ref{fig:miant_arraypatt} shows examples of these configurations and the resulting effect on the realized gain patterns at 600 MHz. Although not shown, the beam pattern undergoes squinting at certain frequencies, where the boresight is no longer the direction of highest gain, but rather splits into multiple peaks $\sim$10$^{\circ}$ from boresight. Additionally, asymmetries form in the direction of adjacent antennas that affect the beam pattern outside the 40$^{\circ}$ beam width, but on average do not significantly affect the pattern within the beam width.

\subsubsection{AMPAs and Receivers}
\label{sec:AMPA_AND_REC}

\begin{figure}[htp]
    \centering
    \includegraphics[width=1\textwidth]{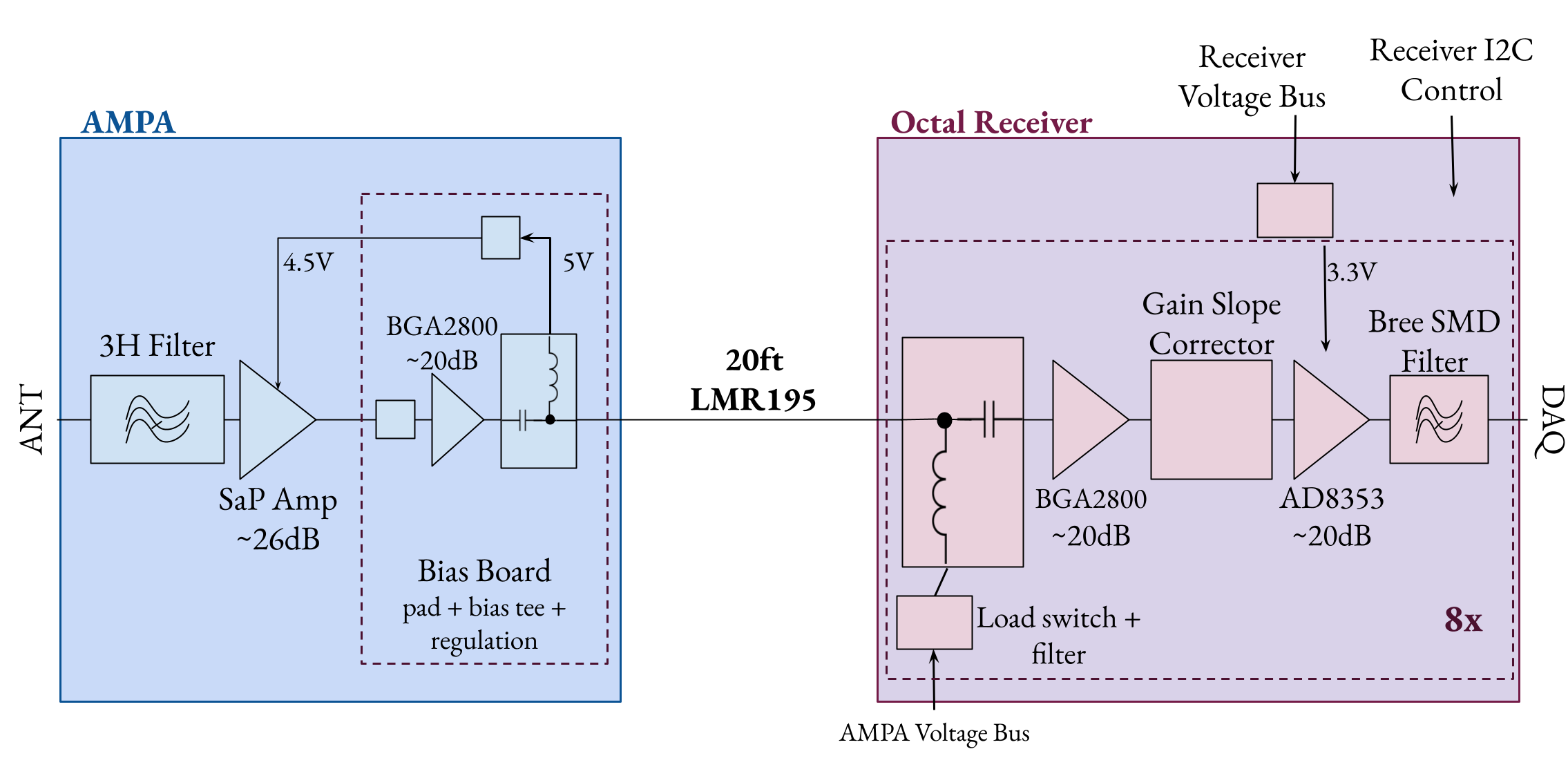}
    \caption{The PUEO MI RF chain block diagram showing amplification and filtering stages.}
    \label{fig:miRFChain_Block}
\end{figure}

The purpose of the RF chain for the MI is to amplify signals received by PUEO's MI antennas to measurable levels with little additive thermal noise and signal distortion. The block diagram for the RF chain is shown below (figure \ref{fig:miRFChain_Block}). Each channel's amplification occurs in two stages: first at its Antenna Mounted Pre-Amplifier (AMPA) and subsequently by an Octal Receiver. The AMPAs are mounted directly on each port of the MI antennas and the Octal Receivers are housed inside enclosures within PUEO's MIE (see section \ref{sec:MIE}). The two stages are connected with LMR195 cables of length $20$~ft, and the Octal Receivers feed signals into the DAQ via RF316 double-shielded cable (see section \ref{sec:MIRFChain_cables} for more details on cabling).

The AMPA is designed to amplify received signals from the antennas without contaminating the waveform with thermal noise. There are three main components to each AMPA: a bandpass filter, an LNA (Low-Noise Amplifier), and a bias board.

\textbf{Bandpass Filter:}
    The first component in the path of the RF chain is a bandpass filter developed by 3H Communication Systems, called the 3H filter. The filter accepts signals between $300$ and $1200$~MHz and rejects other out-of-band sources to protect the following LNA. The filter is designed to have a relatively steep cutoff at $300$~MHz, thereby allowing for rejection of the $280$~MHz satellite communication band at the level of $\geq 20$~dB that would otherwise saturate the amplifiers. Another noise source near PUEO's band is the $1610$~MHz Iridium communications band used on-board for telemetry; thus the filter's $1200$~MHz cutoff can be less steep while still allowing for noise rejection at the level of $\sim100$~dB.
    
\textbf{LNA:}
    After the filter, an LNA developed by SuperApexCo, called the SaP, amplifies the signal by $\sim26$~dB to boost the waveform above electronics-introduced thermal noise levels. This amplifier presents a low noise figure ($<0.6$~dB in band) and provides enough amplification that further signal conditioning later in the chain can occur without noise contamination. The SaP is mounted on a copper heatsink to avoid overheating during flight.
    
\textbf{Bias Board:}
    The final component of the AMPA is the bias board, placed after the SaP in the signal path and connected with an 086-4SNRSN+ handflex SMA cable. The bias board contains a signal path and a power path, separated by an inductor choke. Both pathways are relatively simple in design. The signal path contains capacitors to protect against DC on the signal path, a $2$~dB attenuator to protect the SaP and later stages from stray voltage or power spikes, a BGA2800 $20$~dB amplifier, and a weak $1300$~MHz low-pass filter. Since the signal travels down $20$~ft of LMR195 after this stage, the BGA2800 amplifier was chosen for its high isolation and ability to boost signal well above potential pickup contamination of the long cable. The bias board's power path contains various filtering capacitors to remove any stray RF signal, a power regulator, and a voltage divider, the result of which outputs a steady $4.5$~V to the SaP. Power is supplied to the bias board through the coaxial cable between the AMPAs and Octal Receivers.

These components are housed in enclosures that are designed to minimize connectors between the antennas and the LNAs. They are also designed to  reduce the EMI emitted by the AMPA components that would be picked up by the antennas. A picture of an open enclosure is shown below in figure \ref{fig:miRFChain_Boards}. The entire enclosure is made of nickel-plated aluminum for EMI reduction, and each hole in the enclosure is accompanied by EMI shielding gasketing\footnote{0.02" nickel-graphite-filled silicone}. The inputs are N(m) connectors to mate with the antenna ports, and the output connectors are TNC(f).

\begin{figure}[htp]
    \centering
    \includegraphics[width=.51\textwidth]{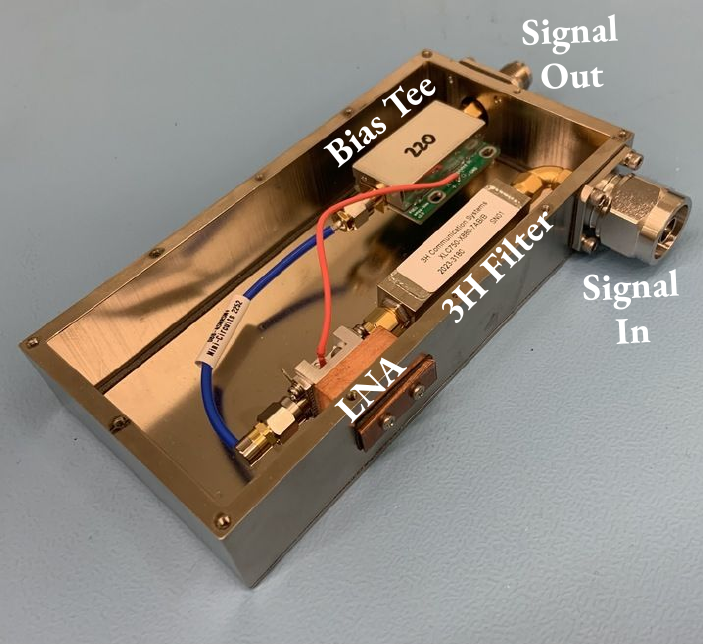}
    \includegraphics[width=.48\textwidth]{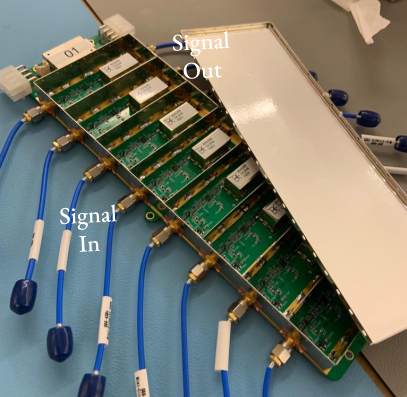}
    \caption{Left: A picture of an open AMPA enclosure with components labelled. Right: A picture of an open Octal Receiver Board with signal input and outputs labelled.}
    \label{fig:miRFChain_Boards}
\end{figure}

The Octal Receiver is designed to provide system housekeeping data, amplify the antenna signals (post-AMPA) to the DAQ's dynamic range, and correct for the gain-slope introduced by the long cabling between stages. Each Octal Receiver Board has one power input channel and 8 input and output signal channels.

The power channel contains a voltage regulator\footnote{MIC29371-3.3} that ensures a constant $3.3$~V is supplied to each receiver channel and an analog temperature sensor\footnote{LMT88} to monitor board level temperatures during flight. Each signal channel contains the power control for the upstream AMPA and a signal path. For the AMPA's power control, a power switch\footnote{MAX4915AEUK+T} is used to output a constant $200$~mA, which is isolated from the signal path via an inductor choke. The signal path includes two $\sim20$~dB amplifiers: a BGA2800 and an AD8353. The BGA2800 was chosen for its high isolation, and the AD8353 was chosen for its high compression point. These features allow for a design that can be used in tandem with the sensitive SaP while also ensuring a wide range of operating powers. Both receiver amplifiers are supplied with $3.3$~V. Much like the AMPA bias board, between each amplifier is a DC blocking capacitor and a power surge protecting $2$~dB attenuator. There is additionally a gain-slope corrector between the amplifiers, constructed using capacitors and inductors, in order to counteract the slope introduced by the intermediate cabling between the AMPA and Octal Receiver. Lastly, after the AD8353 and protective capacitor and attenuator, there is a final custom bandpass filter by Bree Engineering. This filter is designed to have a similar profile to the 3H filter, but is a surface mount. The filter acts as a final signal conditioning, removing any stray out-of-band noise that contaminates the signal after the 3H filter.

Several steps have been taken to reduce cross talk between the tightly packed signals on the Octal receiver boards. Each channel is enclosed by a thin aluminum wall, as shown in figure \ref{fig:miRFChain_Boards}. Within each channel cavity, the wall mounting holes are soldered closed, and the gaps between the SMA edge launches and the enclosure are both sealed with a combination of copper tape and solder. Lastly, the ceiling of each cavity is lined with an RF absorber\footnote{KEMET EFF03}.

\begin{figure}[htp]
    \centering
    \includegraphics[width=\textwidth]{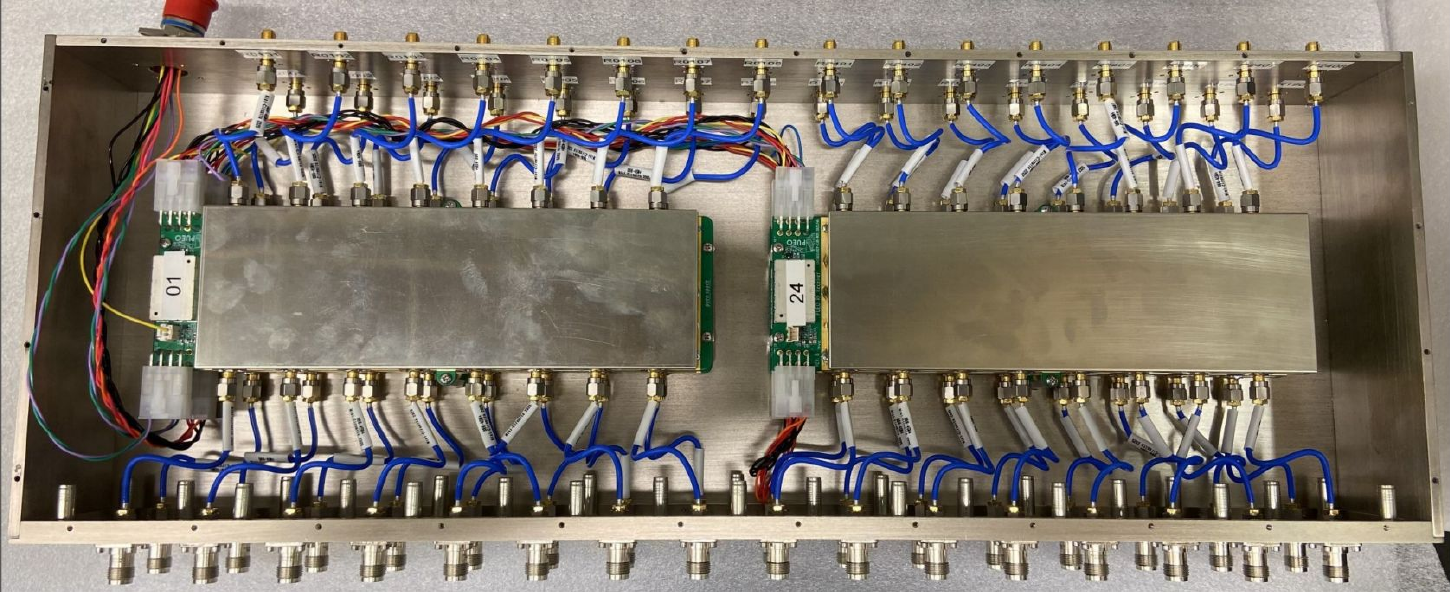}
    \caption{A packed receiver box for the MI antennas before being installed in the MIE. The filters and amplification circuitry are under the metal housings on each circuit board. Signals enter through the TNC wall on the bottom edge of the picture, and flow out the SMA wall on the top edge to the DAQs.}
    \label{fig:miRFChain_RxBox}
\end{figure}

Octal Receiver boards are housed in groups of four inside a Faraday enclosure called an Rx Box, resulting in six boxes total for all 192 channels of the MI. Each box conditions four phi-sectors of antennas, with each stack conditioning one polarization. An image of a fully packed Rx Box is shown in figure \ref{fig:miRFChain_RxBox}.

 The Rx Box walls are also nickel plated to reduce EMI, and the input wall has 32 TNC(m) to SMA(f) bulkhead connectors, each with custom gasketing. There are also press-fit studs installed to mount the Rx Boxes within the MIE. On the output wall are 32 SMA(f) to SMA(f) bulkhead connectors. The input/output connectors are attached to the Octal Receiver Boards using 086-4SM+ handflex cables. The output wall also features a 12-pin power connector\footnote{D38999/20FB35PN} which supplies power to the Octal Receiver Boards, reads out the temperature sensor on the top Octal Receiver on the closest stack, and toggles on/off the Octal Receivers and downstream AMPA channels for each stack layer. Lastly, each lid is covered in absorber to further avoid cross talk between boards.

\begin{figure}[htp]
    \centering
    \includegraphics[width=.49\textwidth]{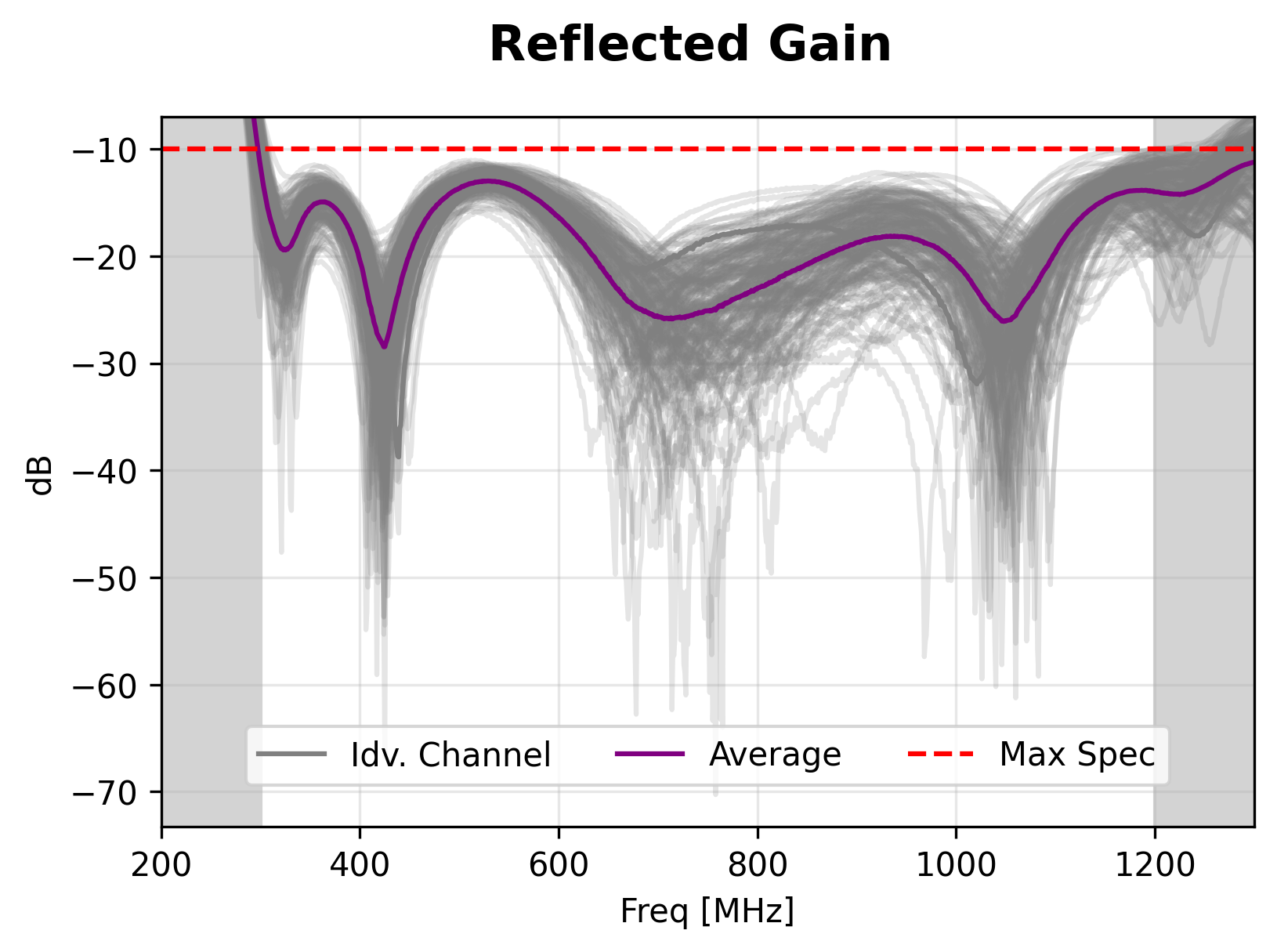}
    \includegraphics[width=.49\textwidth]{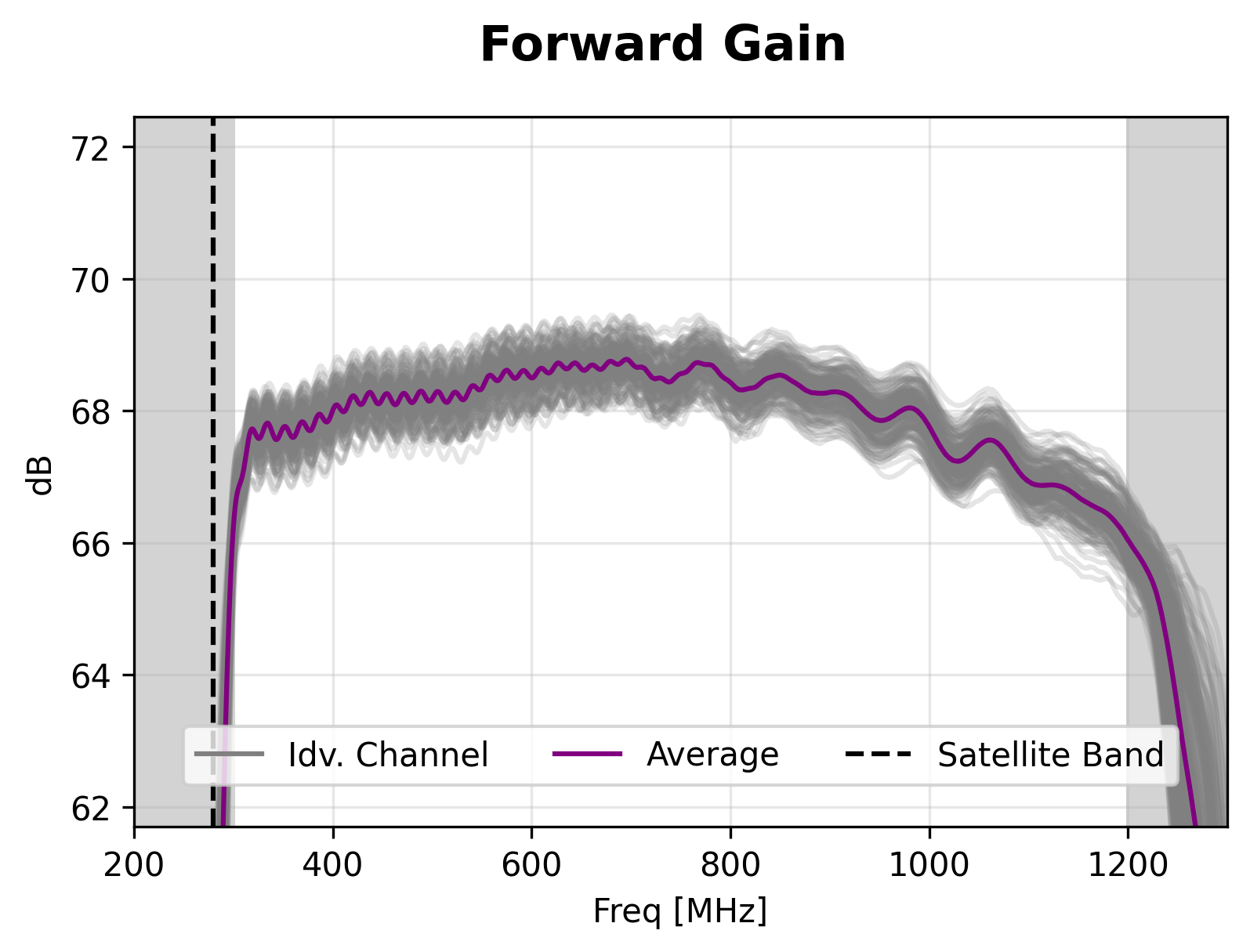}
    \caption{Left: Reflected gain (S11) of all channels as measured in the lab before flight. In comparison with the specification, all channels met the goal for reflected gain across the bandwidth. Right: Forward gain (S21) of all channels as measured in the lab before flight.}
    \label{fig:miRFChain_gains}
\end{figure}

 After construction, all 192 channels (AMPAs and Rx Boxes) were thermal-cycled and stress-tested under flight conditions. This was done to validate the integrity of solder joints, connections, and components. After all channels passed stress tests, the S-parameters were measured. The forward and reflected gains of all 192 flight channels are shown in figure \ref{fig:miRFChain_gains}. These were measured using a Rohde and Schwarz ZNL3 in VNA mode. Each channel was confirmed to meet the reflected gain maximum specification of -$10$~dB to reduce re-emission and self-triggering from the antennas. The forward gains present an average of $68$~dB, with the $3$~dB points at $300$ and $1210$~MHz. The average gain slope in band is $<1$~dB, and the forward gains between channels vary with a standard deviation of $0.28$~dB. This means all 192 channels on the PUEO MI are uniform and have very little signal distortion across the band.

\begin{figure}[htp]
    \centering
    \includegraphics[width=.48\textwidth]{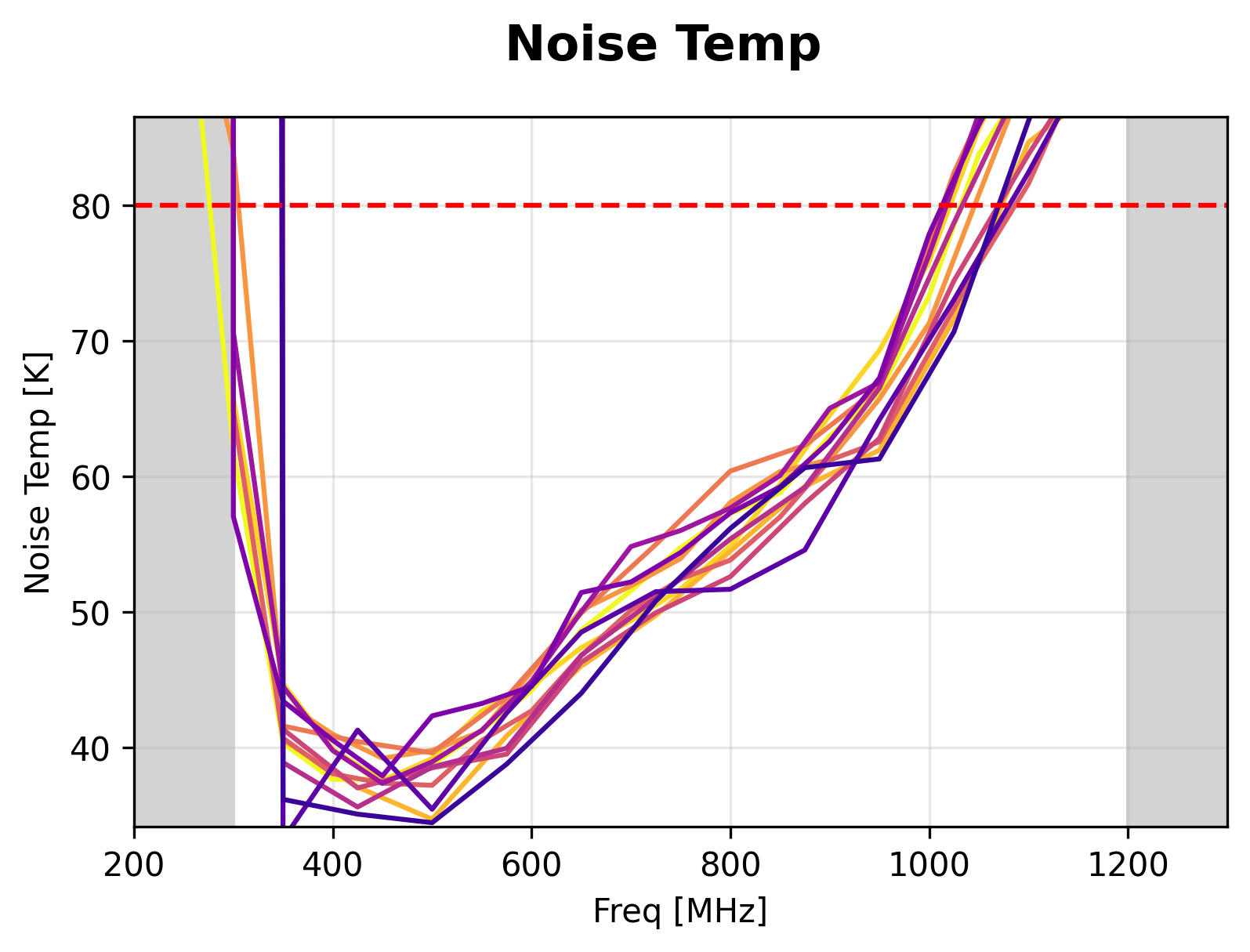}
    \includegraphics[width=.51\textwidth]{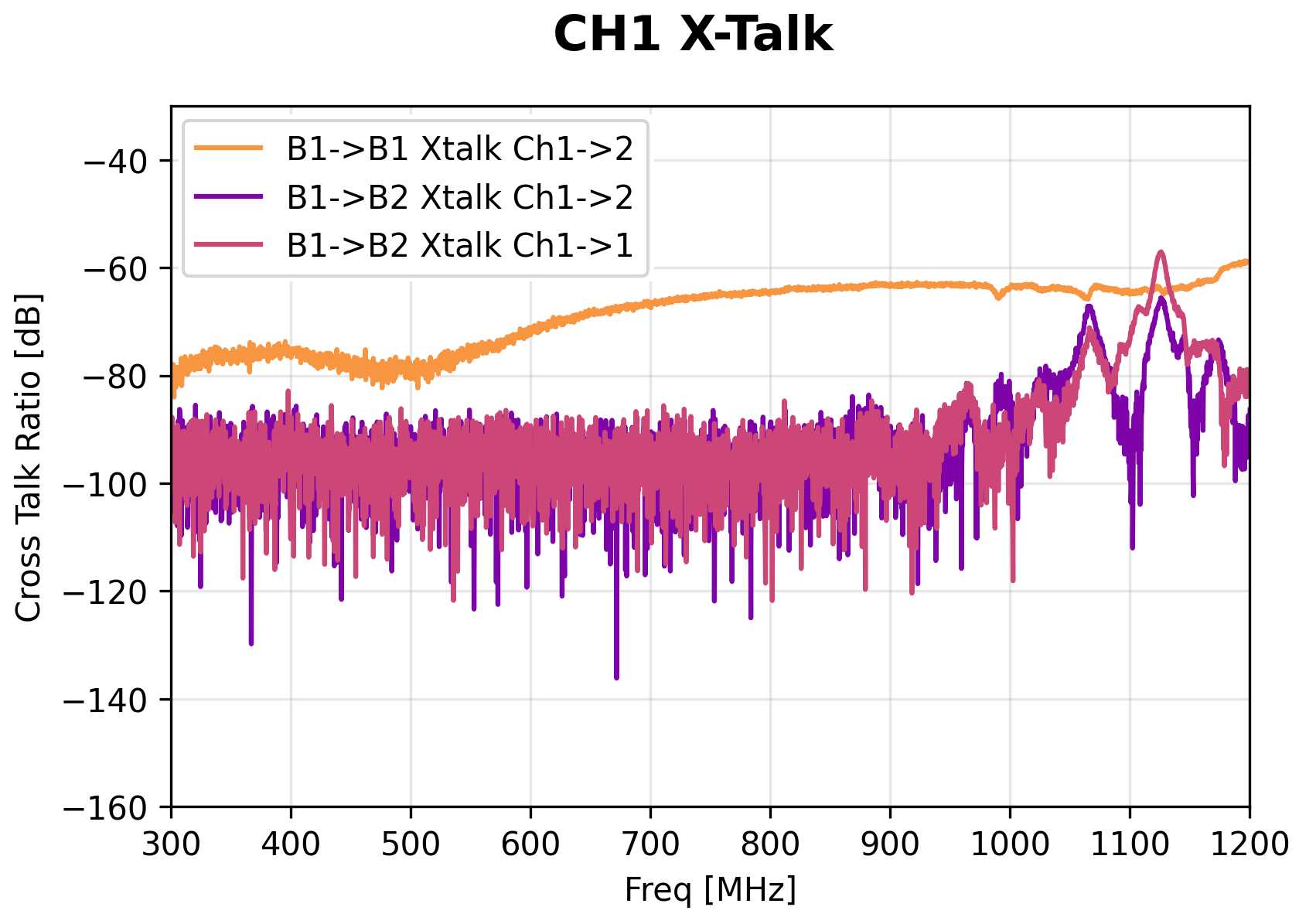}
    \caption{Left: The noise temperature of the full chain over the bandwidth of the MI. Each line is a randomly sampled channel, and the dashed horizontal line is PUEO's maximum specification for noise temperature. The shaded gray regions denote the edge of the PUEO MI band. Right: A sample of cross talk between Rx Box Channels. B1 and B2 denote separate Octal Receiver boards stacked on top of each other, and Ch1 and Ch2 denote channels located next to each other. The yellow curve is therefore the cross talk ratio between neighboring channels on the same board, whereas the pink and purple curves are the cross talk ratios between neighboring channels on different boards.}
    \label{fig:miRFChain_NoiseContam}
\end{figure}

 As discussed, attention is paid to reducing two important sources of background from the chain: thermal noise from the AMPA and cross-talk contamination from the Octal Receivers / Rx Box. Figure \ref{fig:miRFChain_NoiseContam} presents a sample of channel measurements characterizing these two contamination sources. The noise temperature of the chain has a max specification of $80$~K between $300$ and $700$~MHz in order to remain a subdominant contribution to the trigger thermal noise (after the sky and ice). The full chain for the MI presents thermal noise levels well below $80$~K within the MI trigger band. The cross talk ratios have a maximum specification of -$60$~dB to ensure subdominant contributions to the signal. Generally, the cross talk between channels and boards stays below this level, especially within the trigger band.

\subsubsection{Cabling} \label{sec:MIRFChain_cables}

\begin{figure}[htp]
    \centering
    \includegraphics[width=\textwidth]{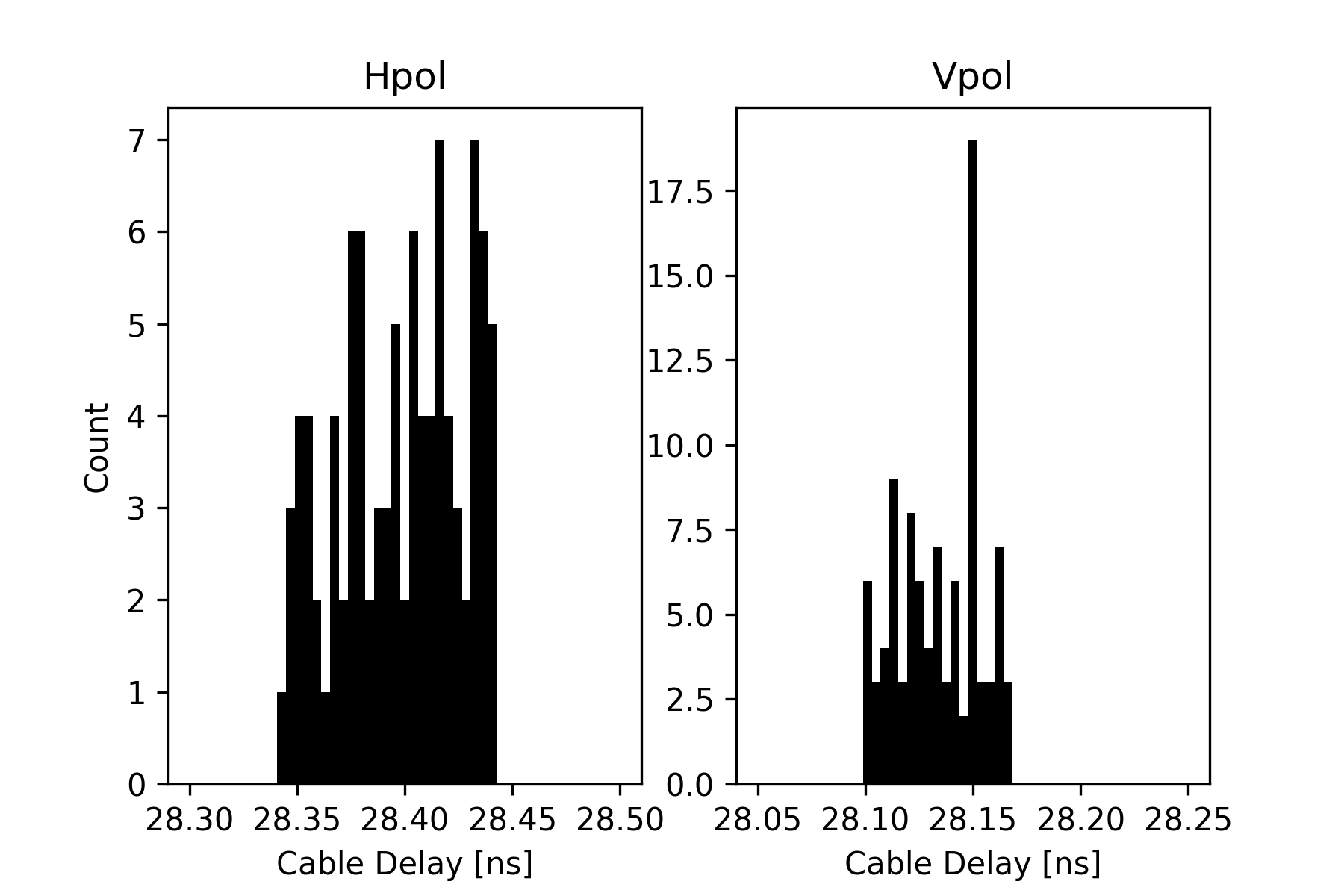}
    \caption{The distribution of the measurements of the cable delays for Hpol (left) and Vpol (right).}
    \label{fig:miRFChain_delays}
\end{figure}

The nature of a beamforming trigger requires precise timing uniformity between channels forming a signal trigger. To avoid channel-by-channel timing difference corrections at the trigger level, the max timing difference between channels in the same trigger must not exceed $100$~ps. Since the trigger works independently between each polarization, this max variation is important between channels for the same polarization, but can be ignored between polarizations. The signal chain requires careful cabling to achieve this precise timing, specifically between the AMPAs and Rx Boxes. Each channel has $20$~ft of LMR195 connecting the two stages. LMR195 is an ideal choice since it is double-shielded, which protects against RFI, and flexible, which allows for the tight packing of channels at the Rx Box input wall. However, few-percent differences in cable length or internal dielectric characteristics could introduce timing variations that degrade the sensitivity, thereby requiring individual cable characterization and validation. Figure \ref{fig:miRFChain_delays} shows the measured cable delays for all 192 MI channels, with a range of $100$~ps for HPol and $67$~ps for VPol. These measurements were obtained channel-by-channel by splitting a picosecond pulse between a Rohde and Schwarz RTO64 oscilloscope and a single flight cable using a T-splitter, keeping the other end of the cable open. By measuring the time delay between the direct pulse and the one reflected off the open end of the cable, the cable delay could be obtained (after dividing the time delay by two to account for the pulse traveling down the cable twice). After processing the data, the resolution achieved on these measurements was $\sim2$~ps.

\subsection{Low Frequency Instrument}
This subsection presents a high-level overview of PUEO's Low Frequency (LF) instrument. A companion paper describing the LF instrument design, deployment system, electronics, and performance in detail is currently in preparation.

\subsubsection{Science Case}
As described in section \ref{sec:RFE}, in addition to the primary Askaryan detection channel, PUEO is expected to be sensitive to the geomagnetic emission produced by air showers sourced from cosmic rays (both direct and reflected off of the ice) and tau neutrinos. Whereas the neutrino-induced Askaryan emission is expected to be predominantly vertically polarized, geomagnetic emission is strongly horizontally polarized due to the orientation of Earth's magnetic field in Antarctica. Compared to in-ice particle cascades, air shower development is substantially more extended, generating characteristically lower frequency emission. The LF instrument is a deployable array of antennas designed to target a lower frequency range than is accessible with the MI in order to provide (i) additional sensitivity to the air shower event classes and (ii) an additional cross-validation and spectra extension/overlap for MI-triggered Askaryan events. The LF instrument did not participate in the trigger logic for this flight.

\subsubsection{Antenna and Array Design}
\begin{figure}[htp]
    \centering
    \includegraphics[width=.99\textwidth]{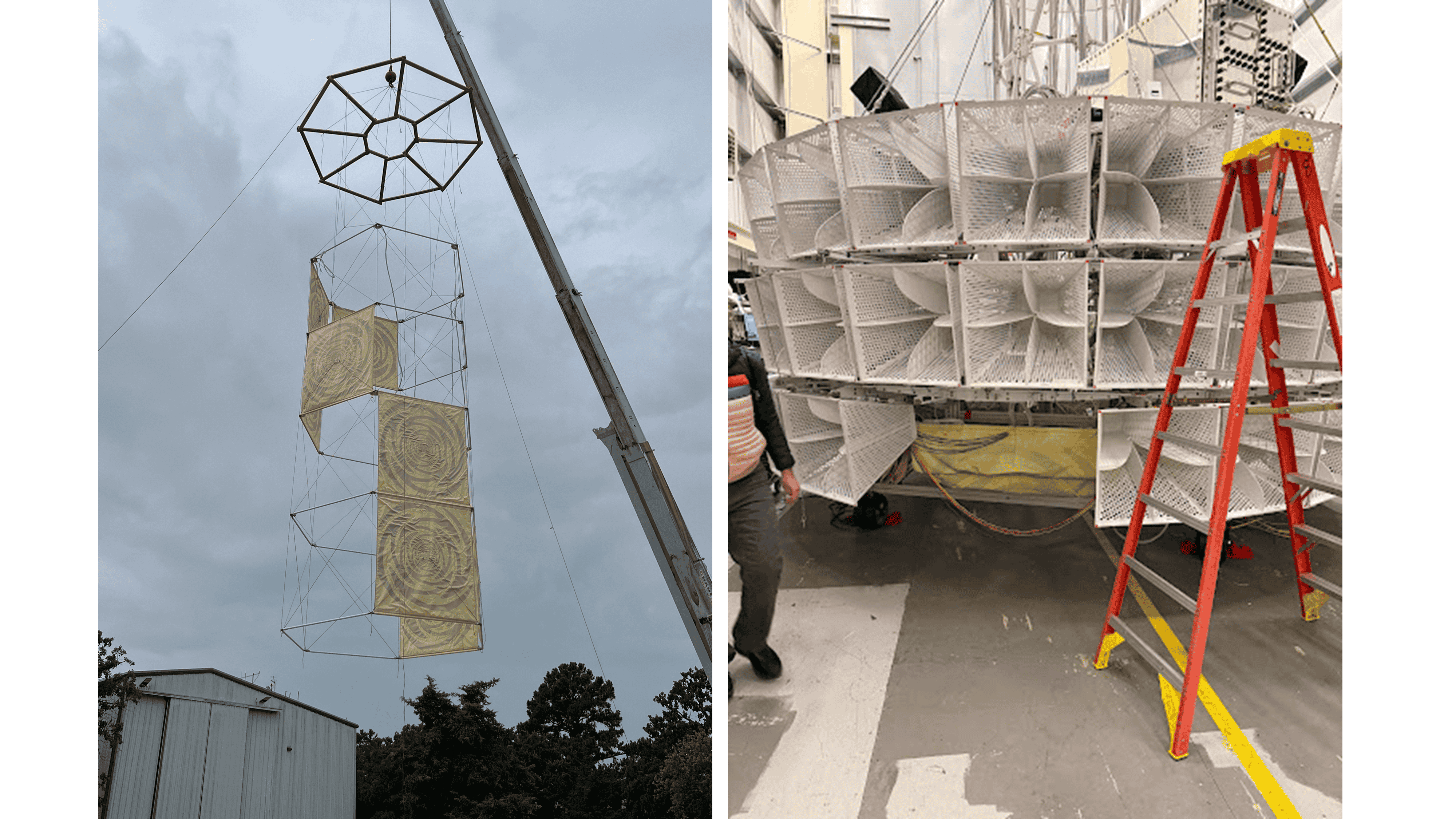}
    \caption{Left: A picture of the deployed LF instrument during the hang test at CSBF in 2025. Right: A picture of the LF instrument stowed inside the payload within the lower rings of the MI, taken during assembly at LDB in Antarctica.}
    \label{fig:LF_gondola}
\end{figure}

\begin{figure}[htp]
\centering
\includegraphics[width=\textwidth]{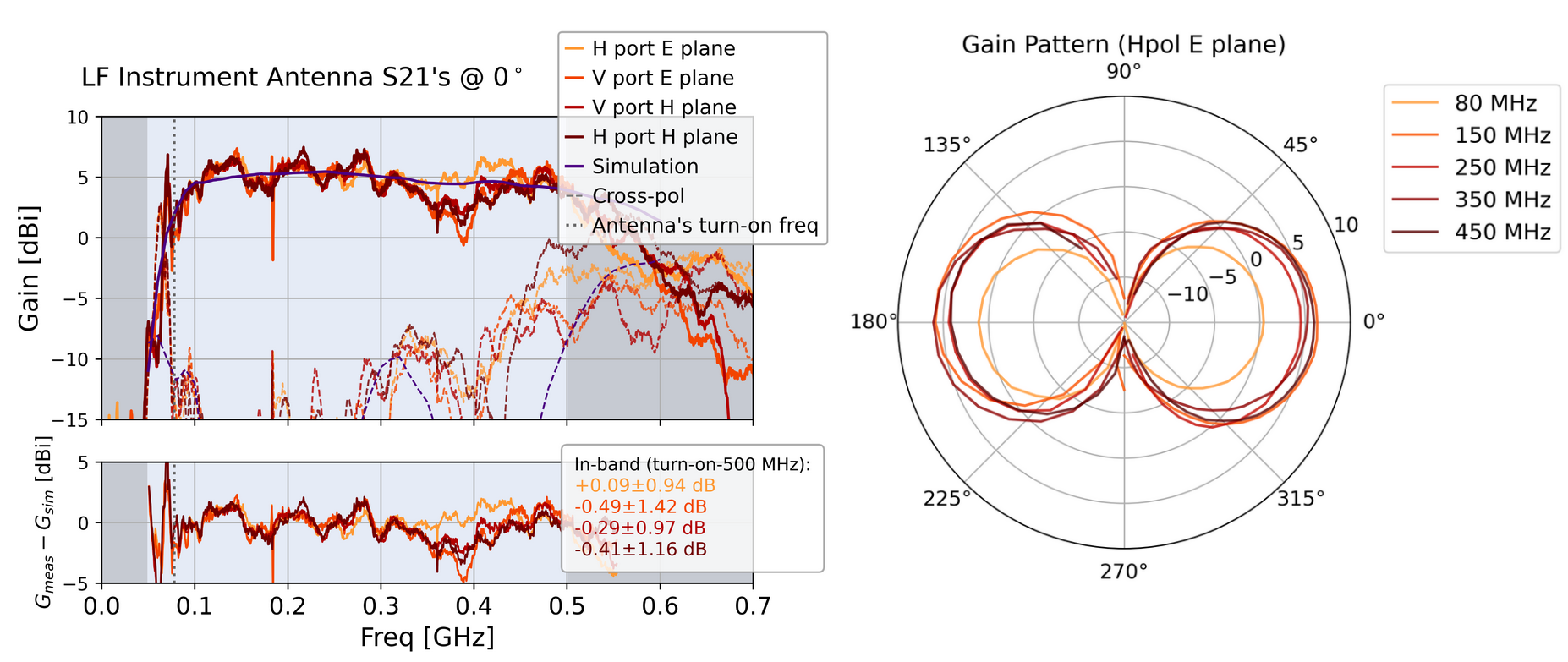}
\caption{Left: The measured and simulated boresight gain of sinuous LF antennas. Right: The measured beam pattern of sinuous LF antennas at different frequencies.}
\label{fig:LF_antenna_measurements}
\end{figure}

Within its array, the LF instrument uses dual-polarized, $1.65$-m diameter log-periodic sinuous antennas, shown in figure \ref{fig:LF_gondola}. To allow for stowage and deployment, the antennas are made of an alloy of nickel, copper, and silver embedded in nylon ripstop fabric and sewn between two polyester sheets. These antennas provide a broadband $5$~dBi gain from $100-500$~MHz, with $\pm3$~dB points at $\pm30^{\circ}$ from boresight, with identical front and rear lobes, as shown in the gain and beam pattern of the LF antennas in figure \ref{fig:LF_antenna_measurements}.

The LF array consists of eight sinuous antennas, arranged on hexagonal rings in four layers, with differing numbers of antennas per layer, in order from top to bottom: 2, 2, 1, 3. The number and size of the antennas were constrained by the available channels in the DAQ RFSoC carrier boards (16, called SURFs), the available space underneath the payload, and the requirement to be sensitive to low frequencies. The antennas are positioned so that the beam patterns have minimal interference with one another while providing a nearly azimuthally symmetric sensitivity. Each ring of the LF gondola structure is made of 1-inch carbon fiber tubes and connected with custom aluminum vertex plates. The rings are connected to one another using colored Dyneema\footnote{See \url{https://www.dyneema.com/}.} rope at each vertex plate, with colors corresponding to the layer number. To prevent twisting of the array, each face has cross ropes between rings. The deployed LF gondola measures $29.5$~ft tall and weighs a total of $\sim100$~lb (including electronics and cabling). A picture of the deployed LF instrument is shown in the left panel of figure \ref{fig:LF_gondola}.

\subsubsection{Stowage and Deployment}
In order to fit within the required launch envelope, the LF must be folded and stowed within the payload. Deployment occurs once at float altitude, $37.5$~km, as determined by GPS monitoring. Before stowing the instrument on the payload, each antenna is folded twice along its horizontal axis with half of the antenna lying on either side of its corresponding bottom carbon fiber tube and the amplifiers, associated cables, and Dyneema following the folds. This ensures compliance with the bottom edge of the launch envelope while minimizing risk of non-deployment. In the stowed configuration, the Dyneema ropes are also daisy-chained to prevent tangling, which also increases the risk of non-deployment. The top ring of the LF is secured to the frame of the gondola via four anchor points located on the spokes of the gondola. Four basket ropes with spliced-in U-channels are fed underneath the stowed LF and held together near the center of the gondola. A routed line to a redundant two-actuator deployment system on the deck level of the gondola is passed through these U-channels and tensioned. Any free cables are loosely attached to the top ring of the LF with cable ties, which come undone under tension during deployment. When the gondola is hoisted, the LF is fully attached to the gondola via anchor points, and secured via the basket ropes. A picture of the LF in the stowed configuration is shown in the right panel of figure \ref{fig:LF_gondola}.

\subsubsection{RF Chain Design and Performance}
\begin{figure}
    \centering
    \includegraphics[width=0.98\linewidth]{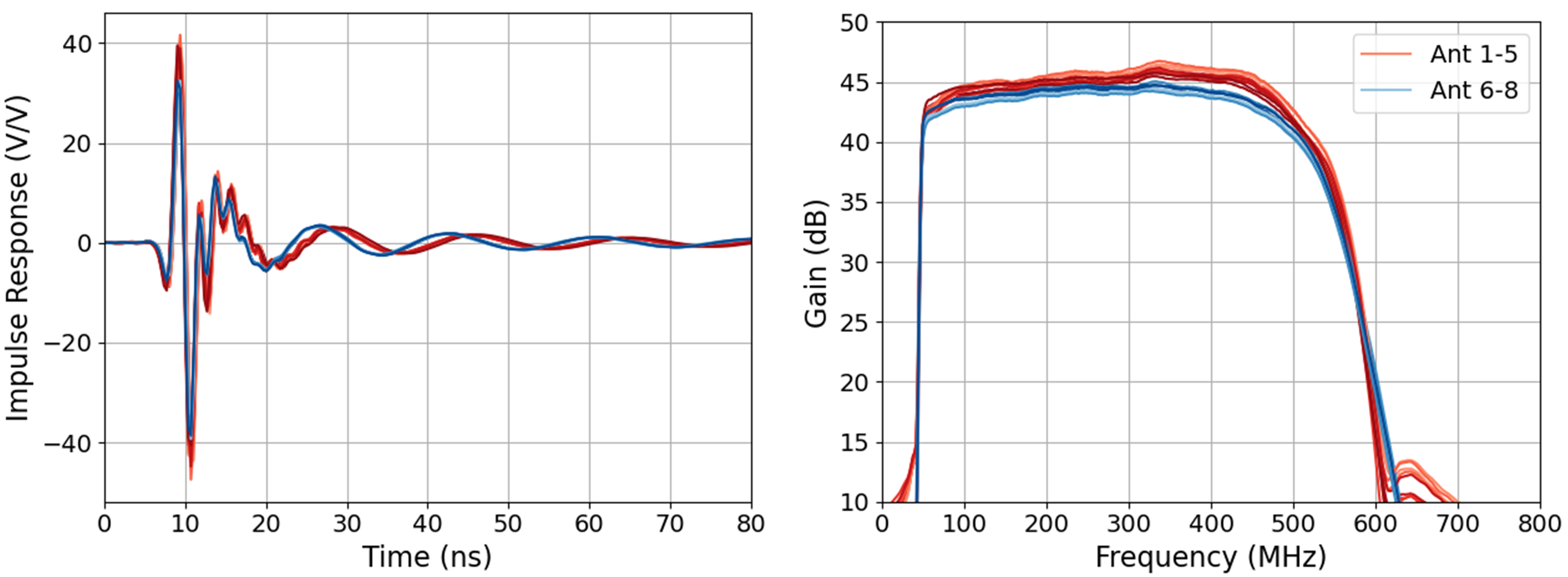}
    \caption{Left: The impulse response of the full chain of the LF antennas. Right: The forward gain of the lowAMPA and receiver board for each LF channel. }
    \label{fig:lowampa}
\end{figure}

The electronics chain of the LF provides $\sim45$~dB of gain between the received signal at the antenna and the DAQ-level digitization, as shown to the right in figure~\ref{fig:lowampa}. The first part of the electronics chain is a 4:1 balun connected to the rear of the antenna (one for each polarization) that converts the antenna impedance to $\sim50~\Omega$ for compatibility with the rest of the chain. The balun connects directly to the antenna-mounted front-end amplifier (lowAMPA), which is mounted inside a nickel-plated aluminum enclosure and bolted through the fabric antenna on the rear-facing side. The lowAMPA is made up of both high-pass and low-pass filters to reduce out-of-band signal, as well as low-noise amplifiers that provide $\sim20$~dB of gain from $50$ to $500$~MHz. White-sheathed LMR195 cables connect each antenna’s lowAMPA to its corresponding vertex plates, where sheathed LMR240 cables of various lengths are routed through the gondola to the MIE. The LF signals are amplified and filtered at a modified MI receiver, described in section \ref{sec:AMPA_AND_REC}, where the MI bandpass filters have been removed. More details on the LF electronics, design, and performance are presented in the companion paper.

\subsection{Main Instrument Enclosure}\label{sec:MIE}

\begin{figure}[htbp]
\centering
\includegraphics[scale = 0.4]{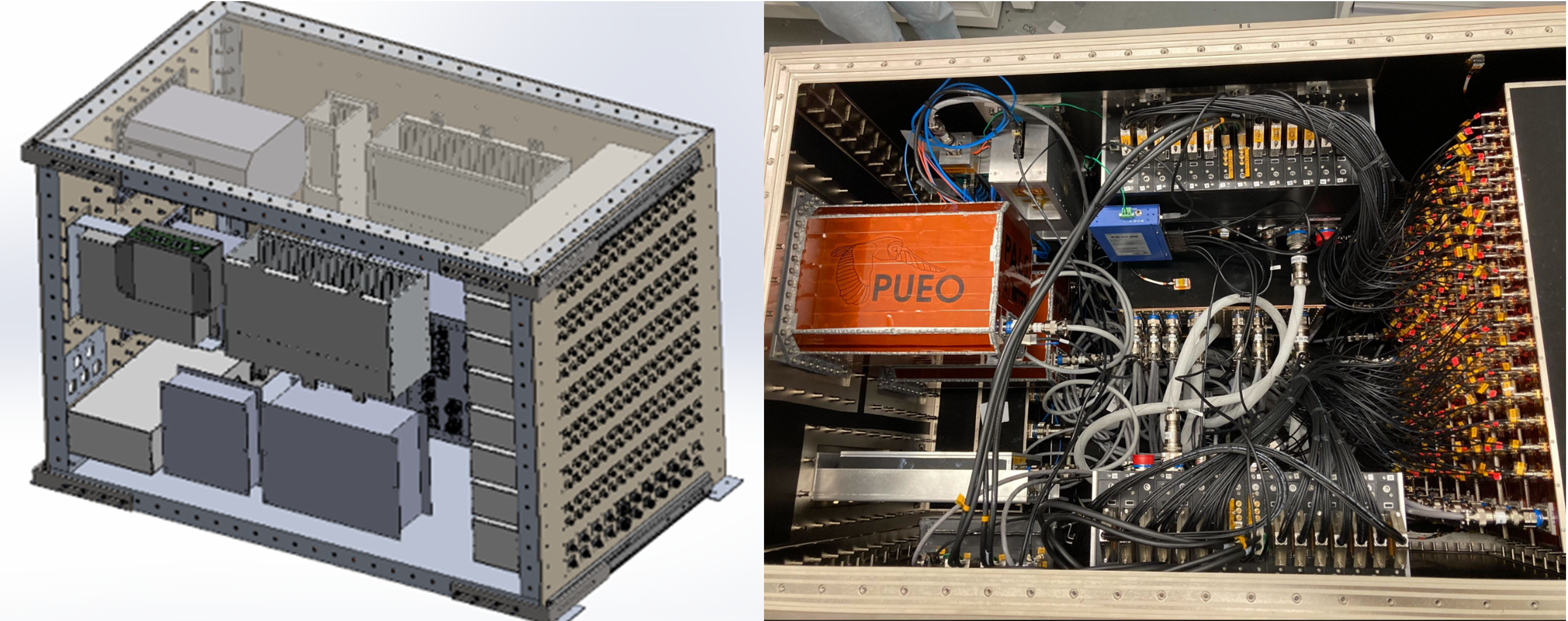}
\vskip 0cm
\caption{Left:  A rendering of the Main Instrument Enclosure showing subsystem enclosure designs mounted to the radiator plates and the thick angle brackets of the frame. The RF-signal faceplate with ports for all MI and LF channels is visible on the right side of this rendering. Right: The top-down view of the MIE as assembled before flight with the lid removed. The RF receiver boxes are mounted on the right side and the hard drive storage enclosures on the left. The gasketing around the frame at the top where the MIE lid is installed is visible. }
\label{fig:mieCad}
\end{figure}

In order to reduce the risk of noise pickup, many subsystems are located in a single RF tight enclosure, the MIE. The MIE is a $34 \times 50 \times 34$~in$^{3}$ modular frame-based enclosure as shown in the design image to the left in figure \ref{fig:mieCad}. The right angle brackets and frame faceplates, or radiator plates, are all constructed from $\nicefrac{1}{8}$~in thick 6061-T6 aluminum. When fully assembled, as shown to the right in figure \ref{fig:mieCad}, with the subsystems and cables inside, the MIE weighs $780$~lbs. All of the radiator plates are coated with electroless nickel and their mounting interfaces include gasketing to achieve an $80$~dB design level of RF attenuation. A small copper tube is instrumented on one radiator plate as a pressure relief for flight, and tuned as a waveguide below cutoff in the MI band. For radiative cooling, silver-coated Fluorinated Ethylene Propylene (FEP) tape from Sheldahl\footnote{See \url{https://sheldahl.com/}.} covers the six outer faces of the MIE. Subsystem RF-tight enclosures are mounted to the inner surfaces of the radiator plates with the thermal compound Arctic Silver\footnote{See \url{https://www.arcticsilver.com/}.}. To improve radiative thermal coupling within the MIE, the inner-facing surfaces of the MIE's outer walls are covered with black Acktar\footnote{Acktar MaxiBlack} and some of the surfaces are coated with black Krylon paint. All signal and power lines (except the RF channels) that interface externally to the MIE pass through filter pins to mitigate the emission of internal electronics noise to the ambient environment. The majority of these filter pins are $200$~V and $10$~A-rated from Spectrum Control with a minimum insertion loss of $65$~dB at $300$~MHz.

\subsection{Data Acquisition}
\subsubsection{Hardware}

The Data Acquisition (DAQ) system is primarily based on the Xilinx Radio Frequency System-on-a-Chip (RFSoC) ZU47DR built into a commercial off-the-shelf (COTS) module from Trenz Electronic (TE0835-2). This module is mounted on a custom printed circuit board (PCB) called the Sampling Unit for RF (SURFv6, or SURFs), which contains power conversion, clock distribution, and the analog signal conditioning. The DAQ is split into two main crates, one for each polarization (Hpol and Vpol), consisting of 12 SURFv6 modules. An additional two SURFv6 modules are present in the Hpol crate to digitize the LF instrument channels, for a total of 26 SURFs servicing 208 analog channels (out of a maximum of 224).

Physically, the DAQ crates are derived from cPCI-Serial
conduction-cooled assembly (CCA) specifications, but using a custom backplane interface. The conduction-cooled assemblies (CCAs) and DAQ crates are custom machined 6061 aluminum with a clear chromate conversion coat per MIL-DTL-5541, class 3, type II Clear, and contact the RFSoCs with thermal paste for heat transfer. Coupling to the DAQ crate is via C/A Design 426C-480UMB-CKP wedgelocks. The outside of the DAQ crates are painted black with Krylon Industrial 5-Ball Int/Ext Ultra-Flat Black paint in order to improve emissivity. The crates are then mounted directly to the MIE radiator plates with Arctic Silver thermal paste. The cPCI-Serial design also allows for a convection-cooled benchtop testing crate using standard IEEE 1101.1 tooling. These crates were used for development and firmware verification when the MIE was unavailable.

Inside each DAQ crate, the SURFs mount on two separate custom backplanes, termed the Radio Array Carrying Kit (RACK). The SURFs then interface with a central triggering unit, the Trigger Unit for RF (TURFv6 or TURF),
using an auxiliary interface board (the Trigger Unit for RF I/O, TURFIOv6 or TURFIO) which consists of a smaller Xilinx XC7A35T field programmable gate array (FPGA). The TURFIO collects data and triggers synchronously from each SURF and delivers the triggers and data to the TURF. The four TURFIOs interface to the TURF using an ePCIe x8 cable (Molex 74546-0801) which also carries commanding, trigger readout requests, and the system clock distribution from the TURF. The TURFIO handles housekeeping data and power control for the SURFs, which is essential in flight for monitoring the temperatures of the units to avoid overheating. 

The TURF is installed in a separate EMI/RFI shielded enclosure, a zinc-plated steel junction box (McMaster Carr 7816K23) and is based on a custom PCB housing an iWave RainboW G35M module based on a Xilinx XCZU19EG Multi-Processor System-on-a-Chip (MPSoC). An aluminum heatsink couples the module to the enclosure via thermal paste, and the module is then mounted to the MIE radiator faceplate.

Both the RFSoC and MPSoC consist of programmable logic (PL) and processing system (PS). The PS uses a minimal Xilinx PetaLinux operating system and a highly compressed standalone Python interpreter (eGenix PyRun) to fit in ${\sim} 32~\textrm{MB}$ of space in the onboard $128~\textrm{MB}$ serial flash storage, leaving the remainder for multiple firmware versions which could be dynamically chosen at startup. The onboard Python software daemons handle initial PL programming, housekeeping, and the communications datapath tuning and Xilinx multi-tile synchronization (MTS). Because the standard Xilinx MTS library required register access that was unavailable, a custom shim library was developed that translated the appropriate function calls.

Each SURF receives 8 analog channels on a mixed D-Sub connector with 8W8 inserts allowing rapid connection and disconnection. The 8 analog channels map to a pair of adjacent phi sectors of a single polarization, covering $30^\circ$ in azimuth, and the SURFs are arranged in the crates in order, meaning that adjacent SURFs are matched to antennas that view adjacent regions in azimuth. The analog signal
path to the RFSoC consists of a high pass filter with a ${\sim} 100~\textrm{MHz}$ $3~\textrm{dB}$ point (${\sim} 50~\textrm{MHz}$ for the LF SURFs), followed by a low-pass filter with a $3~\textrm{dB}$ point at ${\sim} 1350~\textrm{MHz}$ ($20~\textrm{dB}$ at $1500~\textrm{MHz}$).
At this point a separate common calibration signal can be switched into the signal path to provide a known signal to confirm proper digitization and timing. Finally, the input signal is converted to $100~\Omega$
differential using a wideband Mini-Circuits TCM2-43X+ balun.

The overall system clock is a doubly redundant SiTime SiT5157 Super-TCXO operating at $125~\textrm{MHz}$ with $\pm0.5\textrm{ppm}$ stability over operating temperature and $<0.5~\textrm{ps}$ RMS phase jitter. This clock is distributed to the 4 TURFIOs using an HMC6832, and then is divided down to $7.8125~\textrm{MHz}$ and distributed to the SURFs using an LMK01010, resulting in $<10~\textrm{ps}$ clock skew between SURFs.

\begin{figure}[htbp]
\centering
\includegraphics[scale = 0.5]{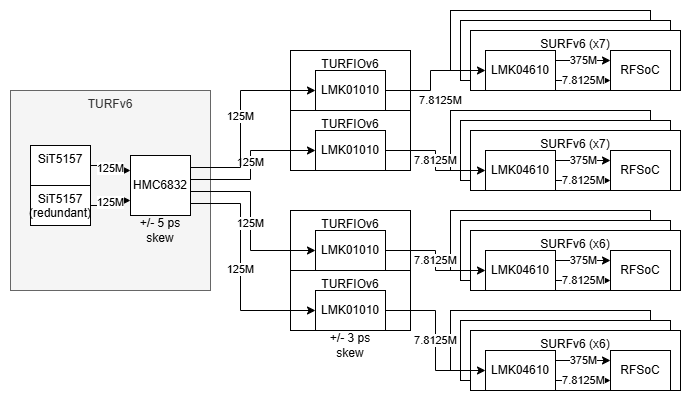}
\vskip 0cm
\caption{ The clock distribution diagram showing system clock forwarding which is designed to keep clock skew as low as possible ($\ll300~\textrm{ps}$) to simplify beamforming and analysis. The main system clock is $125~\textrm{MHz}$, which is divided down to $7.8125~\textrm{MHz}$ and then distributed before being multiplied up to $375~\textrm{MHz}$ to be used as the $1/8$ sampling frequency reference.}
\label{fig:daqClockDiagram}
\end{figure}

The DAQ clock diagram is shown in figure~\ref{fig:daqClockDiagram}. At each SURF, the clocks are multiplied up using an LMK04610 clock synthesizer in a zero-delay feedback mode to a SURF clock of $375~\textrm{MHz}$, used to derive the $3000~\textrm{MHz}$ sampling clock, and the forwarded  $7.8125~\textrm{MHz}$ is then used as the JESD204B-style SYSREF synchronization signal. Clock phase alignment of the derived clocks is established using the synchronous commanding interface from the TURF to the SURF and back. In addition, the RFSoC MTS process is performed with a common target latency that is reachable by all of the RFSoCs, resulting in digitizer synchronization better than 1 sample clock.

\subsubsection{Beamforming Trigger Logic}
\label{sec:beamform_daq}

The trigger is based on a coincidence between the envelope exceeding the threshold of an 8-antenna synthesized beam on two adjacent SURFs. The implementation itself required significant resource optimization to reduce power and meet timing - implementation will be covered elsewhere in detail \cite{pueotriggerpaper} with a brief overview given here.

\begin{figure}[htbp]
\centering
\includegraphics[scale = 0.7]{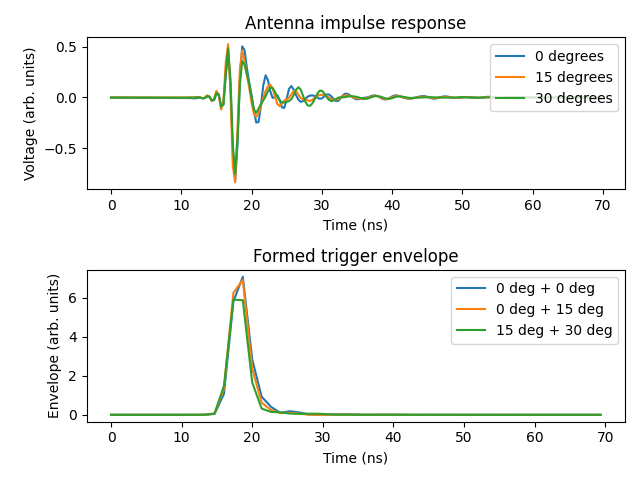}
\vskip 0cm
\caption{Top: Band-limited impulse response sampled at $3~\textrm{GSa/s}$ at off-axis angles corresponding to one and two phi sectors away. Bottom: Effect of summing off-axis signals together on the beamformed trigger envelope. No effect is seen in the trigger envelope other than the power loss due to the off-axis gain.}
\label{fig:beamform_response}
\end{figure}

Prior to beam synthesis, the input signals are conditioned to emphasize the desired signal. Because the antenna gain begins to drop faster off-axis above $700~\textrm{MHz}$ and the Askaryan signal tends to have higher signal at lower frequencies, the signal is first passed through a halfband filter, reducing the bandwidth to $750~\textrm{MHz}$. This also produces a stable off-axis response from the antennas, allowing beamforming to be done with a straightforward add with geometric path delays without amplitude or phase corrections, as seen in figure \ref{fig:beamform_response}.

Following the halfband filter, the signal is then passed through an approximate matched filter using a multiplierless design limited to taps with coefficients as powers of 2. Next, an additional optional radio-frequency interference (RFI) mitigation section was implemented, but not deployed due to thermal concerns in flight.

Finally, the signal is then scaled via an automatic gain control section, followed by a dynamic range reduction to five total bits employed symmetrically ($-15.5\dots15.5$) scaled to an RMS of 4. This significantly reduces the firmware resources required to implement the beam-level triggers while not affecting the overall trigger efficiency \cite{xie2021improving}.

\begin{figure}[htbp]
\centering
\includegraphics[width = 0.9\linewidth]{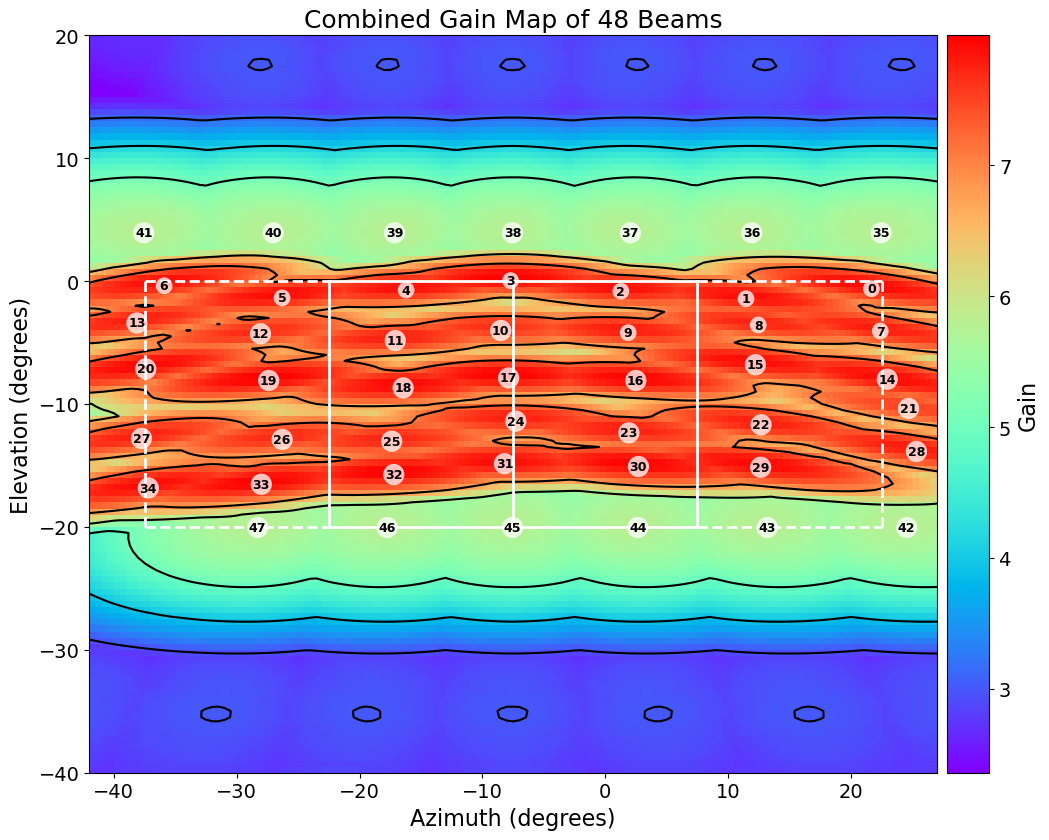}
\vskip 0cm
\caption{Peak gain of all 48 trigger beams as a function of azimuth and zenith, as observed outward from the payload, where the beam centers are indicated by their indices. Lines shown are contours at interval 1. $0^{\circ}$ azimuth is centered on a single $15^{\circ}$ phi sector. Solid vertical white lines shown indicate nominal boundaries of the two phi sectors of antennas used in these beams. Dashed vertical white lines show nominal boundaries of left and right adjacent phi sectors.}
\label{fig:beam_map}
\end{figure}

\begin{figure}[htbp]
\centering
\includegraphics[width = 0.9\linewidth]{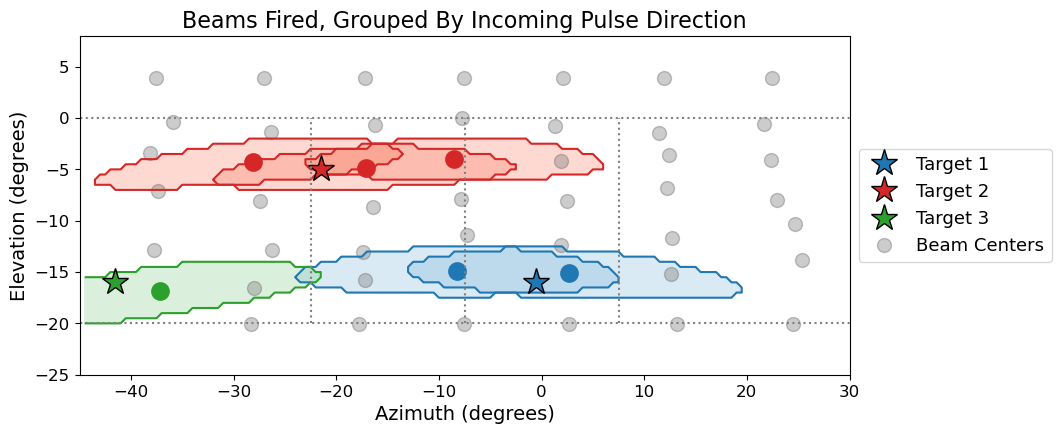}
\vskip 0cm
\caption{The beams triggered by three incoming pulse directions, using the custom beamforming pulsing system as described in section \ref{sec:BeamPulse}. Contours of a gain of 5.5 are highlighted for each triggered beam. }
\label{fig:beam_examples}
\end{figure}

After the signal conditioning, 48 beams are formed, with 35 (0 through 34 in figure \ref{fig:beam_map}) using all 8 antennas in a 5x7 grid covering approximately $0^{\circ} \ldots -17^{\circ}$ in elevation and $-30^{\circ} \ldots~ 30^{\circ}$ in azimuth. The remaining 13 beams are formed with only the 6 antennas in the lower rings. These beams have a significantly wider zenith beamwidth and act to efficiently extend coverage at only a slight loss in overall gain. A map of the combined gain of all beams is shown in figure \ref{fig:beam_map} to illustrate the full coverage. An example of individual beams triggering in response to incoming signals is shown in figure \ref{fig:beam_examples}.

Post-beamforming, a simple trigger envelope for each beam is formed by squaring the resulting signal and computing the $8$-$\textrm{sample}$ sum every 4 samples (equivalent to a $3~\textrm{dB}$ video bandwidth of ${\sim}187.5~\textrm{MHz}$). This envelope is then compared to a threshold value; if any beam is above the threshold the trigger time is then transmitted to the TURFv6. Each beam is assigned to one or more geometric ``octants'' representing which signal arrival directions they are most sensitive to. Each trigger transmits the octants containing a triggering beam as metadata. This combined trigger time and metadata is termed an ``L1'' trigger.

At the TURFv6, the triggers are received and stored at the appropriate time in a rolling window corresponding to $32.768~\mu\textrm{s}$. The SURFv6 also stores all incoming data (without conditioning) in a signal buffer of equivalent size. Once a programmable trigger latency has elapsed (allowing for all triggers to arrive), the instrument trigger status is read out. If L1 triggers in two adjacent SURFv6s are received in geometric octants that share signal arrival direction, an L2 trigger is generated. The trigger validation studies shown in section \ref{sec:TrigVal} reflect the trigger as implemented and flown.

Because RFI for PUEO is mostly directional, the individual SURF sectors can be masked off (``phi masking'') to prevent L2 trigger saturation caused by such RFI. After masking, any individual L2 remaining generates an L3 system-level trigger which is sent out to the SURFv6s to initiate a data readout.

\subsubsection{Communication and Data Flow}

\begin{figure}[htbp]
\centering
\includegraphics[scale = 0.15]{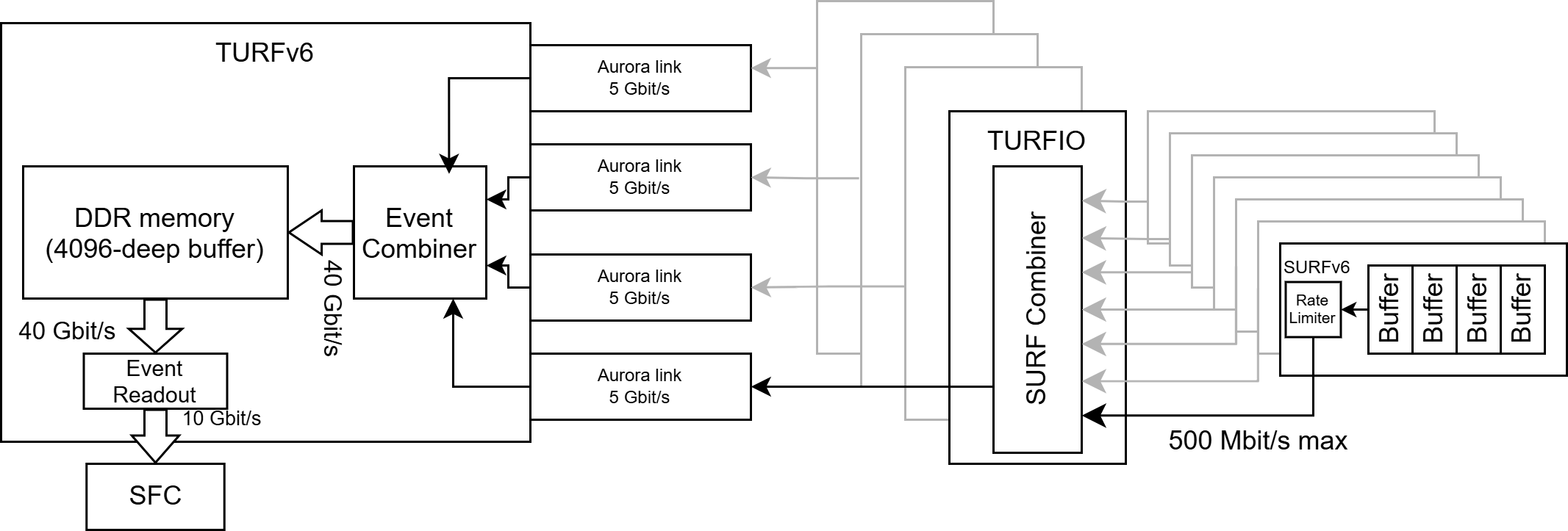}
\vskip 0cm
\caption{The DAQ communication diagram shows the data/commanding path from the SURFs to the TURFIOs and onward to the TURF through the Aurora link, which is then combined and sent to the Science Flight Computer (SFC).}
\label{fig:daq_comms}
\end{figure}

Communication between the TURFv6 and the DAQ crates consists of three main data paths: the trigger and readout request path, the data and commanding path, and the housekeeping path. An overview of the DAQ communication paths is shown in figure~\ref{fig:daq_comms}.

The housekeeping path consists of a serial interface using the sideband signals in the ePCIe cable to a softcore processor (Xilinx PicoBlaze) running in the TURFIO, and then optionally broadcast to all SURFs, which respond with an open-drain multidrop response signal. The SURF housekeeping path communicates directly with the processors onboard each RFSoC and is used for setup and initialization, but then shut down for thermal considerations during normal operation. The housekeeping path monitors board temperature and currents and also allows for power control of individual SURFs. Housekeeping commands are sent from the Science Flight Computer (SFC) as packets encoded with consistent overhead byte stuffing (COBS) \cite{cheshire2002consistent}, and are routed by the TURF to either the desired TURFIO or an internal
TURF housekeeping process.

The data and commanding path uses multi-gigabit transceivers present in the TURF and TURFIO using the Xilinx Aurora protocol, with command responses multiplexed into data responses. The data/commanding path from the SURF operates over a $500~\textrm{Mbit/s}$ serial datapath synchronous to the system clock. The transceivers operate at a user throughput of $5.12~\textrm{Gb/s}$, which is higher than the maximum inbound data rate from the (up to) seven SURFs per TURFIO of $3.5~\textrm{Gbit/s}$, allowing for minimal buffer space at the TURFIO.

The trigger (\textrm{COUT}, from SURF) and event readout request (\textrm{CIN}, to SURF) path is also a $500~\textrm{Mbit/s}$ interface, with a transmit path synchronized to a forwarded clock and the receive path left system-synchronous. These paths allow for clock phase alignment and monitoring by using the tapped delay lines available in the FPGA and RFSoC. The \textrm{COUT} path consists of a $32~\textrm{bit}$-payload with a 12-bit trigger time and an 8-bit metadata (with the remaining bits reserved). The \textrm{CIN} path consists of a $16~\textrm{bit}$-command word for synchronous operations (run start/stop and clock phase synchronization) as well as for high-speed firmware update delivery. The command word is optionally followed by a $12~\textrm{bit}$-trigger time which is used to initiate an event readout request.

Events consist of $1024~\textrm{samples}$ at $12~\textrm{bits/sample}$ - while the RFSoC ADCs themselves are 14-bit, at the frequencies of interest, digitization noise renders the remaining 2 bottom bits essentially useless. The full size of an event at each SURF (with 8 channels each) is therefore $12288~\textrm{bytes}$, and the data path can sustain a throughput of ${\sim} 5~\textrm{kHz}$ - however, the trigger rate is throttled to add a ${\sim} 0.5\textrm{ms}$ delay between readouts (not triggers), limiting the maximum sustained trigger rate at the SURF to ${\sim} 2~\textrm{kHz}$. A 4-deep event buffer allows for back-to-back triggering in the case of bursts of triggers.

At the TURF, the events from each SURF are collected along with event and trigger monitoring information and buffered into DDR RAM with a shared read/write throughput of ${\sim}80~\textrm{Gbit/s}$ and a maximum buffer depth of $4096~\textrm{events}$. These events are then streamed out to the SFC over the $10~\textrm{GbE}$ interface. If the SFC Ethernet link is fully saturated with event data, this would essentially match the $2~\textrm{kHz}$ maximum readout rate from the SURFs.

The TURF also monitors the buffer count available at each SURF and its own outbound buffers and halts triggering when the SURF buffers are full or the outbound buffers near exhaustion. The deadtime when this occurs is accumulated and transmitted in each event. During typical flight operation deadtime was under $1\%$.

Firmware updates via the \textrm{CIN} path utilize the same event readout buffers in the SURF, as they are unused during update operations. Transfer to the RFSoC processors was inspired by the XBERT \cite{hofmann2021xbert} framework, but modified to allow for live readback without clock shutdown and operates inside a Linux operating system. These modifications reduce the effective throughput significantly (to ${\sim}100~\textrm{kB/s}$), as the readout operations must then occur through a secondary low-performance security processor. However, because the \textrm{CIN} path is broadcast to all SURFs simultaneously, this slow update is sufficient.

For ground development on the flight crate, a single SURF was connected via Ethernet to the SFC to allow debugging and update testing prior to deployment. Additionally, multiplexed JTAG access to all SURFs allows for emergency recovery from a failed firmware update using the Xilinx JTAG system console and a custom direct file download mechanism.

\subsubsection{Flight Software Acquisition Daemon} \label{sec:acqd}

The TURF communicates with the flight computer via $10~\textrm{Gbit/s}$ Ethernet over twinax copper ({\fontfamily{cmtt}\selectfont 10GBASE-R}) using User Datagram Protocol (UDP) packets. The UDP/IP infrastructure in the TURF programmable logic is based on the {\fontfamily{cmtt}\selectfont verilog-ethernet} core used in the Corundum \cite{forencich2020corundum} project and is capable of throughput at the full line rate. The Ethernet interface is used for all DAQ interfaces (data, commanding, and housekeeping) and is multiplexed using standard UDP ports. Commanding and housekeeping paths are designed as command/response, allowing for lost UDP packets to be handled with a timeout.

TURF events are split and transmitted in fragments of configurable length to minimize UDP overhead and also to adjust to the maximum transmission unit (MTU) capabilities of the Ethernet interface of the SFC. The total PUEO event size was finalized at $448.5~\textrm{kB}$, and the fragment size utilized in flight was $2048~\textrm{bytes}$, corresponding to $225~\textrm{fragments}$.

Once all fragments for an event are received, the SFC transmits an acknowledgment to the TURF to allow that event buffer to be reused. Acknowledgments also act to restrict the number of events transmitted before the acknowledgment of the first is received, similar to a sliding window protocol. During flight, up to 32 events could be in transmission before event streaming was paused for an acknowledgment. Note that this pause occurs at the readout side, and therefore no deadtime is accumulated in this situation. Events could also be ``no-acknowledged'' (nacked) to force a retransmission; however, this feature was unused during flight.

In addition to receiving events, the flight computer is responsible for configuring operational parameters and ensuring the DAQ is in the correct state. To this end, there are two implementations of the DAQ interface software. A \texttt{python3} module~\texttt{pueo-python}~(\url{https://github.com/barawn/pueo-python}) was used during development and also used in flight for DAQ bringup, initialization, and verification of functional state. This Python module has an abstract hardware interface mechanism, allowing the same modules to be used on the SURFv6 and TURFv6 processors as the SFC. For in-flight data taking as part of the main data-taking daemon \texttt{acqd}, an independent C23 implementation called \texttt{pueodaq} is used, but does not implement DAQ initialization.

Data is taken in runs where each run corresponds to an invocation of the \texttt{acqd} daemon. A run proceeds until either an anomalous condition (e.g. DAQ hang or unexpected condition) is reached, data taking is stopped by command, or a configurable timer expires. This timer can be set via command and takes effect even in the middle of a run.

At the beginning of each run, \texttt{acqd} runs a python script to verify the DAQ is ready to take data, and if it is not, the DAQ initialization procedure is run. Due to increased in-flight power consumption (discussed in section~\ref{sec:thermalpreflight}), the DAQ initialization script was executed far more often than anticipated, and a cooldown feature was implemented and deployed over Starlink in the \texttt{acqd} daemon. This cooldown feature turns off the DAQs at the end of a run and waits a set amount of time before quitting. When \texttt{acqd} quits it is automatically restarted by \texttt{systemd} unless data taking mode is disabled by command.

In order to take data, the DAQ is initialized by aligning the clocks on the TURFIOs and SURFs while clearing errors, semi-autonomously. Next, MTS is executed per SURF and initial trigger thresholds are set before a final check to ensure the system is ready for data taking. Once the DAQ is ready, triggering and data flow are enabled and events start flowing in. In addition to RF triggered events, there is an option for GNSS-originated PPS triggers and for software triggers. During the flight, software triggers ran at 5 Hz.

A thread (or optionally, more, although only one readout thread was used in flight) is dedicated to reading in UDP packets which form event fragments and performing bookkeeping by incrementing atomic counters. A separate thread is used to send acknowledgments, assuming there is a buffer available to receive the data. Testing on the ground before flight indicated that this DAQ and software system supports event rates above $1$~kHz. However, the event rate near $1$~kHz could not be sustained in practice due to limitations downstream in the storage and prioritizer (see section \ref{sec:prio}) throughput, which limited sustained throughput without dropping events to around 300 Hz and 200 Hz, respectively.

In addition to receiving data, the flight software is responsible for setting L1 thresholds for the total of 1152 main instrument beams. This is done using a scaler goal and a servo which adjusts the thresholds in the direction of the scaler goal. The scaler goal used in flight was 666 Hz / beam, which resulted in $\sim$100 Hz global trigger rate in testing with terminated amplifiers.

To constrain the trigger rate to manageable levels, L2 beams can be masked. If the L2 scaler for a single SURF is over 30 Hz, such as in the case of anthropogenic EMI, that SURF is disabled until the trigger rate drops below 15 Hz for more than three consecutive seconds.

In addition to saving waveform data, the flight software also saves a suite of DAQ housekeeping data, including scalers, thresholds, AGC parameters, deadtime counters and other indications of DAQ health.

\subsection{Navigation}
The PUEO navigation suite is designed to deliver \ang{0.05} precision in all three attitude angles (heading, pitch, and roll), as well as \SI{50}{\meter} precision in 3D position. The interferometric pointing resolution is expected to achieve \ang{\sim0.1}, and, as described in section \ref{sec:design}, will yield the smallest uncertainty in the elevation direction which enables differentiation between above- and below-horizon events. These design goals ensure that the event pointing resolution is not limited by the precision of the navigation systems. 

The navigation suite comprises two high-quality inertial navigation systems (INS), a purely GNSS attitude system, a sun sensor array, a magnetometer, and two star trackers. Both INS units have temperature-activated heaters to deliver $\sim10$~W to avoid cold temperatures limiting livetime. The instruments in the navigation suite are summarized in table \ref{tab:nav_instruments} and described below. 

\begin{table}
\caption{The suite of navigation instruments together meets the PUEO design goals.}
\begin{tabular}{| c | l | l | l |}
\hline
Component & Observable & Heading res.& Pitch/roll res.\\ \hline
Boreas & Accelerations, angular velocities, GNSS & \ang{0.006} & \ang{0.005} \\ \hline
CPT7 & Accelerations, angular velocities, GNSS & \ang{0.01} & \ang{0.03} \\ \hline
ABX-Two & GNSS & \ang{0.02} & \ang{0.075} \\ \hline
Sun sensors & Sun & \ang{\sim0.1} & \\ \hline
Magnetometer & Earth's magnetic field & \ang{\sim0.1} & \\ \hline
Star tracker & Magnitude $<6.4$ stars & \ang{0.0006} &  \ang{0.0006}/\ang{0.012} \\ \hline
\end{tabular}
\label{tab:nav_instruments}
\end{table}

\subsubsection{Inertial Navigation Systems}

PUEO utilizes two different INS units: the Advanced Navigation Boreas D90 \cite{boreas}, and the Novatel CPT7 \cite{cpt7}. The Boreas utilizes traditional micro-electromechanical systems (MEMS) accelerometers and digital Fiber Optic Gyroscopes (DFOG); these are very high precision, but are less tested than traditional fiber optic gyroscopes (FOG). The CPT7 is made by Hexagon/Novatel, and utilizes MEMS technology for both accelerometers and gyroscopes. 

Due to the novel nature of the technology on a balloon platform, a dedicated Allan deviation test was performed, where the noise characteristics of the inertial system were measured \cite{ieee_gyros}. The performance specifications of the Boreas accelerometers include a bias instability of \qty{7}{\micro g} and velocity random walk of \qty{17}{\milli\meter\per\second\per\sqrt\hour}. For the gyroscopes, the bias instability is \qty{0.001}{\degree\per\hour} and the angle random walk is \qty{0.001}{\degree\per\sqrt\hour}. The Allan deviation test was performed in the basement of the physics building at the University of Delaware on a heavy optical table over a weekend to minimize activity that might introduce noise into our measurement. These tests validated the specifications for the Boreas accelerometers and gyroscopes to be utilized for the flight. For the CPT7, the accelerometers have a bias instability of \qty{0.075}{\milli g} and a velocity random walk of \qty{0.06}{\meter\per\second\per\sqrt\hour}. The gyroscopes have a bias instability of \qty{0.45}{\degree\per\hour} and an angular random walk of \qty{0.06}{\degree\per\sqrt\hour}. These values were confirmed in the same testing methods as with the Boreas. Both INS units were vacuum- and cold-tested to ensure their suitability for high-altitude operation. To ensure they were operating in a flight-like configuration, they were connected to GNSS antennas using feedthroughs in the vacuum chamber. The INS units were flight-qualified with these tests.

To minimize systematic errors in the inertial systems due to centrifugal forces, the INS units are mounted along the vertical axis of the payload within its central column. They are housed inside custom RF-tight enclosures. These enclosures were verified to block RF signals from the Boreas at the anechoic chamber at the University of Kansas. Two GNSS antennas are mounted on the antenna array at the top of the payload and connect to each unit. There are ethernet ports for ground-based communications and a single flight power and interface cable carrying $12$~V and an RS-232 connection to the MIE.

\subsubsection{Multi-antenna GNSS: ABX-Two}

The ABX-Two, made by Trimble, is a purely GNSS attitude system with three GNSS antennas.
This is the successor to the ADU5, which flew on all four ANITA flights. In the same process as the INS units, the ABX-Two was qualified for flight via vacuum and environmental testing at the University of Chicago. It was also RF-tested, though this system is housed inside the MIE, so it was qualified for RF tightness along with the MIE.

\subsubsection{Sun Sensors}
An array of eight Sun Sensors is installed on the payload and separated by \ang{45} in azimuth. Each Sun Sensor consists of two main components: a $10$~mm $\times$ $10$~mm Duolateral Position Sensing Detector (PSD) \footnote{The PSM2-10 from On-Trak Photonics.} (previously used in ANITA flights) and readout electronics. These are located in custom RF-tight enclosures with a small pin-hole and baffles restricting the azimuthal field of view to \ang{110}, ensuring that the Sun is always visible to two sensors. Each PSD has 4 electrodes and their currents are recorded. Sunlight hits the detector surface at a specific angle and the position of the brightest spot provides the orientation of the payload with respect to the Sun. 

Each current signal is shaped through a 2-stage amplifier and digitized by a fast 4-channel 24-bit ADC \footnote{The ADS1220.} per Sun Sensor. To take into account thermal effects such as the temperature dependence of the current draw, each Sun Sensor has a high-precision calibrated temperature transducer attached directly to the bottom of the PSD \footnote{The AD590.} and a temperature sensor internal to the ADC chip. All six digitized values of current and temperature are combined by an ATmega328 microcontroller to form a data packet, which is then transferred over the RS-232 interface through filter pins to readout electronics located inside the MIE. This eliminates the possibility of analog signal distortion due to traveling through long transmission lines. The readout electronics are responsible for providing power to each sensor and combining the data from all eight sensors to forward on to the SFC via a USB interface. The key component of the readout board is a Teensy 4.1 extension board. 
All eight Sun Sensors performed nominally during the PUEO flight at a rate of $8$~Hz, and their data is currently being calibrated and processed into attitude.

\subsubsection{Magnetometer}
Along with the Sun Sensors, geomagnetic field measurements are used to reconstruct payload pointing and geographical coordinates. 
The PUEO instrument utilizes a fluxgate magnetometer with a resolution of $20 \times 10^{-6}$~G, model 529 Applied Physics Systems, which interfaces with the housekeeping system and was able to deliver magnetic field components on a cadence of $2$~s throughout the flight. This unit is installed on the main deck of the gondola in an RF-tight enclosure. 
It was calibrated on the ground using a 3-axis Helmholtz coil and qualified in the same environmental tests as the other navigation systems.
Data from the magnetometer is currently undergoing post-flight analysis.

\subsubsection{Star Trackers}
Two nominally identical star trackers were developed and flown for redundancy and as a first-of-its-kind technology demonstration on a freely rotating balloon platform. The units are located on the main deck of PUEO in RF-shielded enclosures. Their large white baffles are visible in figure~\ref{fig:payload}, which also illustrates their pointing inclination relative to the payload’s horizontal plane. Each enclosure uses an RF-sealing mesh across the camera baffle aperture, which introduces a regular distortion from the grid in the point spread function (PSF) of observed stars. Each unit uses a ZWO ASI294MM astrophotography camera featuring a 4/3" Sony IMX492 $14$-bit ADC sensor operating $2$ by $2$ binned, yielding $4144$ by $2822$ pixels at a $4.64$~$\mu$m pixel size. The field of view of each unit is $10.8$ $\times$ $7.4^{\circ}$. A VersaLogic Owl single-board computer interfaces with the camera through USB and with the SFC through multimode fiber and RS-422. Each tracker includes a $630$~nm long-pass filter (to reduce scattered blue light), lens\footnote{Sigma $105$~mm f/$1.4$ DG HSM Art lens}, lens controller, and two dedicated $4$~TB SSDs (solid-state drives). Raw ($2$ $\times$ $2$ down-sampled) images were stored in the onboard SSDs, and a small subset of images were downloaded during flight. The trackers were qualified for flight through vacuum and RF-emission testing. Twilight conditions on the ground were used as a surrogate for the expected Antarctic illumination at altitude, but after flight we determined that this underestimated the apparent sky brightness, resulting in a reduction in the number of discernible stars during flight compared to expectation.

There was no on-board real-time image processing or background subtraction. Occasional remote access to the Owl computers through use of the onboard Starlink terminal allowed for in-fight adjustments to the camera parameters based on real time viewing of the downloaded images. However, in addition to a brighter-than-expected sky background, many of the images were also contaminated by a persistent structured optical ghost with a circular pupil/stop-like profile. As the baffle prevents light from the Sun or stray light bouncing off the balloon from directly entering the detector, the most likely explanation is sufficient light bouncing off the black baffle interior and then, via an off-axis path into the lens, illuminating an internal lens component. We compare our post-flight image pipeline that uses a \texttt{python} script for star detection to the results obtained by applying the ESA Tetra3 and Cedar Detect packages to the images. A significant fraction of images can be solved for a much lower investment in per-image CPU time, but we find that the structured optical ghost does not subtract out well, causing a higher rate of solution failure than with the \texttt{python} script.

Unexpected interruptions affected both star trackers during early flight operations. The trackers shared a $12$~V supply with the navigation subsystems and some data-acquisition hardware. Because they were a secondary payload subsystem and had not yet demonstrated reliable real-time pointing solutions, nor was it apparent they were saving any usable data, the star trackers were kept powered off for most of the remainder of the flight as a precaution against further disruptions. Both trackers and their onboard SSDs were recovered successfully, and the archived images are undergoing post-flight analysis.

\subsection{Housekeeping Subsystem}
Various sensors as well as the power system are monitored and controlled by the housekeeping system, which is built around a TM4C123GXL launchpad development board. The housekeeping system interfaces with the flight computer via a USB interface, providing serial communication and an in-circuit debug interface. 

The housekeeping system also records internal and external temperature measurements with $94$ DS18B20 1-wire temperature probes operating in parasitic mode and therefore only needing coaxial cable for readout. To mitigate RF noise pickup, probes are placed inside custom-built copper enclosures with SMA connectors. These signals are sent through filter pins at the MIE interface. This array of sensors is secured to the instrument and payload with conductive epoxy or copper tape. The housekeeping system gathers a temperature profile of the MI antenna array, useful for characterizing the temperature-dependent amplifier behavior. During flight, antenna temperatures varied between $0$ and $50^{\circ}$C as the payload freely rotated. 

The housekeeping commands the receiver board power rails for control over individual phi sectors and rings as a contingency to power down individual channels during flight. Though not needed during flight, this feature was utilized during preflight validation with ground system testing.

\subsection{Science Flight Computer Subsystem}
The Science Flight Computer (SFC) subsystem is responsible for configuring and collecting data from all subsystems, passing data to the storage and telemetry subsystems, and responding to commands. This subsystem includes many ancillary electronics such as DC-DC converters, networking hardware, and media converters. 

\subsubsection{Hardware}

\begin{figure}
  \centering
  \includegraphics[width=5.25in]{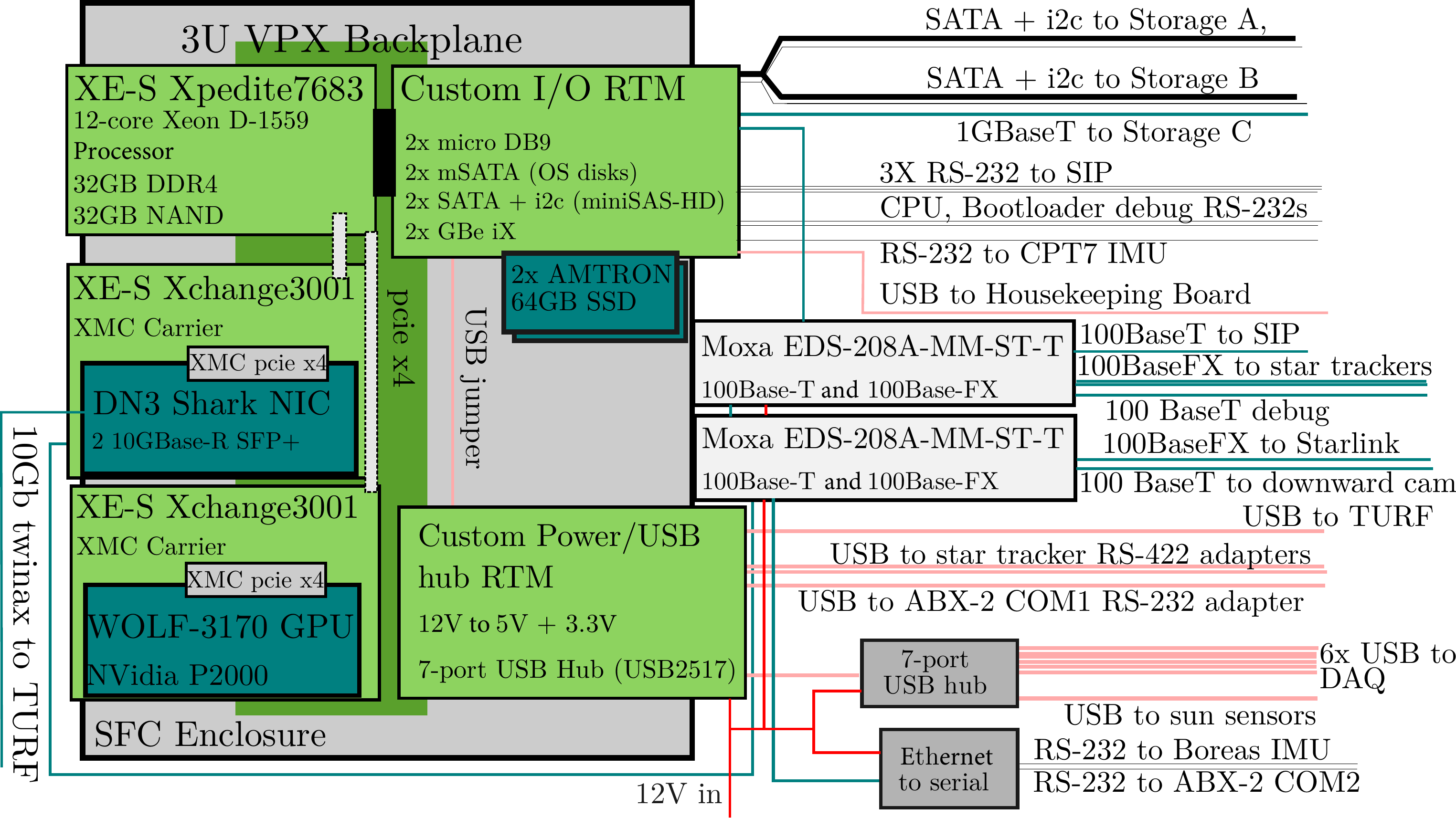}
  \caption{An overview of the hardware comprising the Science Flight Computer (SFC) and connections to other subsystems. See text for details.}
  \label{fig:sfc}
\end{figure}

Figure~\ref{fig:sfc} shows an overview of the SFC subsystem and provides many details on the hardware components and their interfaces. The SFC is built around a 12-core Xeon D-1559 CPU (3U VPX X-ES XPedite7683) with 32 GB of DDR4 RAM as well as $32$~GB of onboard NAND storage, used as a telemetry cache. This unit and associated electronics are housed inside a custom-built conduction-cooled crate with a 3U VPX backplane (Kontron BKP3-CEN03-15.2.9-4). Through the backplane, PCI-Express 3.0 interfaces with a GPU (Wolf-3170 XMC, used for data prioritization) and a network card (EKG DN3-C001-Shark), each on VPX XMC carriers (X-ES XChange 3001). The network card includes a twinax 10 Gb/s connection to the TURF for data acquisition, as well as a connection to the pair of $100$~Mb/s network switches (Moxa EDS-208A-MM-ST-T) using an SFP-to-RJ45 module (Vivotek SFP-1000-CPTX-X1I Industrial Gigabit).

On the other side of the backplane, two custom rear transition modules (RTMs) break out additional peripheral interfaces from the CPU. Two SATA ports are broken out to mSATA $64$~GB SSDs (Amtron  MSM-AK0064GI) that form a RAID1 pair for the OS disk. The SFC connects to the SIP through three of six RS-232 lines, while the others are used for a navigation subsystem and debugging lines to the CPU and bootloader. The CPU has two USB interfaces. One is used to communicate with the housekeeping board, while the other is broken up using a USB hub to fan out to the USB hub inside the TURF, additional navigation subsystems and an additional hub hosting debugging pathways for the TURFIOs of the DAQ. An external (SFF-8644) MiniSAS-HD connector carries two SATA lanes and two I${^2}$C lines connected via a Y-cable to the storage arrays (``StorageA'' and ``StorageB''). Two iX ports provide 1000-BaseT ethernet, one connected directly to the SSD storage array (``StorageC'') and the other connected to the previously-mentioned pair of network switches for redundancy with the network card connection. These switches extend the SFC connections to the star trackers and Starlink through 100 Base-FX multimode fiber and also provide access to ethernet-based telemetry systems inside the SIP. Additionally, a dual Lantronix EDS82100 ethernet to serial device is used as a serial breakout for navigation.

All SFC subsystem power is derived from an input $12$~V line with a custom power converter board which includes an eFuse at a current limit of $8$~A. The input $12$~V and the down-converted $5$~V and $3.3$~V are used by the CPU and the various peripherals. The SFC subsystem uses around $50$~W when idle. This increases to $70$~W due to CPU and GPU load when taking data.

\subsubsection{Software}

The PUEO flight software\footnote{See \url{https://github.com/PUEOCollaboration/pueo-flight-software}.} is a primarily C17-based framework of daemons used to operate the instrument running on AlmaLinux 8. As is typical for AlmaLinux 8, \texttt{systemd} is used to orchestrate the various services to operate the instrument. The daemons share a common custom configuration format based on code generation from configuration specifications as well as standard inter-process communication (IPC) mechanisms with POSIX message queues and \texttt {tmpfs}-exported status information. Each daemon runs as a \texttt{systemd} service, using \texttt{journalctl} for logging and \texttt{systemd} watchdogs. The raw data format (\texttt{libpueorawdata}\footnote{See \url{https://github.com/PUEOCollaboration/libpueorawdata}.}), the DAQ interface (\texttt{pueodaq}) and the storage A and storage B interface (\texttt{libpals}) are implemented as external shared libraries.

A modified version of \texttt{gpsd}\footnote{\url{https://gpsd.gitlab.io}} with added device support communicates with the various GNSS and IMU devices, while \texttt{chrony} is used for network time synchronization. Several instances of \texttt{autossh} are used to create reverse SSH tunnels to known external hosts (when reachable). \texttt{CUDA} 12 is used for GPU programming, along with the proprietary Nvidia driver. Several networking parameters are tuned, including the UDP receive buffer size and the TCP congestion control algorithm.

Below is a summary of the most important daemons that run on the flight computer:

\begin{description}
 \item[\texttt{sipd}]: Manages the serial connections to the SIP, responsible for handling commands sent from the ground and forwarding them to the responsible daemon via message queue. It is also responsible for collating housekeeping-related data and forwarding information to the high-rate serial port used for TDRSS. As \texttt{sipd} receives information from the SIP, it will set the CPU time if more than a minute behind (a feature intended for startup, as the SFC lacks a battery-powered clock).

\item[\texttt{telemd}]: Other daemons send data to be telemetered to its message queue inboxes which are then collated and sent to the shared high-rate inbox on \texttt{sipd}. \texttt{telemd} can be configured via command to decimate any of the streams arbitrarily.

\item[\texttt{hskd}]: Interfaces with the housekeeping board, via its serial interface, requesting and recording housekeeping data as well as responding to commands to turn certain subsystems on and off.

\item[\texttt{storaged}]: Writes all data to the storage arrays. All data to be written to disk arrives to a \texttt{storaged} message queue, where it is repackaged for disk and then fanned out to the available different storage devices. It also manages switching between the different drives used for StorageA and StorageB as well as ensuring that the StorageC NFS mount is present.

\item[\texttt{acqd}]: Collects data from the DAQ and configures the operational parameters controlling the trigger. The thresholds and L2 masking of the DAQ are adjusted on 1-second cadence depending on operational conditions. Data is fanned out to \texttt{storaged} and \texttt{prioritizerd}, although \texttt{acqd} can be configured to skip prioritization and directly send a subset of data to telemetry queues. 
It also optionally can dictate the amplifier state per phi sector. Further details are available in section~\ref{sec:acqd}.

\item[\texttt{navd}]: Acts as a \texttt{gpsd} client to collect data from each GNSS device, store to disk, and send for telemetry. It receives attitude information from the Boreas, CPT7 and ABX-2 and timemark information from the timing GNSS, which tags a subset of triggered events.

\item[\texttt{sund}]: Interfaces with the sun sensor readout board to read out the sun sensors, as well as managing the firmware of the sun sensor board.

\item[\texttt{prioritizerd}]: Implements the online analysis described in section~\ref{sec:prio}. Events are streamed in from \texttt{acqd} and batched out to the GPU, where events are assigned priority flags and are configured to a telemetry probability. In addition to TDRSS telemetry, prioritized events may be saved for Starlink download.

\end{description}

There are various other small daemons (e.g. ensuring Starlink is in the configured power state, transferring Starlink data when Starlink is up,  gathering basic information from the star trackers) that are implemented as bash scripts. LOS telemetry, which is ethernet based, is handled by each daemon which has its own UDP socket and port.

\subsubsection{Prioritizer Algorithms } \label{sec:prio}

PUEO collects data at a higher rate than the throughput of the telemetry. Coupled with the risk of not recovering data drives if the payload terminates in an inaccessible location, a prioritized downlinking approach is implemented. Drawing from previous prioritizer designs (see \cite{BenStruttThesis}), PUEO's prioritizer runs on the SFC GPU and executes low-level analysis strategies to identify interesting, signal-like events. By creating reconstruction maps, coherently summing, and integrating power over the resulting Hilbert envelope, events are assigned priority levels and inserted into a queue for downlinking. The following properties are used to tag events; their threshold values are tuned preflight using ground-based data:
\begin{itemize}
    \item \textbf{Signal-like:} Events that reconstruct below horizontal and have coherently summed Hilbert Envelopes where most of the power is emitted in a short, configurable time period are promoted to high signal-like ranks. There are three ranks, denoting how signal-like the event is:
    \begin{itemize}
        \item \textit{Most Signal-like:} 30\% power in $<$10\% of envelope's time.
        \item \textit{Signal-like:} 30\% power in $<$12.5\% of envelope's time.
        \item \textit{Noise-like:} above horizontal or emit 30\% power in $>$12.5\% of envelope's time. 
    \end{itemize}
    Each signal rank has a tunable downlinking probability, with most signal-like events being preferentially downlinked. However, samples of all event classes are sent to provide an accurate relative sampling of each (signal-like events, calibration events, anthropogenic noise events, etc). This enables a full analysis on the downlinked-and-prioritized dataset only, reducing the risk of failing to achieve the mission goals if the data is not recovered. 
    \item \textbf{Calibration Pulse:} Events that are recorded within the same configurable subsecond window as an expected ground calibration or HiCal (see section~\ref{sec:calibration_hi}) pulse are tagged as likely calibration pulses. More information on calibration pulsing can be found in section~\ref{sec:calibration_hi}. These events have a separate downlinking queue to ensure sufficient calibration events are telemetered.
    \item \textbf{Payload Blasts:} Previous ANITA flights saw various forms of self-triggered noise, called payload blasts \cite{AndrewLudwigThesis}. These flights found that payload blasts appeared on the upper ring or bottom 3 rings of antennas preferentially which saturated large portions of the payload, or otherwise contaminated the waveforms with low-frequency noise. The prioritizer classifies payload blasts in the following ways, with each threshold value configurable during flight:
    \begin{itemize}
        \item \textit{Top Ring Blast:} An event where the largest top ring antenna peak-to-peak was 1.2 times larger than the largest bottom ring antenna peak-to-peak.
        \item \textit{Bottom Ring Blast:}  An event where the largest bottom ring antenna peak-to-peak was 1.5 times larger than the largest top ring antenna peak-to-peak.
        \item \textit{Full Payload Blast:} An event where 25 antennas had an RMS exceeding 200 ADC. 
    \end{itemize}
    An event tagged as a payload blast has a reduced downlinking probability compared to its non-blast tagged counterparts.
\end{itemize}

\subsection{Data Storage}
Three storage vaults were flown, named StorageA, StorageB, and StorageC. StorageA and StorageB are identical arrays of Helium hard drives while StorageC is an SSD array. Each of the three devices is designed to be removed from the instrument enclosure externally, in case a rapid partial recovery was necessary.

\subsubsection{StorageA and StorageB}

These vaults each consist of six 22 TB Helium-filled Hard Drives (Western Digital WD221KRYZ). ANITA had previously demonstrated that such sealed hard drives could be operated in vacuum with suitable thermal management~\cite{ANITA3}. These drives were connected to the SFC via a custom SATA multiplexing circuit board. This PCB includes power rails and one SATA input and nine SATA outputs, six of which are used in flight. Only a single drive is powered, and its corresponding SATA path is activated at any given time. I$^{2}$C GPIO expanders control high-speed differential switches to implement the SATA path. StorageA and StorageB are connected to the SFC using a MiniSAS-HD Y-cable. MiniSAS-HD carries four SATA pairs, two of which are used to carry the I$^2$C lines. This arrangement allows the drives to enumerate at SATA3 speeds, significantly more than the drive maximum transfer speed. 

The enclosures, one for each of StorageA and StorageB, include vibration-damping for the hard drives, protecting them from shocks when the payload lands. Unlike other subsystems these enclosures are mounted to the MIE wall with quick-release plates. The units were thermal-vacuum tested and all of the helium drives functioned nominally during the flight. The enclosures survived termination and the data from the drives has been recovered with no issues found on the $55$~TB recorded.

\subsubsection{StorageC}

This vault consists of five $16$~TB $2.5$'' SATA SSDs (three TeamGroup EX2 and two Mushkin Source HC; nearly identical inside other than firmware) connected to an Exaviz\footnote{Formerly Axzez.} Interceptor single board computer, built around the Raspberry Pi Compute Module 4. The Interceptor runs Debian 12 and exports the five SATA disks via NFS to the flight computer over a 1000BASE-T connection. The flight computer round-robins writing data between the SSDs so that they are usually idle. Similar to the other data vaults, this enclosure is designed for quick-release. It includes heatsinking of the Exaviz computer, the SSDs, a power-regulation circuit board and EMI gasketing. This enclosure also survived termination and the data from these drives, redundant with the others, was clean of corruption.

\subsection{Power}
Long-duration stratospheric balloon flights launched from high-latitude sites benefit from nearly continuous solar illumination, enabling sustained PV power generation throughout the mission. PUEO uses an azimuthally symmetric solar power system, supplemented by onboard batteries, to provide both primary and backup power. The solar panels are stowed during launch, covering the bottom two rings of antennas as shown on the left side of figure~\ref{fig:gondola_dims}. Shortly after launch, they are deployed via a rigging system driven by two redundant actuators powered by dedicated $24$~V batteries. A successful ground deployment of the panels is shown in figure~\ref{fig:payload}.

The solar power system consists of eight PV arrays distributed around the payload. 
Each PV array consists of six $30$-cell Maxeon Gen III solar panels manufactured by SunCat Solar, with the cells laminated onto a lightweight honeycomb panel structure. Within each array, the panels are configured as two parallel strings of three panels in series, with isolation diodes to prevent reverse-current flow during dormant periods. Similar solar array designs have extensive flight heritage on long-duration balloon experiments such as ANITA, SuperTIGER, and XL-Calibur \cite{GORHAM200910,ABARR2021102529,Binns_2014}, demonstrating their reliability in high-altitude balloon flights. The PV system consistently provided the payload's required maximum power of $1200$~W during the flight.

Power generated by the PV arrays is regulated by a $60$~A TriStar MPPT charge controller manufactured by Morningstar Corporation. 
The controller manages charging of the payload battery system and distributes power to the payload electronics. The charge controller operates in \textit{float mode}, resulting in an average battery voltage of $58$~V. Temperature compensation is performed using a remote thermistor, allowing the charge controller to reduce the battery voltage as temperature increases, and increase it as temperature decreases.

The battery system consists of eight $12$~V, $20$~Ah rechargeable sealed lead-acid batteries produced by Power Sonic Holdings LLC (model PS-12200). These are the same battery model flown on SuperTIGER and XL-Calibur \cite{Binns_2014,ABARR2021102529}. For PUEO, the batteries are arranged as two sets of four batteries wired in series, providing two $48$~V bus systems, with each set placed inside an RF-tight battery box. The two boxes were refurbished from those flown on the ANITA-IV instrument \cite{ANITA4}. With no PV input, the payload bus voltage is $48$~V; during flight, with the charge controller operating, the bus voltage floats to approximately $58$~V, as described above.

From the bus voltage, all payload power is regulated down via DC-DC converters inside a single RF-tight enclosure located in the MIE. This power box houses the MPPT charge controller, relays, and the housekeeping circuit board for monitoring and control. Three Flex Circuits BMR456 bus converters, rated at $420$~W,  provide $12$~V for the Data Acquisition (DAQ) crates and the flight computer system. Vicor V48C5T100BL DC-DC converters paired with a uRAM2T22 MicroRAM provide $4$~V and $5$~V for the RF AMPAs, including LF, and the RF receivers. The power control electronics, relay circuits, and Science Stack are powered from a $24$~V rail supplied by a Vicor V48C24T75BL. The remaining payload electronics, including the navigation, data-storage, and GPS systems, are powered from the $12$~V rail of a Vicor V48C12T150BL. Starlink is supplied with direct bus voltage switched by a relay. Crydom $10$~A and $5$~A relays switch power to individual subsystems. Relays and power converters are controlled by either the Science Stack, the housekeeping board, or, in some cases such as the RF system, both.

\subsection{CSBF Science Instrument Enclosure}
CSBF provides the SIP, which routes all telemetry (bidirectional) between the MIE and the ground station equipment. It provides bias voltage for the telemetry antennas such as TDRSS and collects NASA-specific mission-essential monitoring data such as GPS coordinates. CSBF also provides the Science Stack, modular electronics that manage the bridge between command and control processing (and other flight support services) and the MIE. The Science Stack was placed inside a custom Faraday enclosure and mounted to the inner static plate/wall of the MIE. A Science Stack command, i.e., turn on the flight computer, is sent via the CSBF interface to the payload, which receives and parses the command and forwards it appropriately to the science stack, which then sends the voltage pulse to the power enclosure relay. On most payloads, the SIP is housed inside an enclosure provided by NASA and CSBF. As on the ANITA flights, the SIP was placed inside a dedicated nickel-plated RF-tight enclosure which utilized EMI fingerstock and gasketing in sealing the lid and walls \cite{GORHAM200910}.

\section{Telemetry}
\label{sec:telem_hi}
Telemetry is the only means of communication between the payload and the ground teams during flight. It is critical to ensure that the telemetry data stream operates under all conditions and is detailed enough to enable the diagnosis of anomalies that occur during flight.

\begin{figure}
\centering
\includegraphics[width=0.8\textwidth]{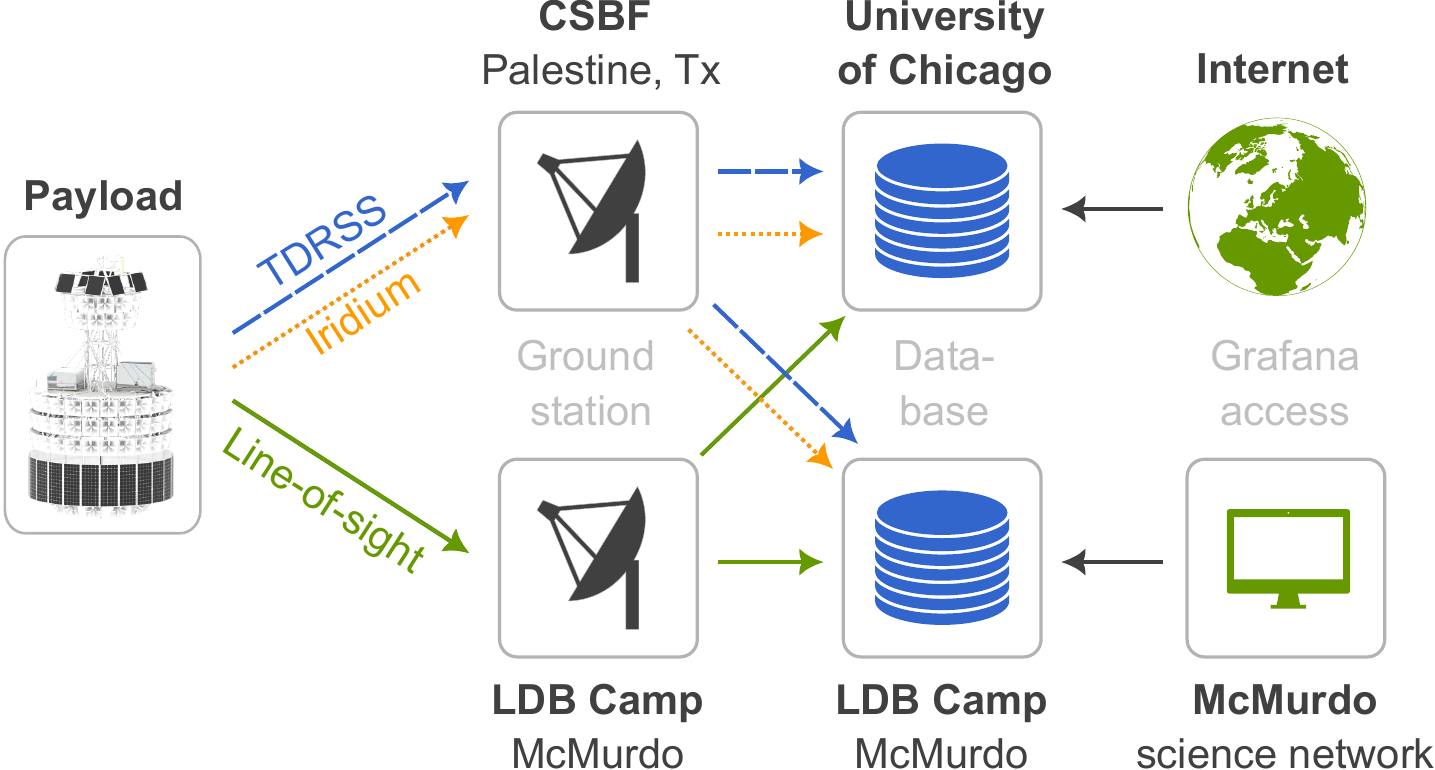}
\caption{Overview of the telemetry data flow from the payload to insertion of monitoring information into databases. The commanding uplink flows from the Ground Stations to the payload.}
\label{fig:telemetry-overview}
\end{figure}

The telemetry system collects information about the status of the payload and its subsystems, organizes this information into several data streams, and sends these to ground systems using three distinct wireless communication links: line-of-sight (LOS) radio, Tracking Data and Relay Satellite System (TDRSS), and Iridium. Figure~\ref{fig:telemetry-overview} summarizes the design of the telemetry system. All telemetry information is collected by the SFC and formatted into a corresponding packet. We implement several packet types to represent the different telemetry data. During flight, we exercised predefined commands sent from the ground, to reset subsystems, including the star trackers, the Starlink system (discussed in sec~\ref{sec:starlink}), and the DAQ system. Telemetry bands used for downlink are all outside the PUEO instrument band.

Telemetry packets are sent via serial interface to the CSBF SIP which restructures the data stream and routes it via one of three attached communication links. During early stages of the flight, telemetry is sent via LOS transmission to the telemetry building at LDB with a throughput of up to $10$~Mbps. When LOS is lost, the payload telemetry utilizes two lower-throughput links: i) TDRSS at $6$~kbps; and ii) Iridium at a maximum of $248$~bytes every 30 seconds. The TDRSS link is limited to periods with satellite visibility. Prioritized events are downlinked in periods of LOS or TDRSS only due to the required bandwidth; Iridium is not sufficient to support full events. During TDRSS, prioritized, downlinked events are limited to waveforms that participate in the L2 trigger, which reduce the event file size from $\sim500$~kB to $\sim50$~kB. 

The SIP is also able to receive command packets from the ground, over Iridium or TDRSS. These short binary-encoded commands are relayed to the flight computer and diverted to the appropriate software daemon to respond. CSBF LOS commanding frequencies are in the PUEO band and were not used during flight.

\begin{figure}
\centering
\includegraphics[width=0.7\textwidth]{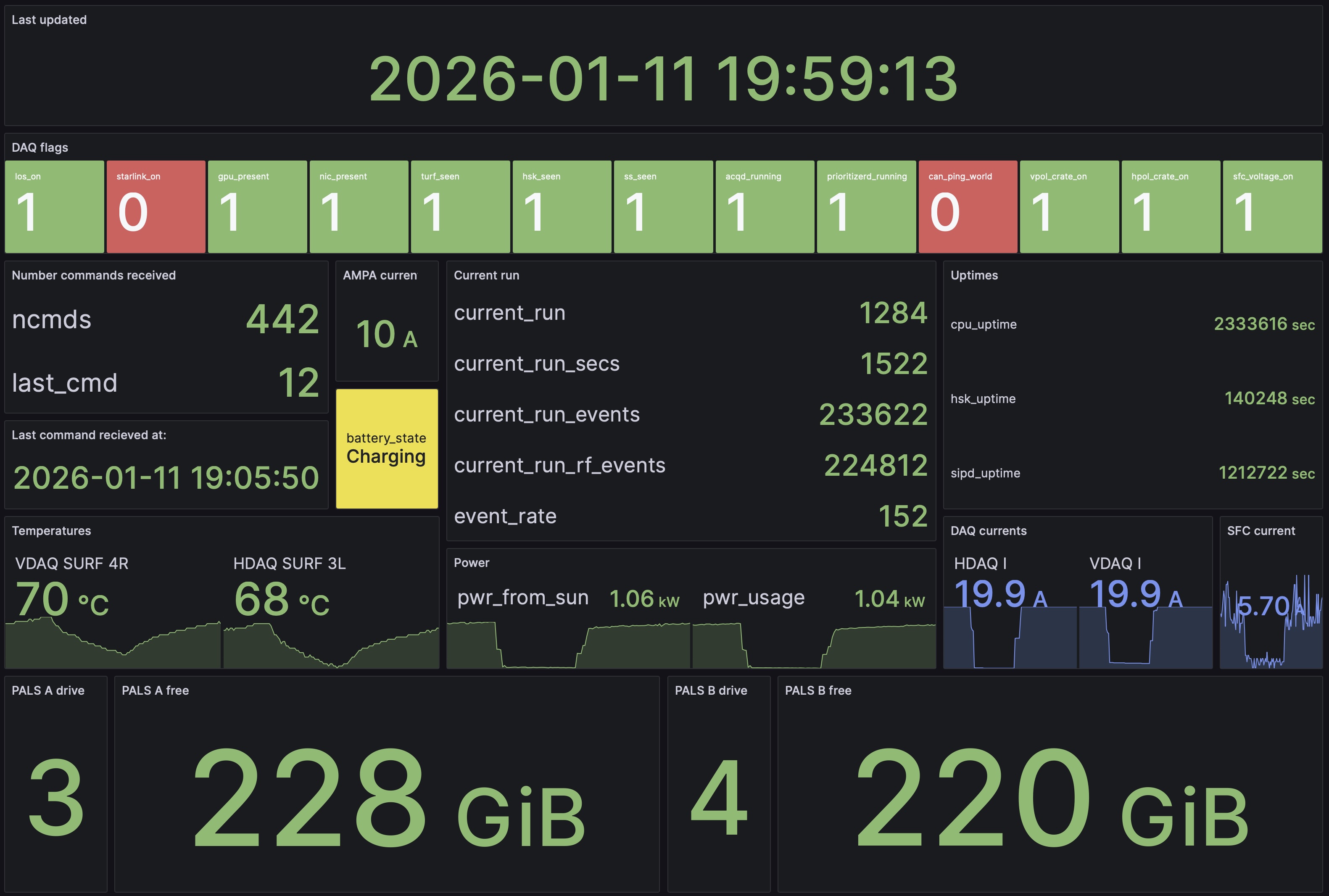}
\caption{An example of a Grafana webpage used to monitor the payload during flight, displaying high-level status information including power consumption, free disk space, and temperatures.}
\label{fig:grafana-example}
\end{figure}

During PUEO's flight, the ground station at the CSBF facility in Palestine, TX received the telemetry data from the TDRSS and Iridium links and ingressed the data into two primary \texttt{postgresql} databases. The monitoring databases were hosted at LDB and at the University of Chicago, replicated in two separate locations to ensure redundancy in the event of network outages. \textit{Grafana}, an open-source data visualization, querying, and monitoring platform, was used to query the database and display time-series data on a number of webpages like the one that appears in figure~\ref{fig:grafana-example} \cite{Grafana:2026}. These acted as the primary source of information, and all telemetry options performed nominally for the duration of the flight.

\subsection{Starlink Telemetry Subsystem} \label{sec:starlink}
As a secondary telemetry option, PUEO utilizes a Starlink High Performance V2 dish, placed on the top of the payload similarly to the other telemetry antennas. This dish interfaces with the SFC network using a 100-BASEFX media converter to go to multimode fiber, preventing potential EMI leakage into the MIE. A custom power circuit board, housed within an RF-tight box near the dish, takes as input the bus voltage from the batteries and converts it to the Starlink Power over Ethernet voltage ($50$~V). During flight, the MI antenna RF chain was powered off for periods of Starlink operation, as ground-based RFI measurements of the dish showed potentially strong in-band emission. Flight performance throughput was nearly $30$~Mbps at peak, and averaged $\sim10$~Mbps.

\section{Preflight Validation}
\label{sec:preflight}
High-altitude balloon payloads operate in near-vacuum environments, typically around $\sim4$~torr. All electronics and cables, including the MIE, were subjected to vacuum tests, which included thermal control to simulate the radiation environment of flight. Additionally, the EMI mitigation strategies were validated preflight using one of the MI antennas to ensure payload subsystem enclosures did not produce RF pickup. Successful operation of the beamforming trigger was also validated using a custom impulsive signal generator.

\subsection{Environmental Chamber}
\begin{figure}[h] 
\centering 
\begin{subfigure}[b]{\textwidth} 
\centering 
\includegraphics[width=0.99\linewidth]{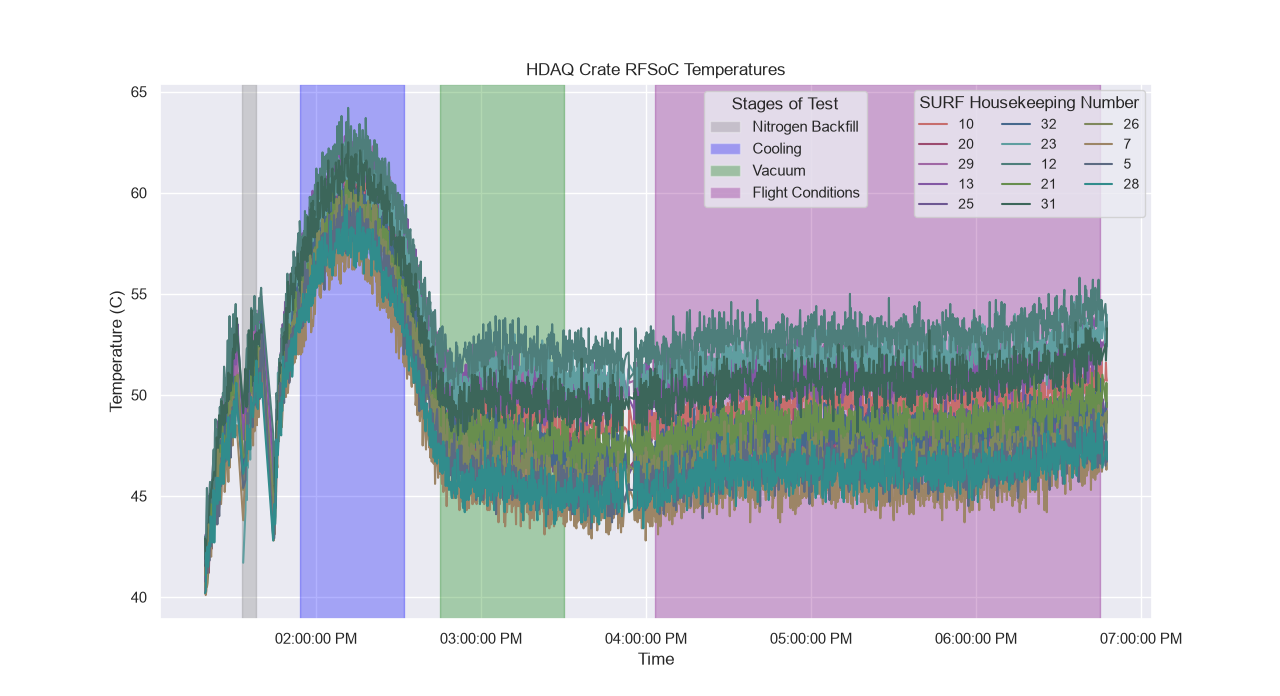} 
\caption{Horizontal polarization RFSoC temperatures} 
\label{fig:hdaqbemco} 
\end{subfigure} \vspace{0.5cm} 
\begin{subfigure}[b]{\textwidth} 
\centering 
\includegraphics[width=0.99\linewidth]{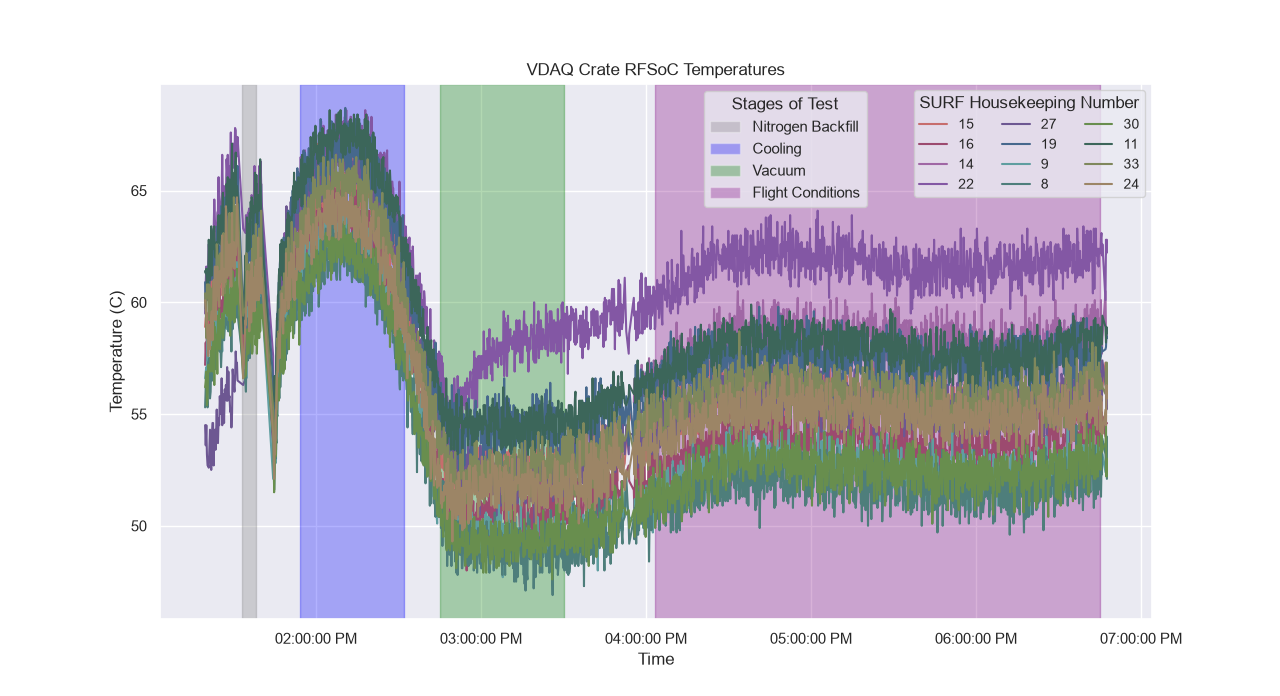} 
\caption{Vertical polarization RFSoC temperatures} 
\label{fig:vdaqbemco} 
\end{subfigure} 
\caption{The temperatures as measured at each of the RFSoCs during the vacuum testing of the MIE in the BEMCO chamber at CSBF.} 
\label{fig:bemcodata} 
\end{figure}

The MIE, navigation subsystems, Starlink power box, battery boxes, actuators for deployables, AMPAs, and external MIE cables were subjected to multiple vacuum tests in the CSBF BEMCO vacuum chamber in Palestine, TX. The operating conditions were chosen in consultation with CSBF engineers to simulate typical conditions in Antarctic balloon-borne flights. The CSBF BEMCO chamber holds liquid nitrogen ($78$~K) in its surrounding walls for thermal control of the radiative environment of the test equipment. We cycled the MIE through various environmental configurations, qualifying the system for operation down to $\sim-40^{\circ}$C while being held at pressures of $\sim3$~torr. The range of temperatures, $-40^{\circ}$C to $-20^{\circ}$C, corresponds to the radiative temperatures of the unpowered MIE in the Antarctic environment with direct and albedo solar radiation. We operated the instrument with simulated workloads in the SFC and the DAQ crates to match the predicted power consumption in flight. Typical MIE temperatures measured during the test, such as the SURF RFSoC temperatures shown in figure \ref{fig:bemcodata}, were within the limits for all subsystems. A replacement Network Interface Card (NIC) on the SFC was later qualified separately at cold temperatures in a smaller vacuum chamber test before flight due to issues with cold-temperature operation with the original model. 

\begin{figure}[htp]
    \centering
    \includegraphics[width=0.98\textwidth]{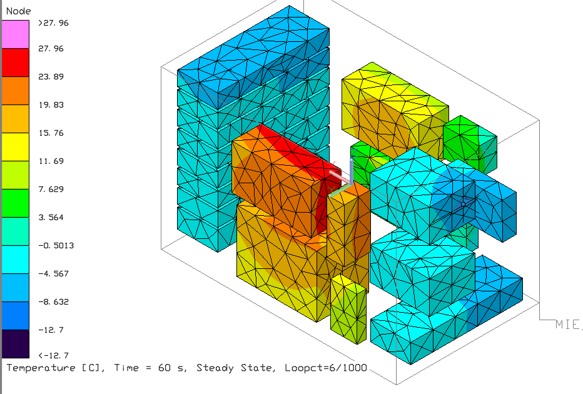}
    \caption{The temperature profile from the steady-state solution of the MIE subsystem enclosures as computed in a Thermal Desktop simulation. In this simulation, the MIE is in a configuration where the Sun faces the side of the MIE with the SFC, located on the bottom left face.}
    \label{fig:MIEThermalDesktop}
\end{figure}

A thermal simulation of the MIE with internal power consumption of individual subsystems was developed with ANSYS Thermal Desktop software. We assumed a solar flux at the payload of $1354$~W/m$^{2}$, an Earth albedo of 0.6, and Earth infrared radiation intensity of $221.5$~W/m$^{2}$. The results show that the MIE remained close to room temperature on average, with the DAQ crates and SFC enclosure reaching $28^{\circ}$~C. A temperature profile from this simulation is shown in figure \ref{fig:MIEThermalDesktop}. 

The CSBF vacuum test DAQ crate temperatures as measured with the housekeeping system showed comparable temperatures to this simulation (just above room temperature values). Together with the internal RFSoC temperature measurements from this vacuum test, these tests and simulations validated the thermal conductance of the heatsinking from the DAQ and SFC enclosures to their internal electronics components. However, as discussed in section \ref{sec:flight_thermal}, the power consumption of the DAQ crates was higher during flight than anticipated, to achieve the best instrument sensitivity, resulting in increased RFSoC temperatures and requiring operation in a duty cycle.

\label{sec:thermalpreflight}
\subsection{Instrument EMI Measurements}
\begin{figure}[htp]
    \centering
    \includegraphics[width=0.98\textwidth]{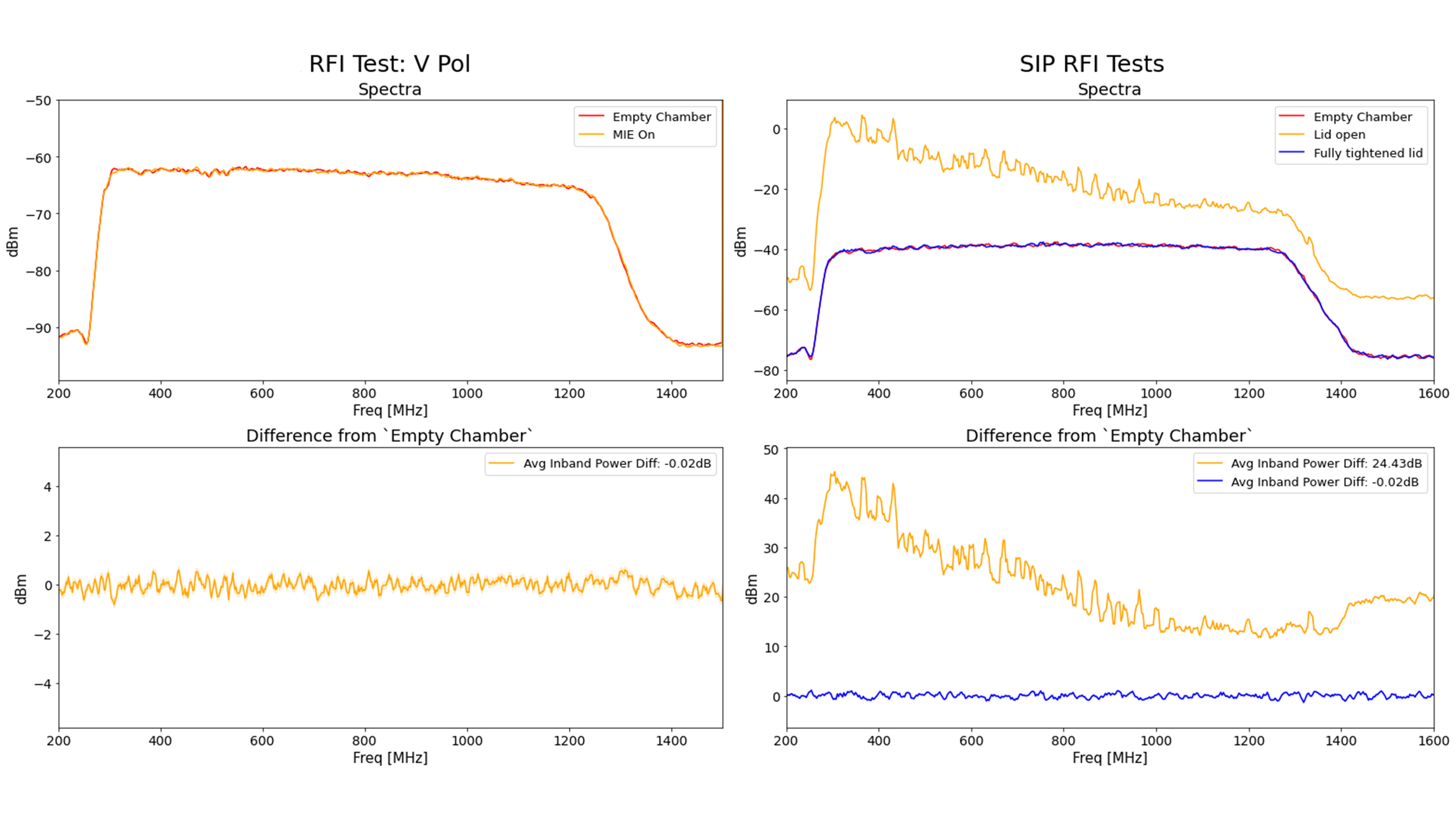}
    \caption{RFI measurements of the MIE (left) and SIP (right) in the RF shielded tent. The empty chamber configuration ensures a proper baseline measurement of the sealed tent. The MIE measurements in the vertical polarization are shown and a similar result was obtained for measurements in the horizontal polarization. The SIP measurement was taken before and after tightening its enclosure lid (right column), illustrating the effectiveness of the RF seal strategies, which reduces the EMI emission from the enclosure.}
    \label{fig:RFISIP}
\end{figure}

Various strategies are used to reduce RF/EMI leakage from PUEO's onboard electronics. As was used on ANITA, a shielding effectiveness benchmark of $\sim 80$~dB is minimally required, which we achieve through gasketing, nickel-plating, and the implementation of in-line filter pins for passing DC and non-RF signals between enclosures across the payload. Validation of these strategies was performed at LDB using an RF shielding tent\footnote{RFT-09 Faraday Tent by Select Fabricators}. Each electronics enclosure was placed in the tent and a main instrument antenna and RF chain were instrumented inside. The output was fed to a Rohde and Schwarz ZNL3 spectrum analyzer, used to monitor the continuous wave (CW) environment, and to a Rohde and Schwarz RTO64 oscilloscope, used to monitor the impulsive environment. Figure~\ref{fig:RFISIP} shows the data from the preflight testing of the MIE and SIP with the setup using the ZNL3, illustrating the effectiveness of the mitigation strategies. All enclosures emitted negligible CW and impulsive EMI in PUEO's band.

\subsection{Beamforming Pulser}
\label{sec:BeamPulse}
As described in section \ref{sec:beamform_daq} and in ref.~\cite{pueotriggerpaper}, the trigger system enforces a coincidence of an 8-antenna synthesized beam between two triggering units. A 16-channel RF beamformer system to generate simulated science signals was developed to validate this trigger. This RF beamformer signal generator box, called the Beamformer Test Set (BTS) was used to produce controlled and repeatable simulated science signals for trigger testing.

\subsubsection{Overview of the Beamformer Test Set}

\begin{figure}[h!]
    \begin{subfigure}{0.5\textwidth}
    \centering\includegraphics[width=1.1\linewidth]{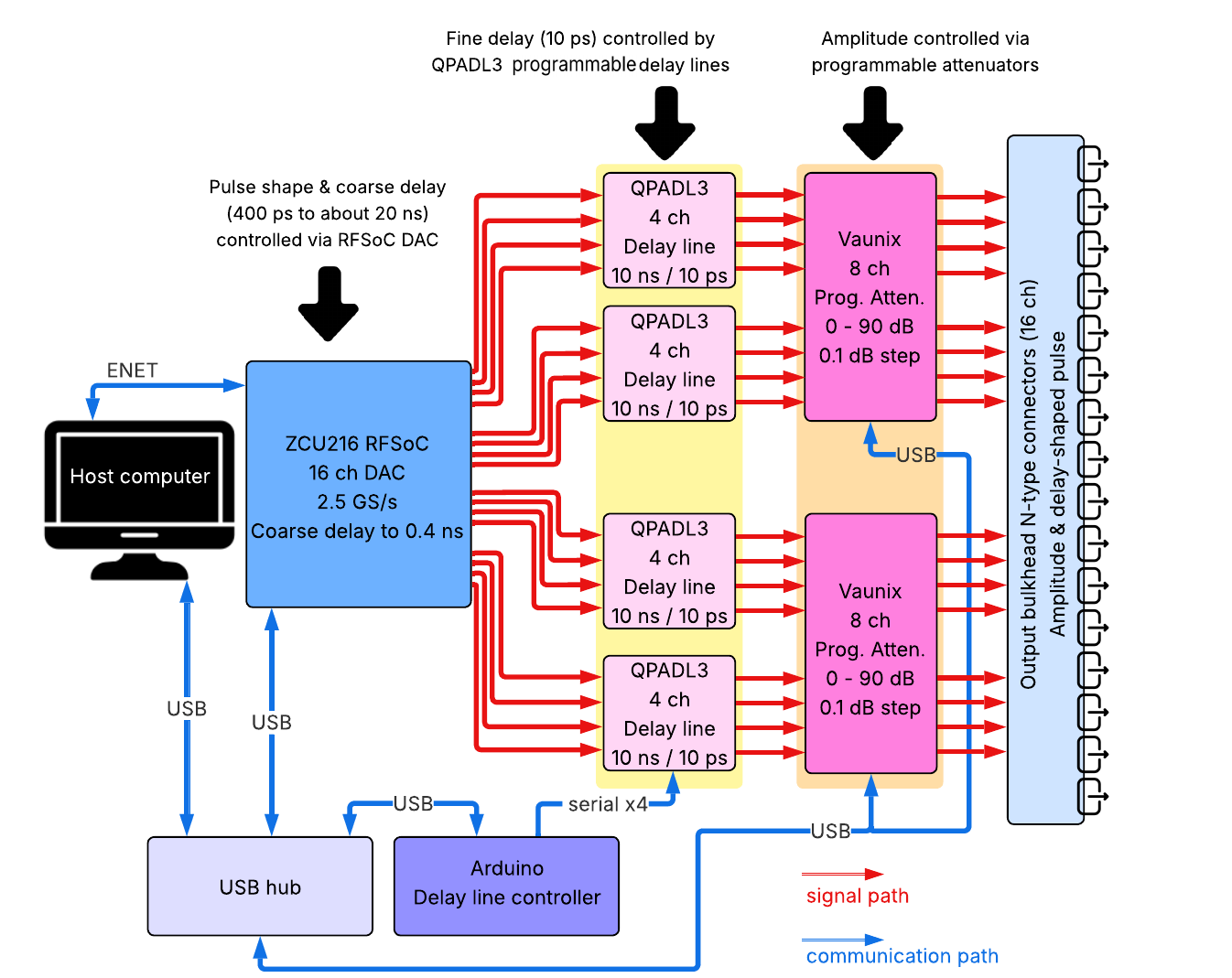} 
        \caption{Block diagram of BTS }
    \end{subfigure}
    \begin{subfigure}{0.5\textwidth}
    \centering\includegraphics[width=.9\linewidth]{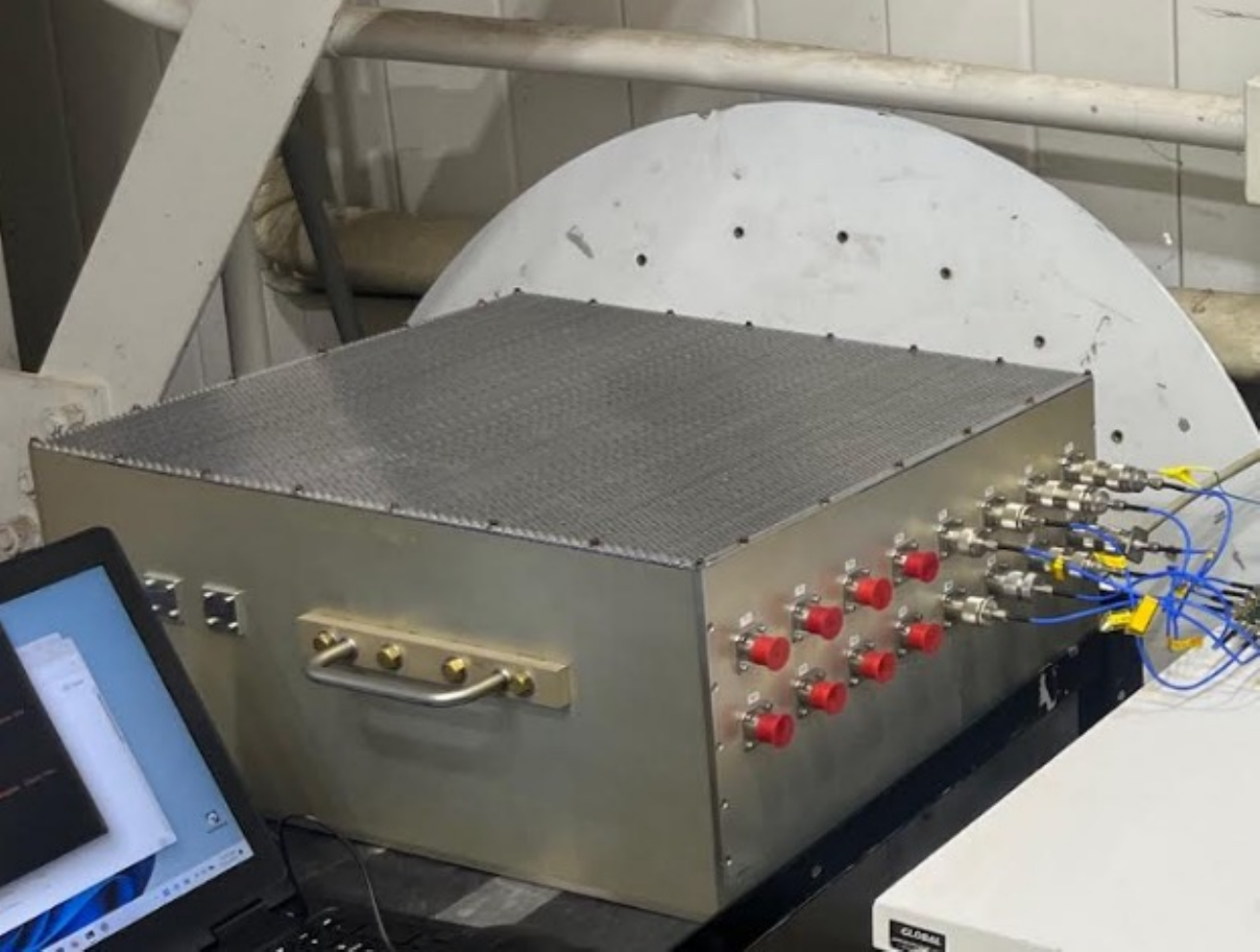}
        \caption{BTS setup}
    \end{subfigure}
    \caption{Left: A block diagram of the components of the BTS is shown. Right: A picture of the BTS and its control laptop is shown. Only the right-most eight channels are connected. The lid of the BTS is a mesh plate that allows airflow cooling while maintaining RF shield integrity.}
    \label{fig:rfbox}
\end{figure}

The BTS is a dedicated signal generator and distribution platform designed to emulate the temporal and spectral characteristics of expected neutrino-Askaryan events. A block diagram of the BTS is shown on the left in figure~\ref{fig:rfbox} and its enclosure is shown on the right. The system architecture is based on a Xilinx ZCU216 RFSoC evaluation board (ZCU216), which serves as the primary digital-to-analog signal generator. The ZCU216 generates programmable RF waveforms that simulate radio impulses from UHE neutrino interactions.

\subsubsection{Signal Generation and Testing Procedure}

The BTS routes the ZCU216 signals to one of four fine delay modules, which provide $10$~ps precise timing adjustments channel by channel. This functionality is essential for reproducing realistic wavefront arrival times across multiple antenna channels, allowing comprehensive validation of the trigger timing logic. The signals pass through independent attenuator modules, for a total of 16 output channels corresponding to 4 phi sectors. This provides amplitude control for each channel for systematic studies of trigger thresholds, dynamic range, and channel-to-channel gain variation. While signal amplitude is adjusted digitally from the ZCU216, these variable analog attenuators allow for much finer control and preclude degradation of signal fidelity at low amplitudes. This system is also instrumented with an external edge pulse and tunable delay to allow for synchronized triggers when testing the trigger. 

\begin{figure}[h]
    \centering
    \includegraphics[width=0.99\textwidth]{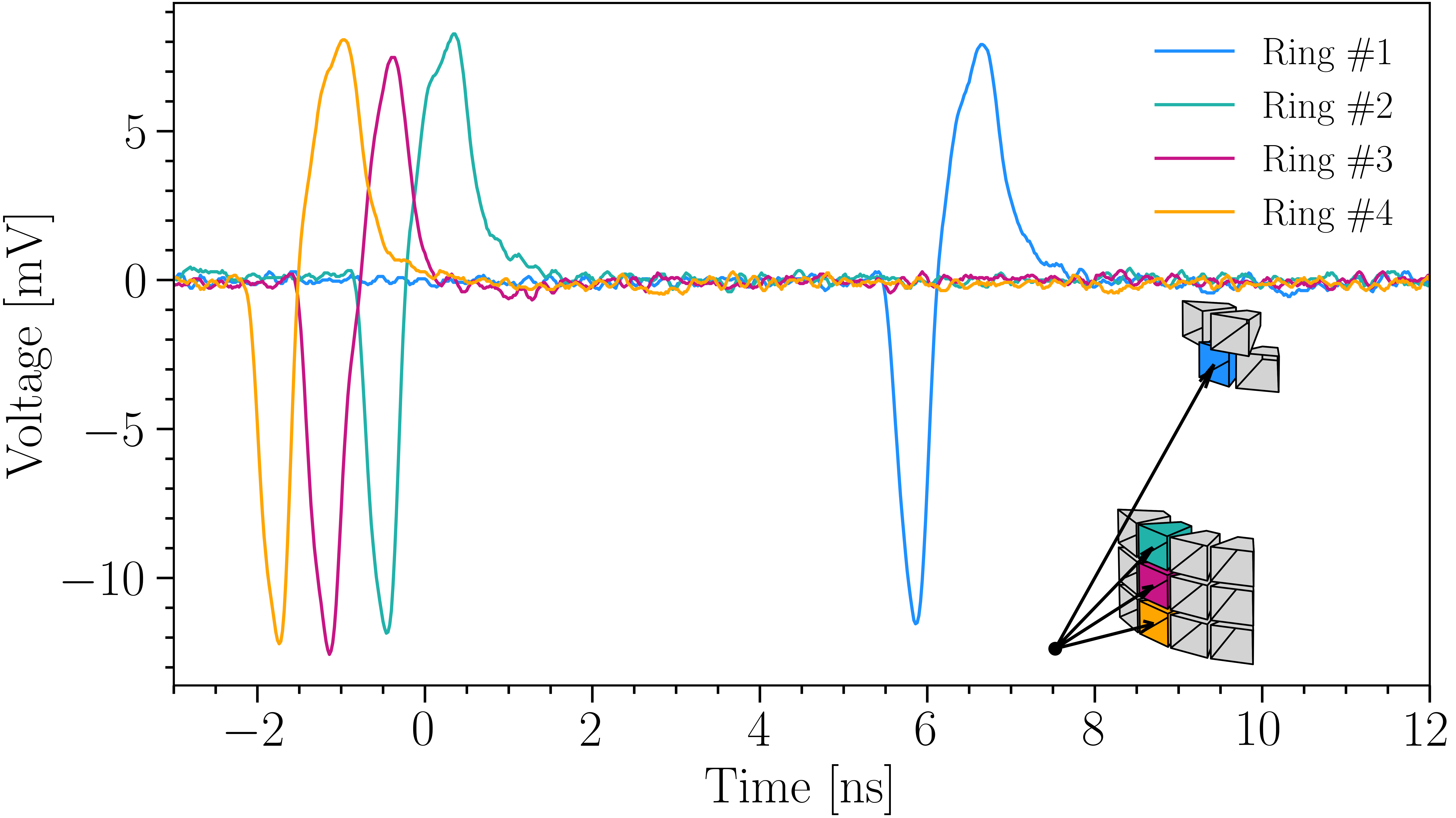}
    \caption{A measurement of the BTS signals injected into the SURF channels. The relative delays are mapped in color with the SURF-channel antenna locations, which are shown in the bottom-right inset image of the section of the payload.}
    \label{fig:antenna_geo} 
\end{figure}

All operations can be controlled through commands issued from the host computer, which is shown to the right in figure~\ref{fig:rfbox} along with the enclosure of the BTS. The BTS produces signals where the arrival angles are specified as viewed geometrically by the real payload. The antenna geometry for four neighboring phi sectors is shown on the inset of figure~\ref{fig:antenna_geo}. Using the relative positions of each antenna, delay signals for the 16 channels can be simulated by providing the azimuth and elevation angles; 4 of these 16 channels are shown as measured with an oscilloscope in figure~\ref{fig:antenna_geo}. After delay adjustment, channel amplitudes are set using the two 8-channel attenuator modules to match the desired SNR and probe trigger threshold behavior. The BTS is also programmed with the duration and repetition rate for a controlled test mode. Thus, the trigger efficiency, latency, and false trigger rates are measured.

\subsection{Trigger Validation}
\label{sec:TrigVal}
\begin{figure}[htb!]
    \centering
    \includegraphics[width=0.99\linewidth]{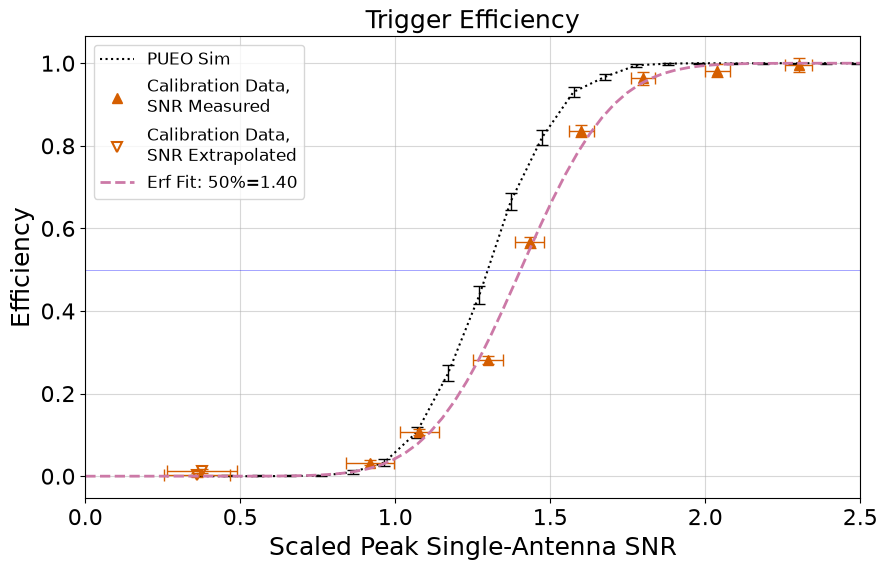}
    \caption{Measurements of the trigger efficiency using the BTS described in section~\ref{sec:BeamPulse}. In keeping with ANITA's standard, SNR here is defined as the peak-to-peak height of the signal divided by twice the noise RMS. The calibration data shown in orange was measured with the recovered flight hardware post-flight, in a lab environment. An error function is fit to the data and used to find the 50\% efficiency point at 1.4 SNR. The black curve shows the ideal efficiency curve as predicted from a full Monte Carlo simulation of the experiment (called \texttt{PUEOSim}) which includes the RF and trigger chain.}
    \label{fig:TriggerValidation}
\end{figure}

The BTS was used to validate the trigger functionality at LDB prior to launch, and again post-flight to further characterize the trigger behavior in a lab environment. Twelve channels corresponding to the lower three antennas of four phi-sectors  were connected to the BTS to form a full instrument trigger. Two of the 16 BTS channels did not function as expected, so to validate the four-phi sector trigger, we selected only the lower three antenna rings. BTS pulses were directly injected in place of the antennas and passed through the full RF chain. The resulting trigger rates matched expectations, and the results of these tests are shown in figure~\ref{fig:TriggerValidation}. A more detailed discussion of this measurement and of the design of the trigger system can be found in \cite{pueotriggerpaper}. Compared to ANITA-IV at $50$\% efficiency, this is an improvement on SNR by more than a factor of 2. To date, this is the most sensitive beamforming trigger built for the detection of UHE neutrinos. 

\section{Calibration}
\label{sec:calibration_hi}
\subsection{Ground Calibration} \label{GroundPulsers}
The ground calibration systems are designed to emit Askaryan-like pulses with precise and tunable timing. This enables calibration of the MI and LF antennas' phase centers and the payload orientation, and provides a useful signal-like dataset for building analysis strategies. There were three sites for ground calibration: LDB during launch, South+200, and Taylor Dome. The LDB site allowed for system tuning within the LOS telemetry period while the other sites provided well-isolated pulses used for precision calibration of the payload geometry, pointing resolution, and trigger efficiency.

\begin{figure}
    \centering
    \includegraphics[width=0.98\linewidth]{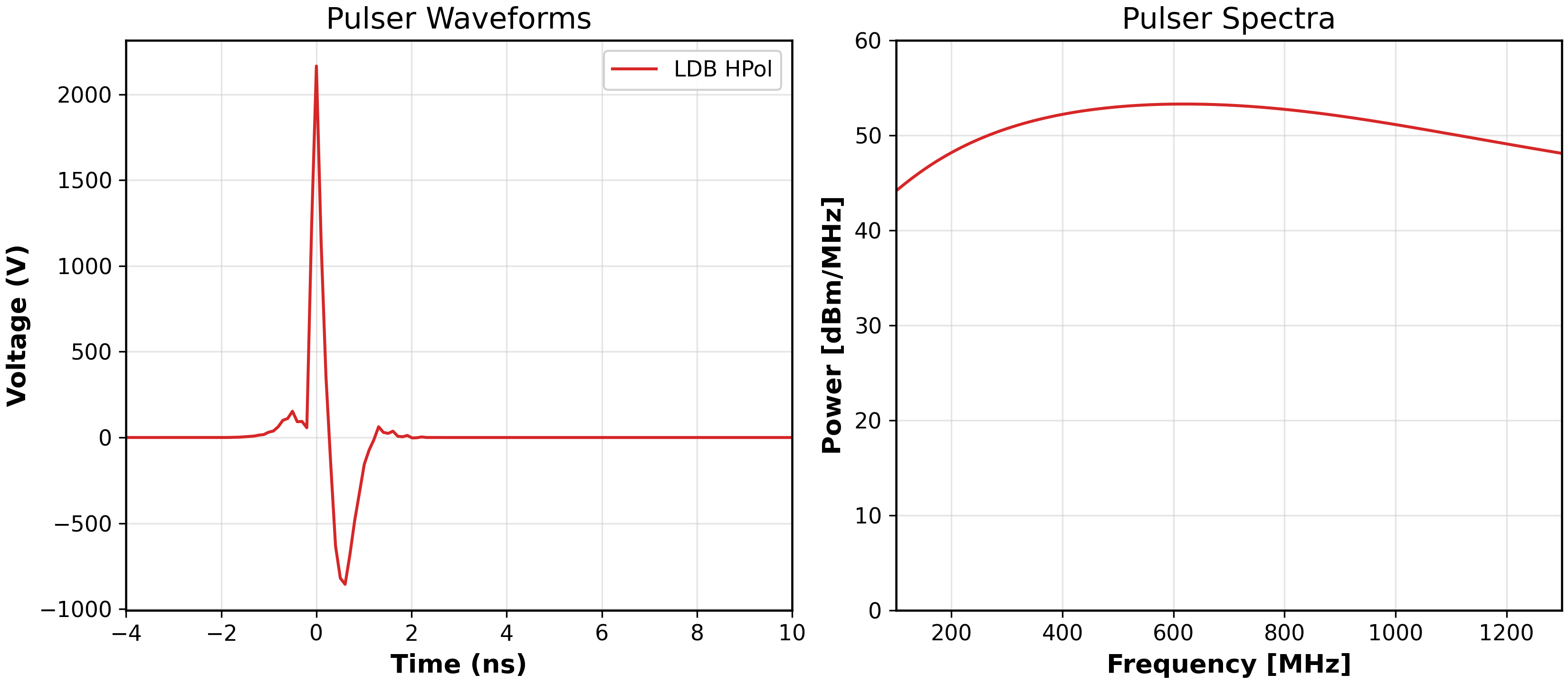}
    \caption{Output of the high voltage pulser used for the ground calibration station at LDB (HPol channel) in the time domain (left) and power spectrum in the frequency domain (right). These measurements are the average of 100 waveforms as measured with a scope at the output of the pulser, and do not include the antenna or cable effects.}
    \label{fig:gndcal-pulser-time-freq-domain}
\end{figure}

A list of the ground calibration systems is shown in table~\ref{tab:groundcal_userdelays}, including a description of the antenna that was deployed at each site. The equipment is modular and redundant to alleviate the risk associated with the uncertainty on payload flight trajectory. The pulses and spectra as measured at the output of each system are shown in figure~\ref{fig:gndcal-pulser-time-freq-domain}. Each setup has distinct emission. Overall, they share common features: high-voltage ($>1$~kV, stable to $2$~\%) and fast ($\sim3$~ns) signals, to mimic the expected characteristics of the science signal. 

\begin{table}
    \centering
    \caption{Summary of the calibration campaign including information on the transmitting antenna, polarization, pulser model number, user-defined time offset, and number of events identified in a preliminary search.}
    \begin{tabular}{ccccc}
       Site  & Antenna / Pol.& Pulser & Time Offset (ms)  & No. Events \\
       \hline
        LDB &  Toyon / Vpol & CPS3 & 900 & 19357\\
        LDB &  Toyon / Hpol & UH & 200 & 20451\\
        LDB & LF Horn / H\&V & FPM 2-1NM20 & 900 & N/A\\
        Taylor Dome & Dipole / Vpol & FPM-10-1PNP & 750 &  20918 \\
        Taylor Dome & Bicone / Hpol & FPG 6-1PNK & 500 & 14705\\
        S+200 & Dipole / Vpol & FPM 6-1PNP & 750 & 90723\\
        S+200 & Bicone / Hpol & F0512 & 0 & 13936\\
    \end{tabular}
    \label{tab:groundcal_userdelays}
\end{table}

\begin{figure}
    \centering
    \includegraphics[width=0.99\linewidth]{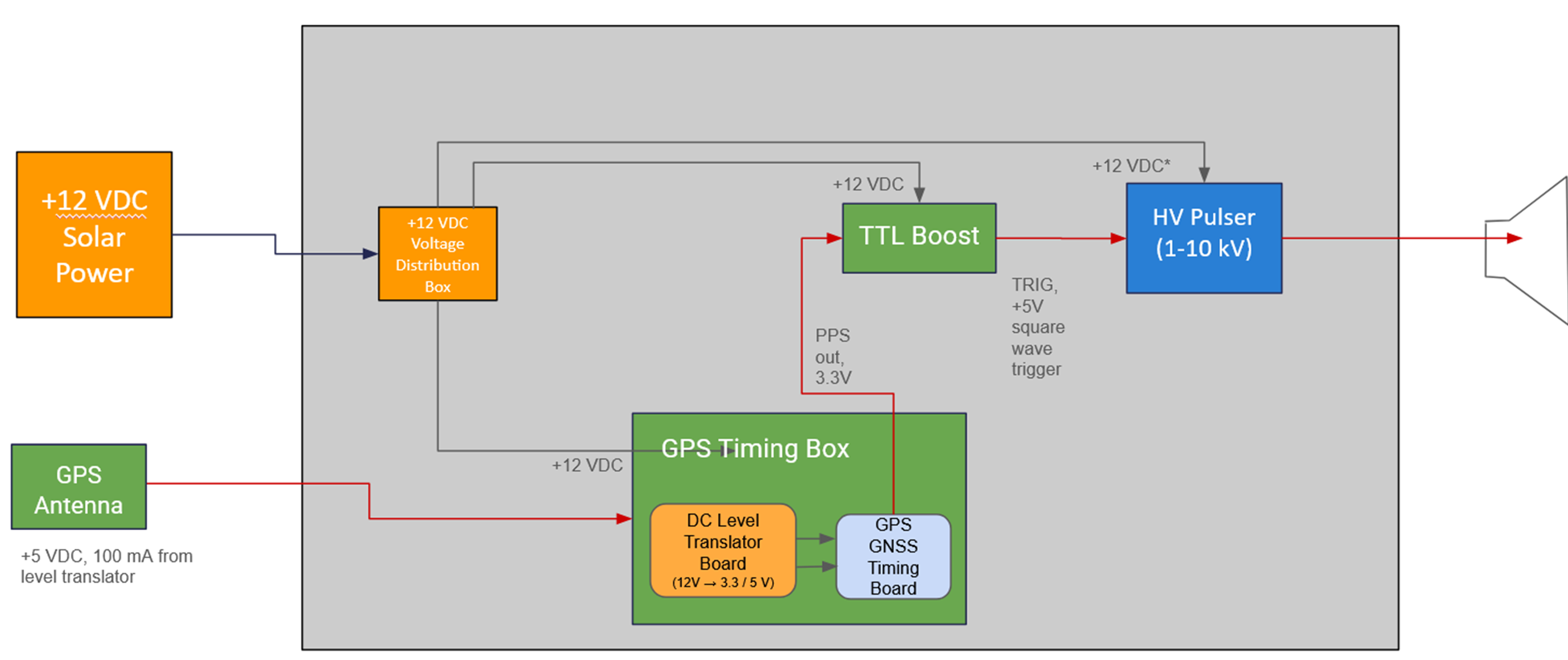}
    \caption{A diagram of the electronics boxes deployed to the ground calibration sites. The GPS timing box generates well-timed pulses synchronized with the GPS second and offset by programmed delays. These pulses are then boosted from 3.3V to 5V in order to trigger a high voltage pulser.}
    \label{fig:ugh}
\end{figure}

\begin{figure}
    \centering
    \includegraphics[width=0.99\linewidth]{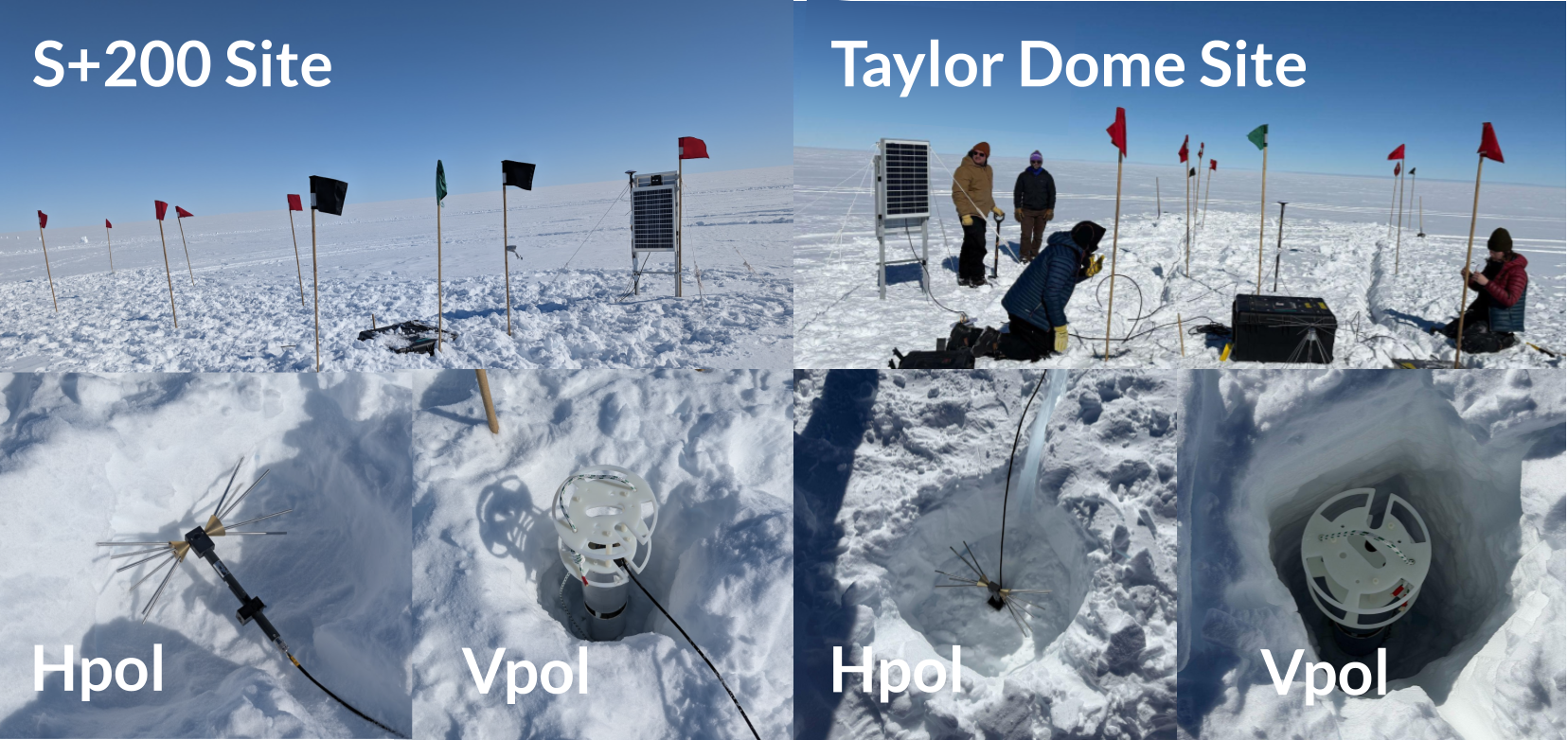}
    \caption{Pictures of the ground calibration systems installed at South+200 and Taylor Dome sites with polarization of the antennas labelled.}
    \label{fig:remote-gndcal-setup}
\end{figure}

Each ground calibration station is synced to the GPS second with a triple-band GNSS Antenna\footnote{The ArduSimple AS-ANT3B-CAL-L1256-SMATNC-01.} using a dedicated receiver chip\footnote{The ublox ZED-F9T-00B.} inside of an electronics box, with an expected time mark resolution $\sim10$~ns. The programmable output signal is tuned with an offset from the GPS pulse-per-second (PPS) by a fraction of a second, called a subsecond hereafter. The subsecond time corrects for the insertion delay per box, which was measured before deployment. Table~\ref{tab:groundcal_userdelays} shows the subsecond offset for each station and polarization. The Taylor Dome and South+200 stations consisted of two pulser boxes, a standalone solar power system, and the pulsing antennas which were buried under the surface snow to avoid the harsh ground conditions, as shown in figure~\ref{fig:remote-gndcal-setup}. A system diagram of the electronics boxes is shown in figure~\ref{fig:ugh}. The LDB site consisted of three electronics boxes and three channels such that both polarizations of PUEO MI antennas and a single polarization of a horn antenna within the LF bandwidth generated pulses. A synchronized trigger was formed for two of these boxes such that the vertical polarization feed on the MI antenna and the LF horn antenna aligned in the subsecond timing of the pulses. The LF horn was physically rotated midway through the pulsing period to ensure transmission of both polarizations. Ground calibration pulses have been found in the flight data from all station antennas.

\subsection{HiCal-3}
\begin{figure}
    \centering
    \includegraphics[width=0.98\linewidth]{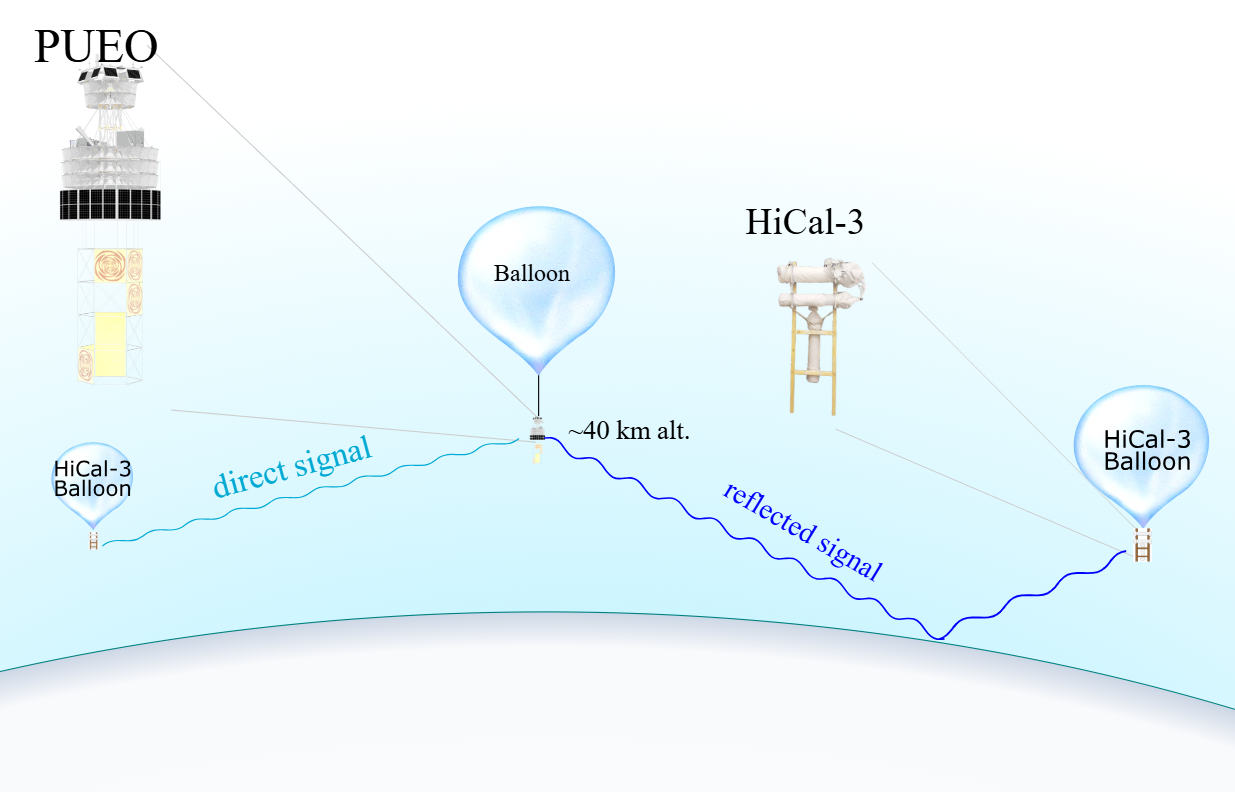}
    \caption{A diagram showing the reflected and direct signals emitted from the HiCal-3 payloads and detected by PUEO. }
    \label{fig:hicalcartoon}
\end{figure}

To provide an independent calibration dataset for the PUEO MI, including to study the polarity of signals reflected off the surface of the ice, balloon-borne hand-launched small payloads (High-altitude Calibration, HiCal) were deployed after the PUEO instrument. These two calibration payloads, called HiCal-3a and HiCal-3b, were attached to their own, smaller high-altitude balloons and float independently, see figure~\ref{fig:hicalcartoon}. The HiCal-3 instruments are designed similarly to the HiCal-2 project for ANITA-IV \cite{PROHIRA201960} and will be discussed in more detail in a companion paper.

\begin{figure}
    \centering
    \includegraphics[width=0.98\linewidth]{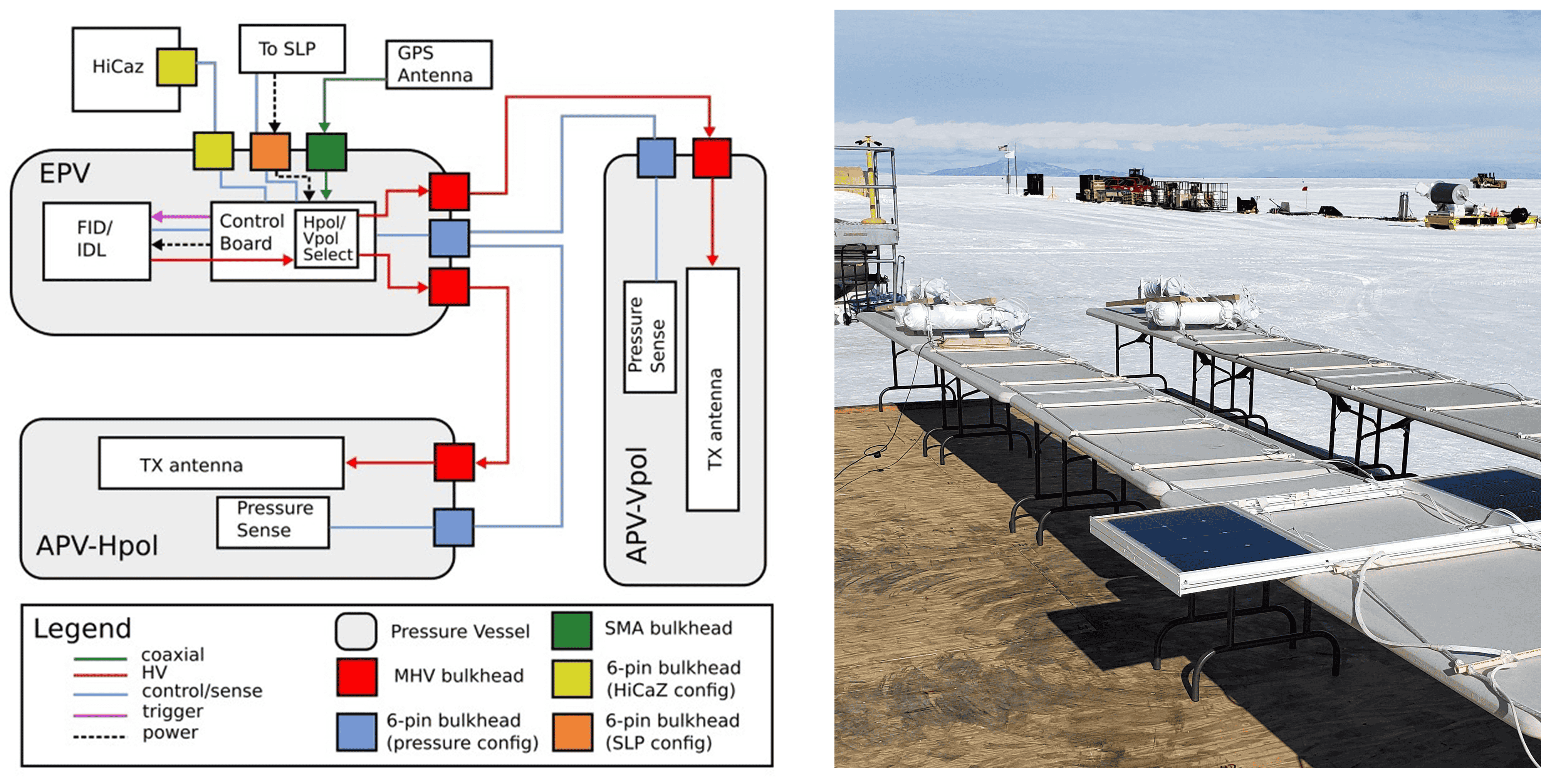}
    \caption{Left: The schematic of the HiCal-3 payloads showing the 3 vessels, antennas of each polarization, and the GPS and pulsing transmitter. Right: A picture of both HiCal-3 payloads and flight trains during pre-launch checkout and setup at LDB. At the end of the tables sit the vessels, and a solar panel is visible in the foreground. }
    \label{fig:hicalprelaunch}
\end{figure}

Each HiCal-3 payload contains a solid-state RF transmitter housed within a lightweight, acrylonitrile butadiene styrene (ABS, a common, impact-resistant thermoplastic polymer) vessel, as well as electronics and GPS. The RF signal chain then splits towards two additional ABS vessels that contain two identical dipole antennas, oriented orthogonally, one horizontal and one vertical.  The HiCal-3 vessels are sealed with the ambient air at local pressure and tested in a vacuum chamber before launch. The pressurization of the vessels is necessary to prevent arcing of the high voltage transmitters at high-altitude low pressures of $\sim4$~torr. A picture of the HiCal-3 payloads preparing for launch is shown in figure~\ref{fig:hicalprelaunch}.

In addition to providing a real-time verification of PUEO triggering once the ground calibration systems are out of range, HiCal-3 improves upon the previous HiCal missions in three important respects. First, each payload emits in both polarizations. Second, the pulse-per-second signal triggers are synchronized to GPS to within $\sim$10 ns. Finally, custom solid-state signal generators with significant frequency content up to 500 MHz are used.

\begin{table}[htpb]
\centering
\caption{Statistics of HiCal-3 identified events in the PUEO data sample including the distance range between PUEO and the HiCal-3 payloads.}
\begin{tabular}{c|c|c}
 & HiCal-3a & HiCal-3b \\ \hline
PUEO events triggered & 19829 & 3410 \\
Range & 253-584 km & 643-735 km \\
\end{tabular}
\label{tab:HC3stats}
\end{table}

\begin{figure}[h]
    \centering

    \includegraphics[width=0.99\textwidth]{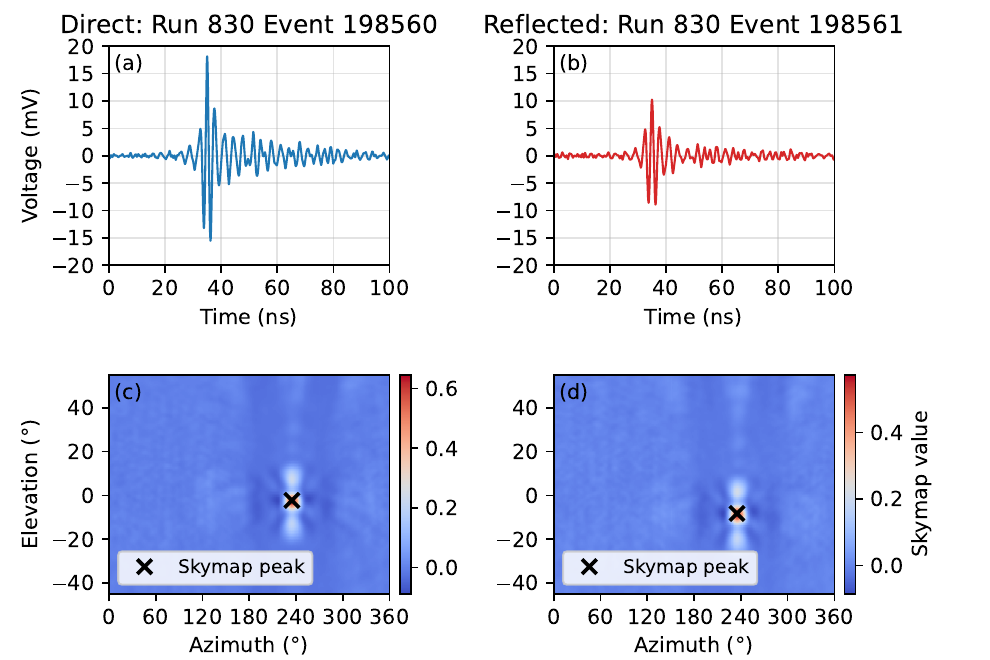}
    
    \caption{The coherently summed waveforms (top row) and their reconstructed pointing directions (bottom row) as seen by the PUEO MI for a HiCal-3a direct (a, c) and reflected (b, d) event pair. Time of flight and reconstruction identify this event pair as originating from the HiCal-3a pulser. The reconstruction and coherent sum are performed using preliminary estimates for the PUEO antenna phase centers from photogrammetry.}
    \label{fig:hical_reconstruction}
\end{figure}

Both HiCal-3 payloads were launched within a two-day window after PUEO aboard $1.128$~MCF high-altitude balloons. HiCal-3b had a total flight time of 5 days and 2 hours.  Unfortunately, a leak developed in the balloon, resulting in premature termination and, consequently, a more limited range relative to PUEO. The 12-day HiCal-3a flight was the longest of the HiCal missions, although a thermal issue with the solar panels arose that required a duty cycle to fully recharge the batteries. There are $\sim 23$K triggered events that trace back to HiCal-3 payloads, found in the PUEO flight data using the subsecond trigger time. An example of an interferometrically-reconstructed trigger pair from the HiCal-3 payloads is shown in figure \ref{fig:hical_reconstruction}.

\section{Flight and Performance}
\label{sec:flight_and_perform}
\subsection{Flight Path and Livetime}
\label{sec:flight_thermal}
\begin{figure}
\centering
\includegraphics[width=0.99\textwidth]{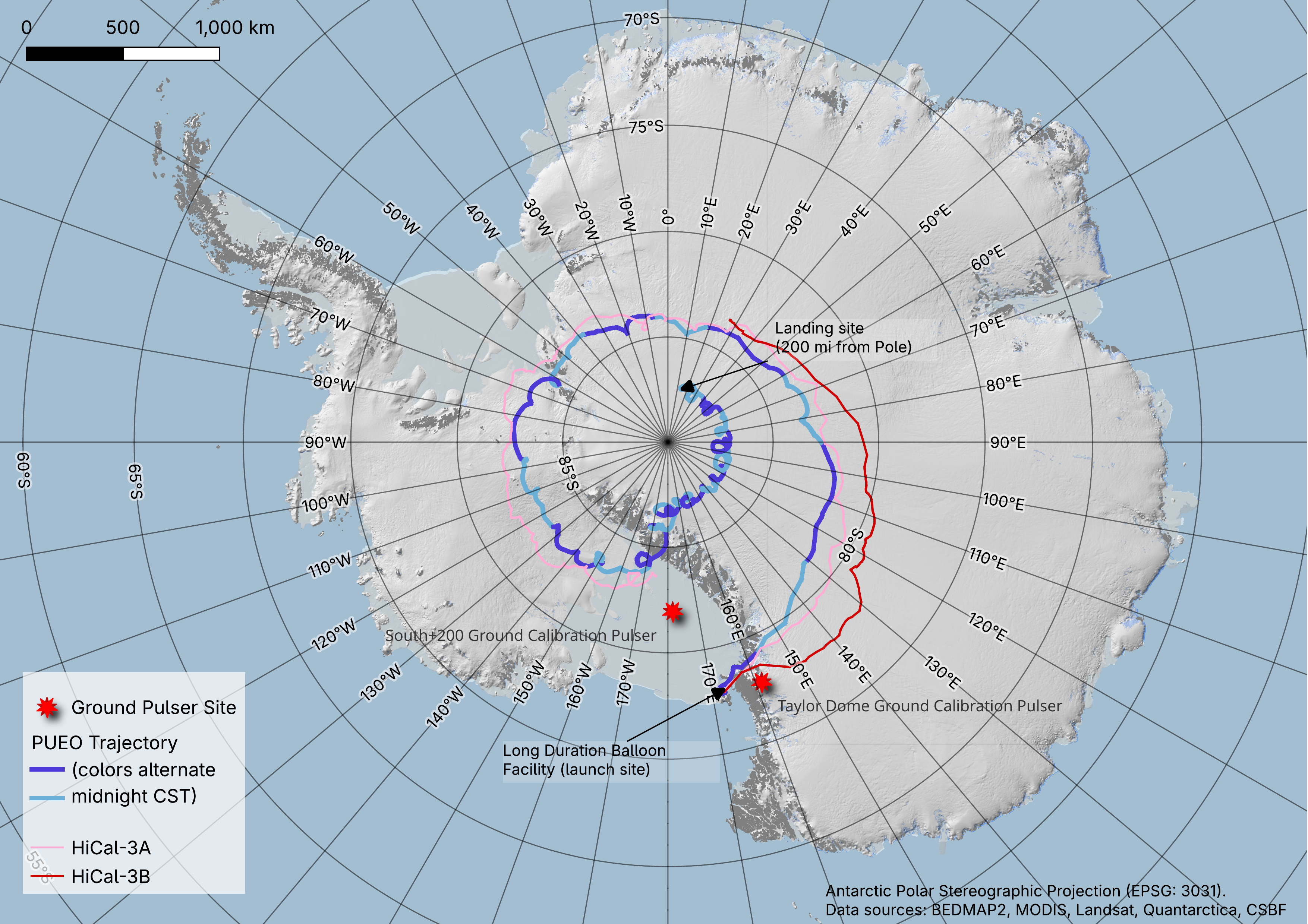}
\caption{ The flight paths for PUEO and the two HiCal-3 payloads. Alternating colors on the PUEO flight path indicate days. Ground Calibration sites at Taylor Dome and South+200 are shown as red stars.}
\label{fig:path}
\end{figure}

PUEO launched on the 19th of December of 2025 at 17:00 UTC and flew for 23 days and 10 hours, terminating on the 12th of January at 02:00. The flight path of PUEO and the HiCal-3 payloads appears in figure~\ref{fig:path}.

\begin{figure}
    \centering
    \includegraphics[width=0.8\linewidth]{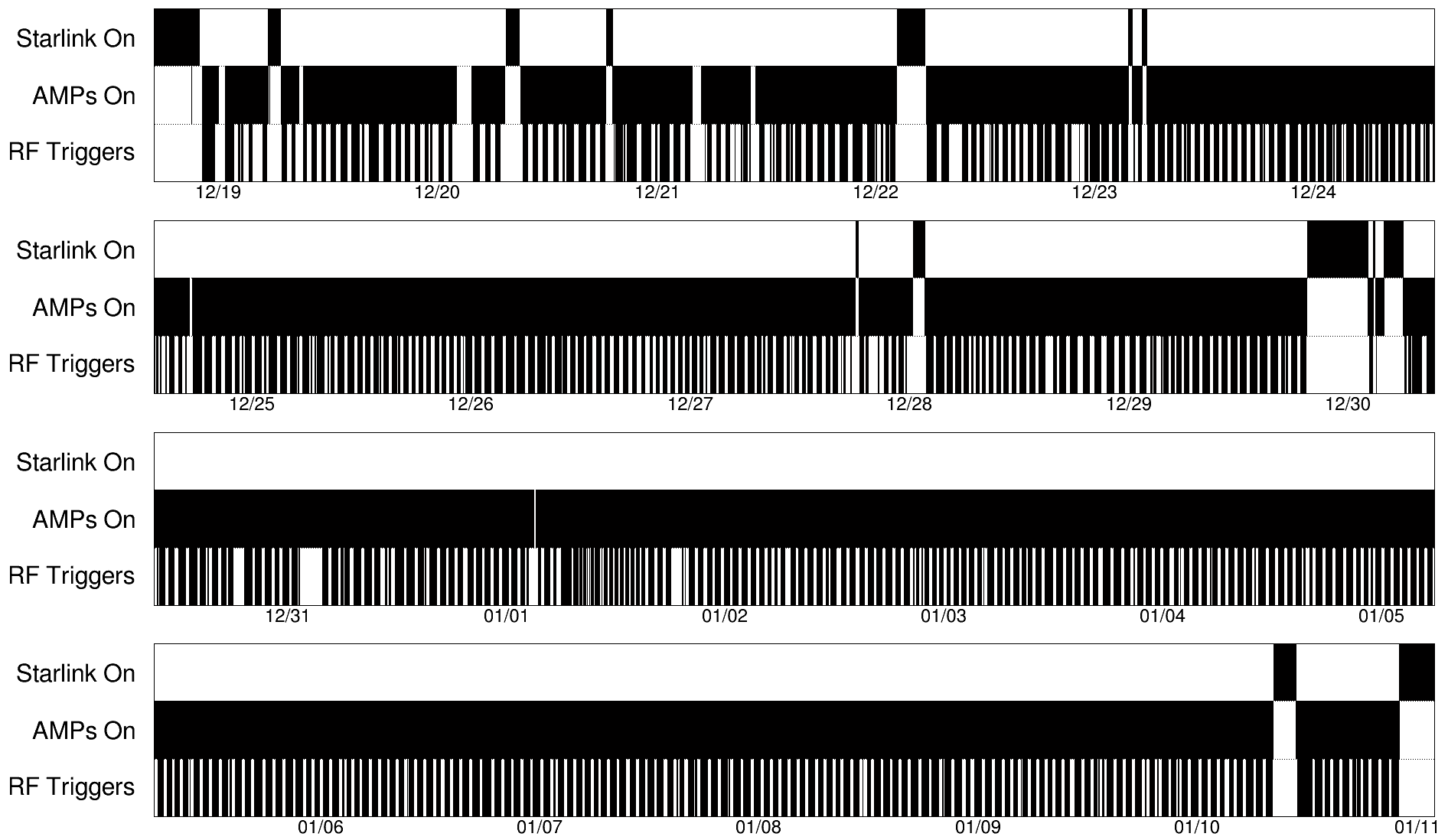}
    \caption{A time series plot of the Starlink activity, AMPA (RF front-end) status, and the status of radio triggers. The noticeable AMPA off periods coincide with Starlink activity, chosen to safeguard the electronics from the nearby RF emission of the Starlink dishy. Periodic RF trigger gaps are visible during data-taking pauses to mitigate the thermal impact of the beamforming trigger. }
    \label{fig:livetime}
\end{figure}

At the time of the flight, the SURF power consumption was higher than the levels simulated and tested due to additional processing required to achieve the specified sensitivity of the interferometric trigger.  As a result, the RFSoC temperatures approached their operating limits.  The duty cycle of the instrument was managed during the flight to keep the RFSoC temperatures below 80$^\circ$C,  with the RFSoC clocks and processors turned off during the recovery portion of the cycle.  This resulted in an overall average duty cycle of $\sim55$\%, resulting in a livetime of approximately 12 days and 21 hours. Future flights of PUEO will ensure a more robust thermal solution allowing for a longer livetime. During periods of Starlink telemetry, data-taking was paused to safeguard against potential catastrophic impact to the RF electronics near the antennas, i.e. the AMPAs. This can be seen in the time series plot of figure~\ref{fig:livetime}. Periods of DAQ system cooldowns are interspersed where the AMPAs are kept powered on, but the beamforming trigger system was paused. 

\begin{figure}
\centering
\includegraphics[width=0.99\textwidth]{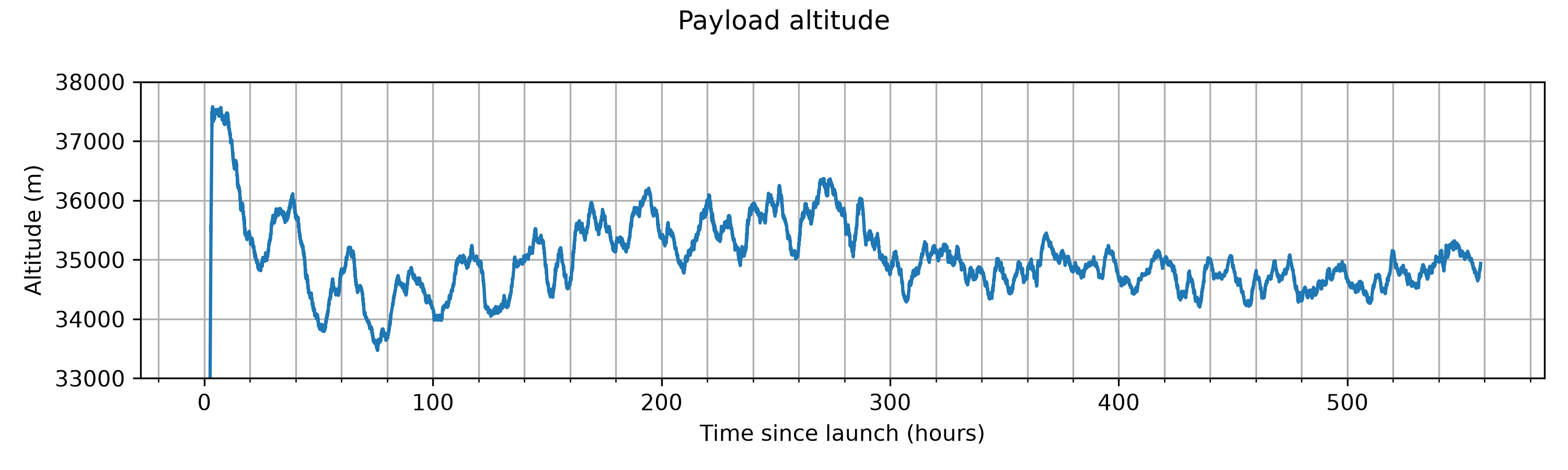}
\caption{Altitude profile after the payload reached float. The diurnal cycle is present in the altitude measurements. The mean altitude during the flight was $35$~km.}
\label{fig:altitude}
\end{figure}

\subsection{In-flight Pulsing}
\begin{figure}[htbp]
    \centering
        \includegraphics[width=\linewidth]{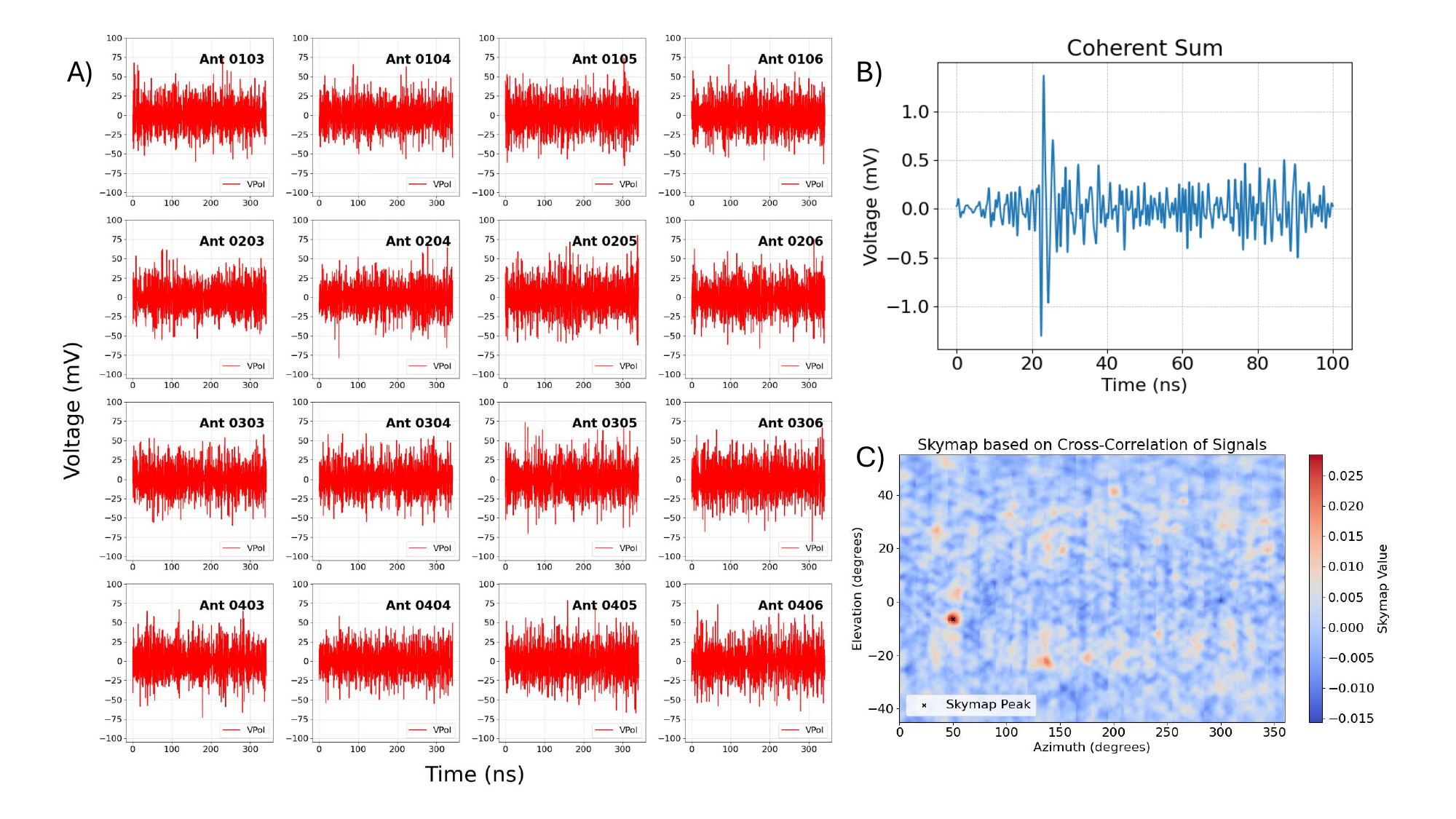}
        \caption{An example calibration signal recorded by the PUEO payload, originally emitted from the dipole calibration antenna at the Taylor Dome site. A) Waveforms of the 4 phi sectors, 3 through 6, of the triggered event. B) Coherent sum of the waveforms shown in A using the photogrammetry determined antenna positions to calculate the time delays. C) A skymap of the event based on cross-correlations of the antennas shown in A, with the PUEO coordinates of azimuth on the horizontal axis and the elevation along the vertical axis, where the colorbar is showing the calculated correlation value.}
        \label{fig:calpulserTD}
\end{figure}

During the flight, more than 200,000 calibration pulse signals were recorded from the ground calibration systems described in section~\ref{GroundPulsers}. An example waveform view of one of these calibration pulses from Taylor Dome is in figure~\ref{fig:calpulserTD}. Ground pulser events like these are picked out from the precise GPS subsecond timing bins, creating a sample of signal-like identified calibration events that can be used for calibration and guiding analysis choices. The coherent sum of the vertically polarized channels and the interferometric image (or skymap) are also shown for this event. The skymap is calculated using cross-correlations of the vertically polarized channels. The reconstruction method shown includes antenna positions from the photogrammetry measurements. Final reconstruction will be done using more precise locations determined using ground calibration events. 

\subsection{Navigation}
\begin{figure}
\includegraphics[width=\linewidth]{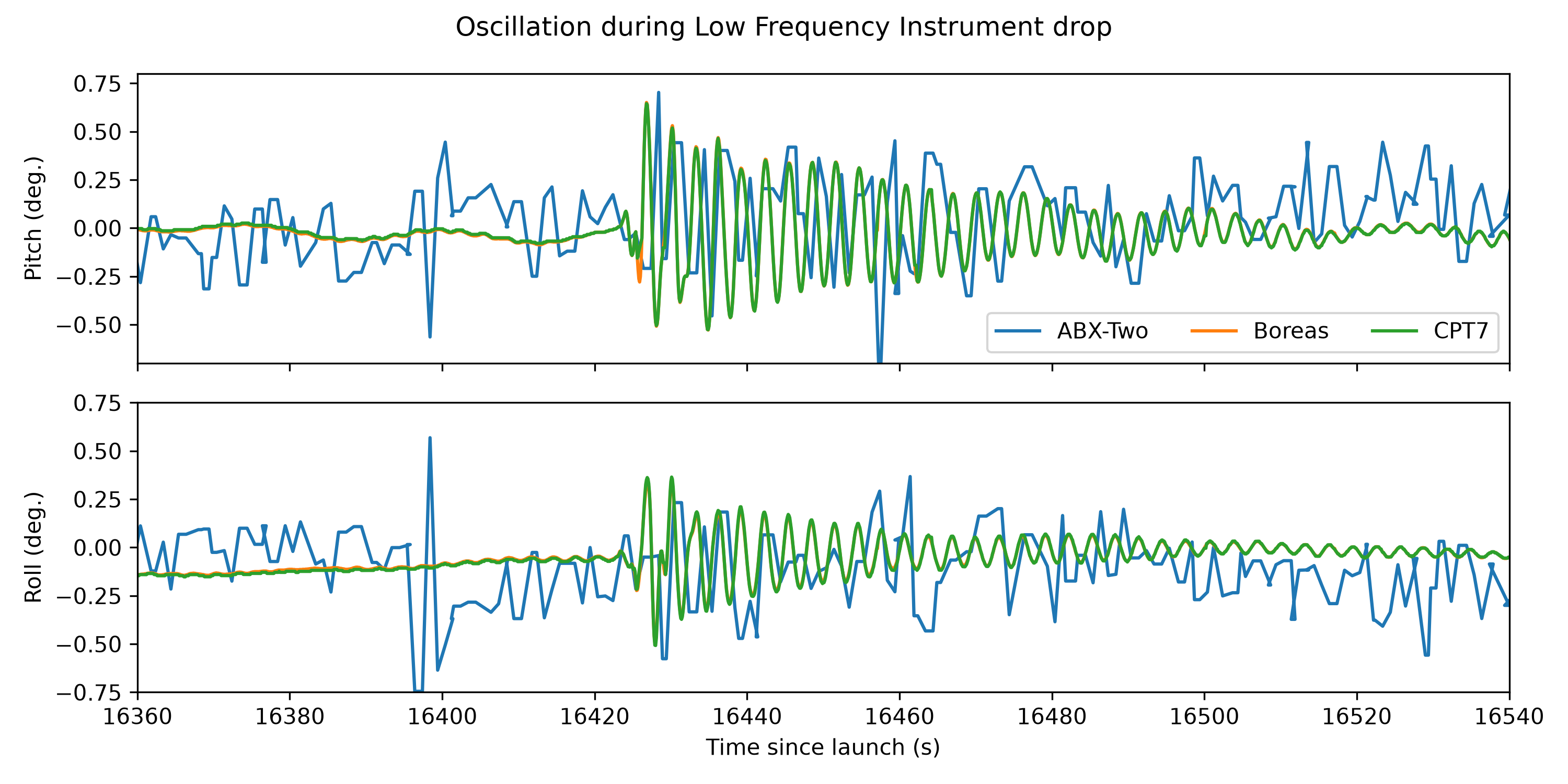}
\caption{The LF drop oscillation as seen by the navigation sensors.}
\label{fig:imu_lf_drop}
\end{figure}

During the first \qty{\sim12}{\hour} after reaching float altitude (about \qty{18}{\hour} after launch), the navigation subsystems performed nominally. Just after reaching float, the low-frequency system was deployed, and pitch and roll oscillation induced by this was captured very clearly as shown in figure \ref{fig:imu_lf_drop}.

\begin{figure}
\includegraphics[width=\linewidth]{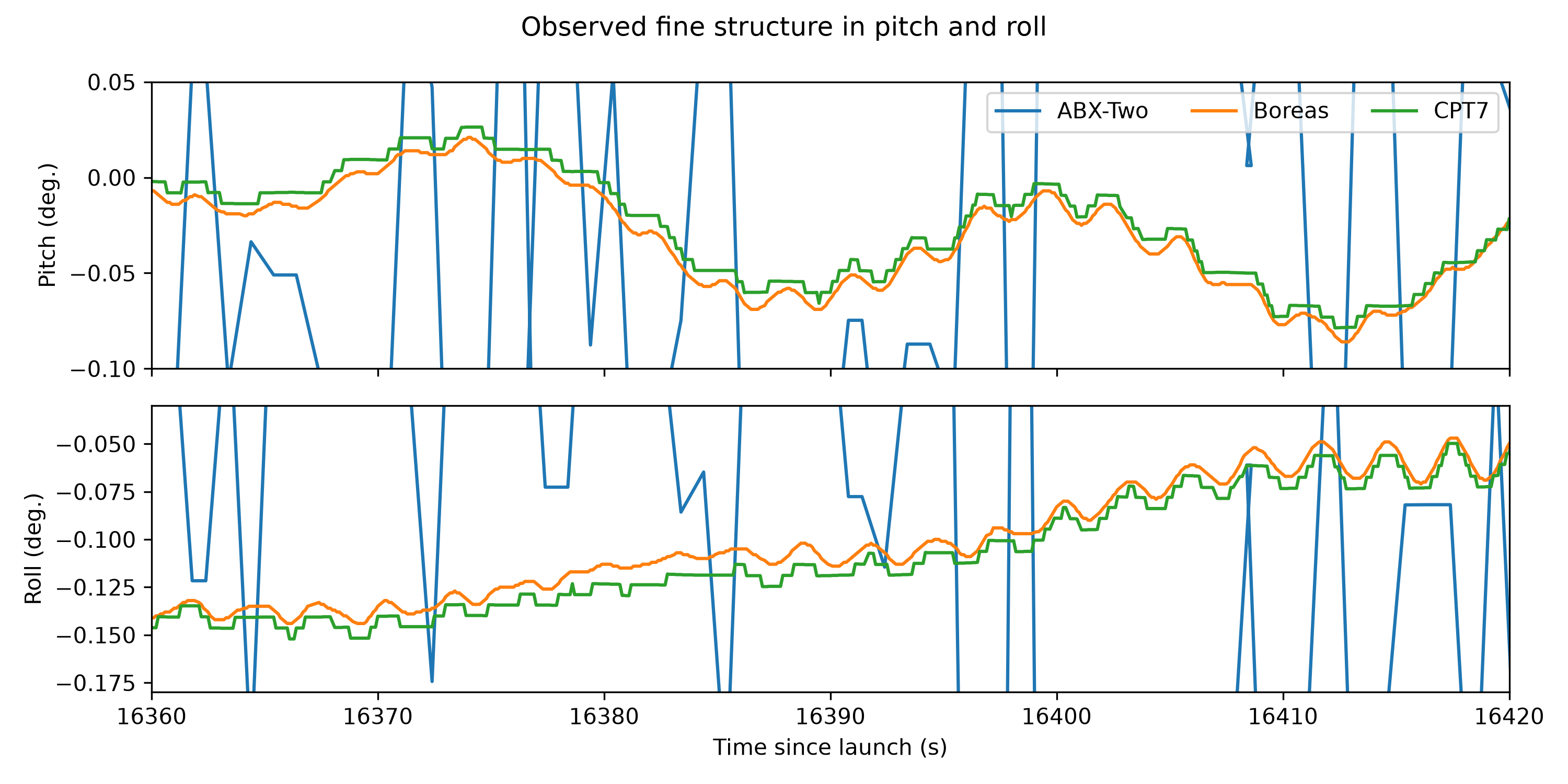}
\caption{Time series data recorded of the attitude fine structure after reaching float altitudes in the pitch (top) and roll (bottom) for three of the navigation subsystems: the ABX-Two, the Boreas, and the CPT7.}
\label{fig:attitude_fine_structure}
\end{figure}

The inertial sensors were also able to capture very high resolution oscillation in pitch and roll after this, perhaps due to the natural flexing of the payload gondola. These have modes with angular deviations on the order of \ang{\sim0.1} and \ang{\sim0.01}, which are lower than the noise in the GNSS solution from the ABX-Two, as seen in figure \ref{fig:attitude_fine_structure}.

During the early periods of flight, instability on the navigation power rail caused system brownouts that forced power resets of several navigation systems; these caused a degradation in their performance, as the preflight calibrations were lost when the units performed new calibrations during boot once they were brought back online.
In the case of the Boreas, there was complete GNSS failure shortly after power instabilities. The CPT7, instead, encountered difficulty achieving a stable attitude solution, likely due to the angular velocity of the payload introducing too high of an error into the startup calibration of the gyroscopes. The CPT7 was seemingly able to achieve a stable solution by running its recalibration routine during a period of low spin, but the exact precision of this is still to be determined.

We can still produce a high-precision solution from the purely GNSS source, the ABX-Two. As a measure of its quality, we compared the solution from the ABX-Two to the solution from the Boreas during the initial period of the flight. We assume their error to be uncorrelated and Gaussian.
This allows us to calculate an upper limit of their standard deviation by calculating the difference between their measurements and assuming $$\Delta\ge\sqrt{\sigma_{ABX-Two}^2 + \sigma_{Boreas}^2}$$.

Due to the very high resolution of the Boreas, we can see that the payload is stable in pitch and roll, within the \ang{0.05} specification. Before the power issues degraded its performance, the standard deviation of pitch was \ang{0.037} and of roll was \ang{0.036}. This informs our ability to reduce the noise in the measured pitch and roll by the ABX-Two through a simple exponential smoothing technique. We use the equation
\begin{equation}
 s_t = \alpha x_t + (1-\alpha)s_{t-1}        
\end{equation}
where $s_t$ is the smoothed value of pitch/roll, $x_t$ is the measured value of pitch/roll, $s_{t-1}$ is the smoothed pitch/roll at the previous time step, and $\alpha$ is the smoothing factor.
We have chosen $\alpha=0.0005$, which for data with a \qty{1}{\hertz} rate corresponds to a time window of about \qty{33}{\minute}.
This means we assume the payload is stable in pitch and roll on a timescale of about \qty{30}{\minute}, and thus this smoothed data is insensitive to any impulsive events. While this is a reasonable assumption, it will be verified by the CPT7 pitch and roll data during the entire flight, which retained its high precision despite losing accuracy. PUEO did not drop any ballast during flight, which would have affected pitch and roll. 

\begin{figure}
\includegraphics[width=\linewidth]{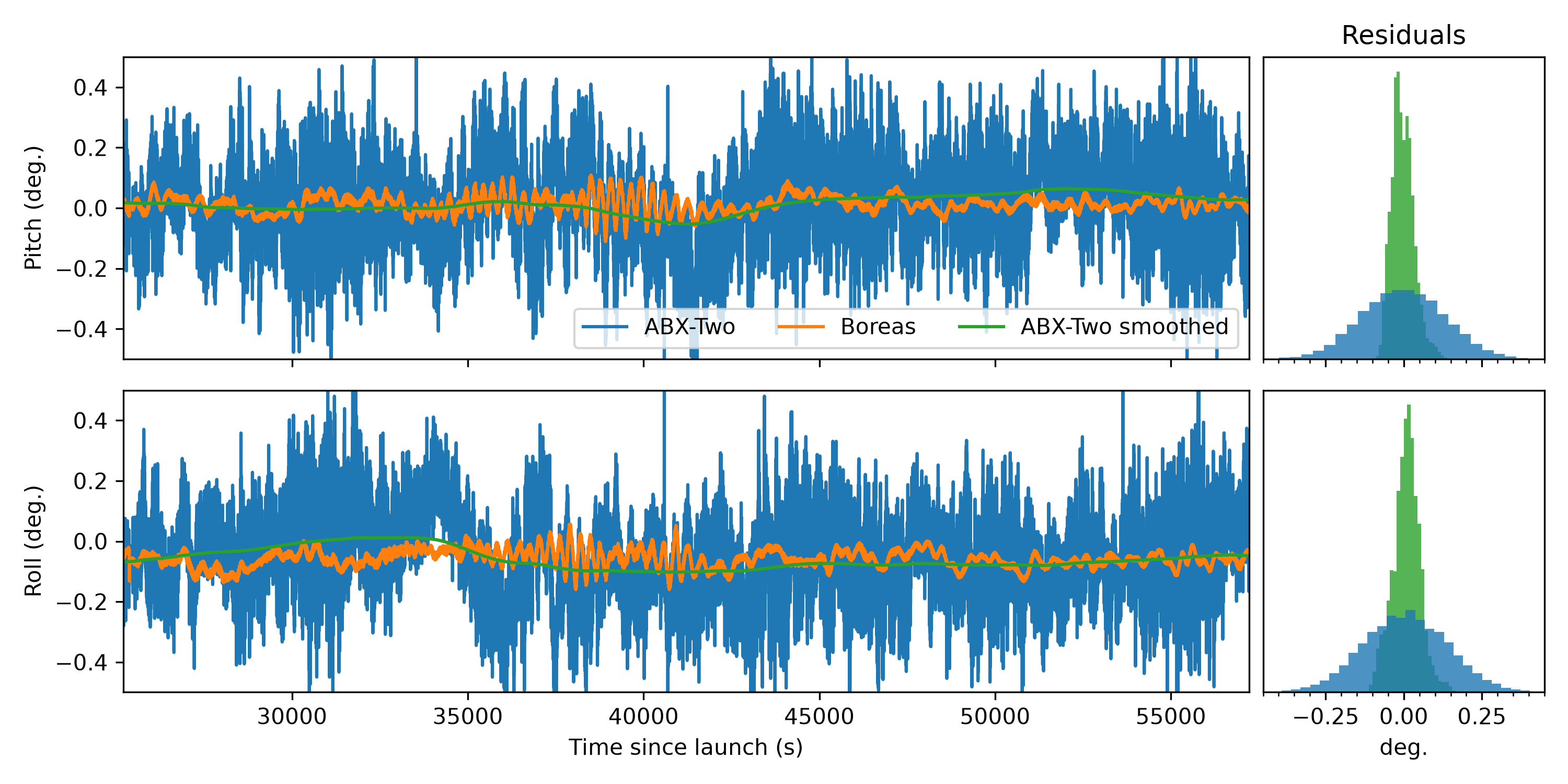}
\caption{The smoothed ABX-Two data compared to the unsmoothed data, as well as the Boreas. On the left, the time series is plotted, and on the right the residual between the ABX-Two data and the Boreas is collapsed into a single histogram. The smoothing significantly improves the residual: for pitch, the standard deviation improves from \ang{0.135} to \ang{0.035}, and for roll from \ang{0.139} to \ang{0.043}. Smoothing preserved the mean difference between ABX-Two and Boreas data of \ang{<0.01}.}
\label{fig:abxtwo_smoothed_comparison}
\end{figure}

The smoothed ABX-Two pitch and roll data are compared to the Boreas in figure \ref{fig:abxtwo_smoothed_comparison}. Based on the small mean difference and $\sigma<$\ang{0.05} for both angles, this is a good representation of the payload tilt. When this smoothing is applied over the entire flight, we find a standard deviation for pitch of \ang{0.015} and for roll of \ang{0.024}. From these values, we infer that the pitch and roll remain stable.

\begin{figure}
\includegraphics[width=\linewidth]{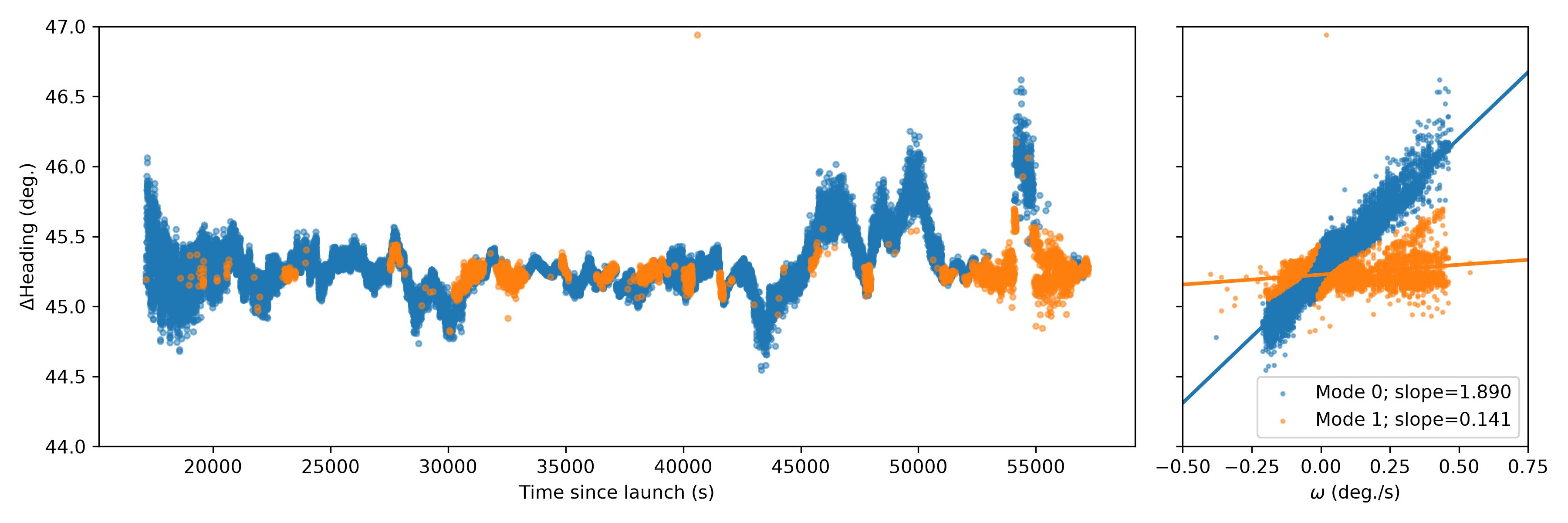}
\caption{The time series data of the difference in heading between the Boreas and ABX-Two shows spin-rate dependence. We have identified two behaviors thus far. About \qty{70}{\percent} of this time period is spent in Mode 0, which has a steeper slope, and \qty{30}{\percent} is spent in the shallower Mode 1.}
\label{fig:abxtwo_spin_dependence}
\end{figure}

Due to the free rotation of the payload, the heading cannot be smoothed in the same way as pitch and roll. To assess the resolution of the ABX-Two heading data, we have compared it to the Boreas during the first period of the flight when all navigation systems were performing nominally. Investigations of the worse-than-expected agreement between the units show what seems to be a spin-rate dependence in the heading value of the ABX-Two. There seem to be (at least) two modes, each being a linear correlation between the spin rate of the payload and the difference between the heading of the ABX-Two and the Boreas. Similar comparisons between the ABX-Two and the CPT7 lead us to believe this is an effect of the ABX-Two, and not either of the IMUs. In Figure \ref{fig:abxtwo_spin_dependence}, we show this correlation.

This behavior may be explained by a timing issue with the data acquisition system for the ABX-Two. This effect would be difficult to disentangle from a systematic spin-rate-dependent error in the ABX-Two.

\begin{figure}
\centering
\includegraphics[width=0.6\linewidth]{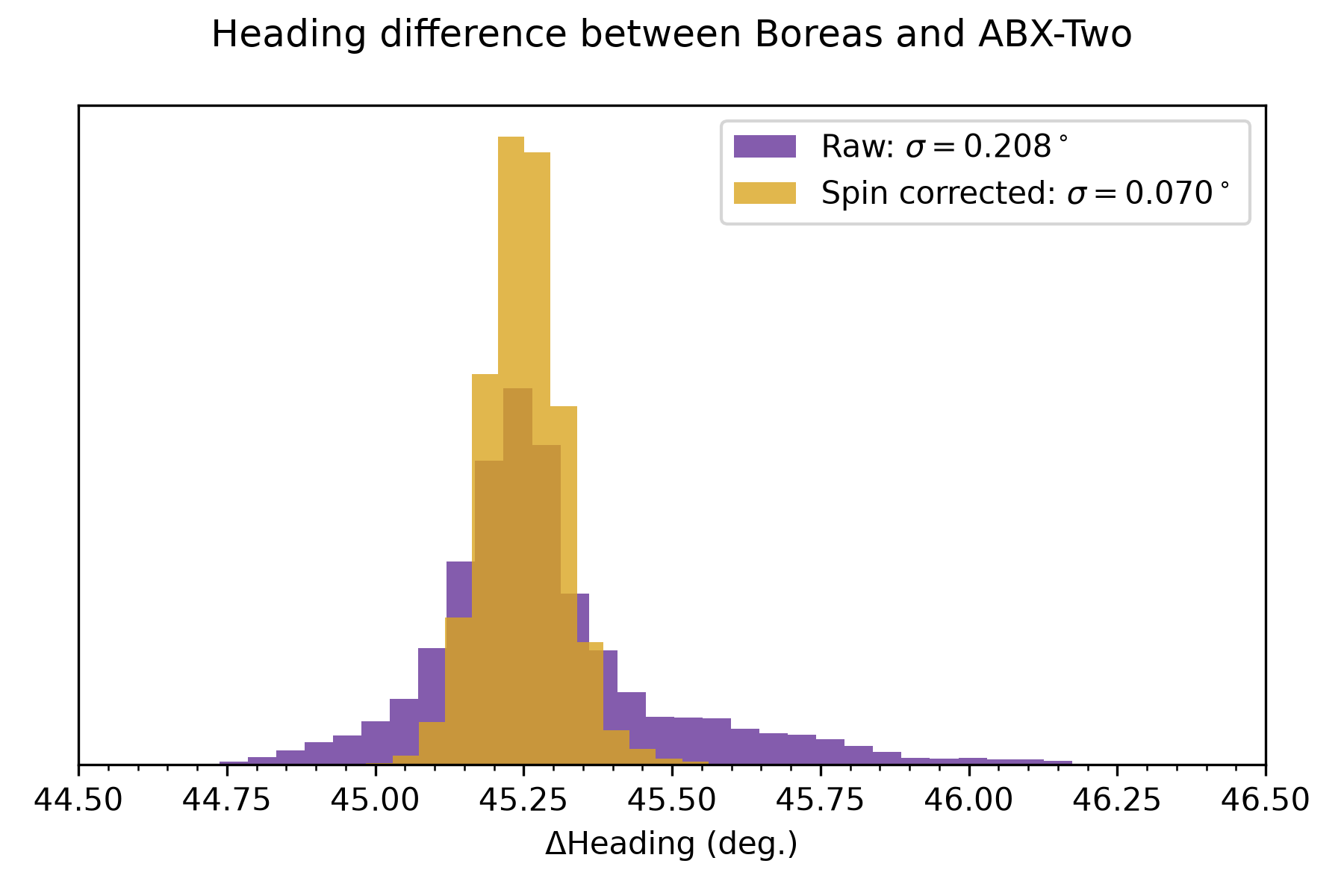}
\caption{The difference in heading between the Boreas and the ABX-Two after applying the spin-dependent correction to ABX-Two heading based on the modes identified. We find the resolution of the heading difference improves to \ang{0.07}.}
\label{fig:dheading_histo}
\end{figure}

Since there is a clear linear structure to these modes, we have corrected for them with a spin-dependent correction of $\text{Heading} = \text{Heading} - \tau \cdot \omega$, where $\tau$ is the appropriate slope based on the mode.
The result of this correction is shown in Figure \ref{fig:dheading_histo}; we achieve a standard deviation of \ang{0.070}, compared to \ang{0.208} before the corrections. 
Assuming the error of the two heading sources adds in quadrature, we can take this value as the upper limit of the error of an individual unit; thus, with this correction, the error of the ABX-Two heading is close to the specification of \ang{0.05}.

\section{Conclusion}
We have presented the design, construction, and flight details for the Payload for Ultrahigh Energy Observations (PUEO). PUEO's instrument design optimized its ability to characterize the ultrahigh-energy neutrino flux through both Askaryan emissions from neutrino-induced cascades in the ice and neutrino-produced tau lepton decays in the atmosphere, while working within the constraints of a NASA long-duration balloon payload. The payload took core design inspiration from the ANITA missions and implemented key improvements including, but not limited to: a more sensitive phased array trigger, smaller antennas allowing for more channels to be included in the payload envelope, the addition of a dedicated low-frequency instrument deployed post-launch, and improved antenna gains and RF front-end electronics. 

PUEO completed its successful 23-day flight above Antarctica in January 2026.  With a livetime of approximately 55\%, it recorded 114,046,211 RF triggered events.

\section*{Acknowledgments}
PUEO is a NASA Pioneers mission funded by NASA grants \#80NSSC20K0775 and \#80NSSC20K0925.
We would like to particularly thank the staff at the Columbia Scientific Balloon Facility, the National Science Foundation, and the Antarctic Support Contractor, whose talented and tireless efforts made PUEO possible.
UChicago additionally thanks the Kavli Institute for Cosmological Physics and the University of Chicago Research Computing Center.
R. Scrandis is supported by NASA FINESST award 80NSSC25K7301.
UCL is supported by the UK's Science and Technology Facilities Council.
D. Besson and D. Seckel thank the NSF for support through the IceCube EPSCoR Initiative (Award ID No. 2019597).
D. Kullgren is supported by an NSF GRFP award DGE1255832.
The Ohio State University thanks the Ohio Supercomputer Center for computational resources.
PUEO data will eventually be made publicly available through the High Energy Astrophysics Science Archive Research Center (HEASARC), a service of the Astrophysics Science Division at NASA/GSFC.

\bibliography{thebib}
\end{document}